\documentclass[superscriptaddress, aps, longbibliography, twocolumn, prx, nofootinbib]{revtex4-1}

\usepackage{graphicx}
\usepackage{bm}
\usepackage{hyperref}
\hypersetup{colorlinks=true,citecolor=blue, linkcolor=blue}
\usepackage{physics}
\usepackage{xcolor}
\usepackage{enumitem}
\usepackage{amsmath, amssymb}
\usepackage[normalem]{ulem}
\usepackage{natbib}
\usepackage{wasysym}
\usepackage{mathrsfs}

\newcommand{\avgP}{\ensuremath{\mathcal{P}}}
\newcommand{\avgPthree}{\ensuremath{\mathcal{P}^{(3)}}}

\newcommand{\avgshadownorm}[1]{\ensuremath{\| #1 \|_{\rm sh, avg}}}
\newcommand{\shdn}[1]{\ensuremath{\| #1 \|_{\rm sh}}}

\newcommand{\eye}{\ensuremath{\mathbb{I}}}
\newcommand{\kraus}[1]{K_{#1}} 
\newcommand{\effect}[1]{E_{#1}} 
\newcommand{\mb}[1]{\boldsymbol{#1}} 
\newcommand{\dket}[1]{\ensuremath{|#1\rangle\!\rangle}}
\newcommand{\dbra}[1]{\ensuremath{\langle\!\langle #1|}}

\begin{document}

\title{Learnability transitions in monitored quantum dynamics \\ via eavesdropper's classical shadows
}

\author{Matteo Ippoliti}
\affiliation{Department of Physics, University of Texas at Austin, Austin, TX 78712, USA}
\affiliation{Department of Physics, Stanford University, Stanford, CA 94305, USA}
\author{Vedika Khemani}
\affiliation{Department of Physics, Stanford University, Stanford, CA 94305, USA}

\begin{abstract}
Monitored quantum dynamics---unitary evolution interspersed with measurements---has recently emerged as a rich domain for phase structure in quantum many-body systems away from equilibrium.
Here we study monitored dynamics from the point of view of an eavesdropper who has access to the classical measurement outcomes, but not to the quantum many-body system.
We show that a measure of information flow from the quantum system to the classical measurement record---the {\it informational power}---undergoes a phase transition in correspondence with the measurement-induced phase transition (MIPT). This transition determines the eavesdropper's (in)ability to learn properties of an unknown initial quantum state of the system, given a complete classical description of the monitored dynamics and arbitrary classical computational resources. 
We make this learnability transition concrete by defining classical shadows protocols that the eavesdropper may apply to this problem, and show that the MIPT manifests as a transition in the sample complexity of various shadow estimation tasks, which become harder in the low-measurement phase. We focus on three applications of interest: Pauli expectation values (where we find the MIPT appears as a point of optimal learnability for typical Pauli operators), many-body fidelity, and global charge in $U(1)$-symmetric dynamics. 
Our work unifies different manifestations of the MIPT under the umbrella of {\it learnability} and gives this notion a general operational meaning via classical shadows.
\end{abstract}

\maketitle

\section{Introduction \label{sec:intro} }

Recent advances in our ability to address and read out individual degrees of freedom in many-body quantum systems have motivated interest in new types of dynamics where the role of the observer is central. In these {\it monitored dynamics}~\cite{skinner_measurement-induced_2019,li_quantum_2018, li_measurement-driven_2019, potter_entanglement_2022, fisher_random_2022} the observer's measurements shape the evolution of the system and drive it to sharply different possible ensembles of late-time states. These `measurement-induced phase transitions' (MIPTs) thus define a new paradigm for phase structure in open systems away from equilibrium.
At the same time, these technological developments have raised the salience of {\it quantum state learning}---the general problem of characterizing properties of unknown, potentially complex quantum states with as few measurements as possible~\cite{huang_predicting_2020, elben_randomized_2023}. 
In this work we connect these two threads by formulating MIPTs as learnability transitions within the framework of {\it classical shadows}, a leading practical approach to state learning. 

The canonical formulation of the MIPT is in terms of a phase transition in the entanglement properties of ensembles of quantum trajectories~\cite{skinner_measurement-induced_2019, li_quantum_2018, li_measurement-driven_2019}. In the standard setup, the system evolves through random circuit dynamics composed of local unitary gates interrupted by local projective measurements with probability $p$. As the measurement rate $p$ is tuned, the ensemble of late-time trajectories undergoes a phase transition from a {\it disentangling} phase in which trajectories display area-law entanglement (at high $p$, corresponding to frequent measurements) to an {\it entangling} phase in which trajectories display volume-law entanglement (at low $p$, corresponding to infrequent measurements)\footnote{
Models of dynamics with additional structure, e.g. fermionic Gaussian systems~\cite{cao_entanglement_2019,fidkowski_how_2021,fava_nonlinear_2023} or systems that obey a symmetry~\cite{bao_symmetry_2021,agrawal_entanglement_2022,majidy_critical_2023}, may exhibit different phenomenology and richer phase diagrams. Here we focus on the more generic scenario of non-Gaussian dynamics without symmetries.}.
While the $p=0$ and $p=1$ limits are transparent, the existence of a robust volume-law phase at any finite measurement rate is, {\it a priori}, surprising. While local unitary gates can only generate entanglement at the boundaries of a subsystem, disentangling measurements act everywhere in the bulk: a (na\"ively) imbalanced competition which should always favor the area-law phase. A key insight for understanding the stability of the volume law phase was furnished in Refs.~\cite{choi_quantum_2020, gullans_dynamical_2020}, which posited that the volume-law phase can be understood as a dynamically generated random code in which the correlations between two subsystems are hidden in highly non-local degrees of freedom, inaccessible to local measurements. 
This naturally leads to two complementary information-theoretic perspectives on the MIPT: (a) the \emph{coding} perspective and (b) the \emph{learning} perspective, discussed below. The latter is the focus of this work.

\vspace{-10pt}
\subsection*{Coding}
\begin{figure*}
    \centering
    \includegraphics[width=\textwidth]{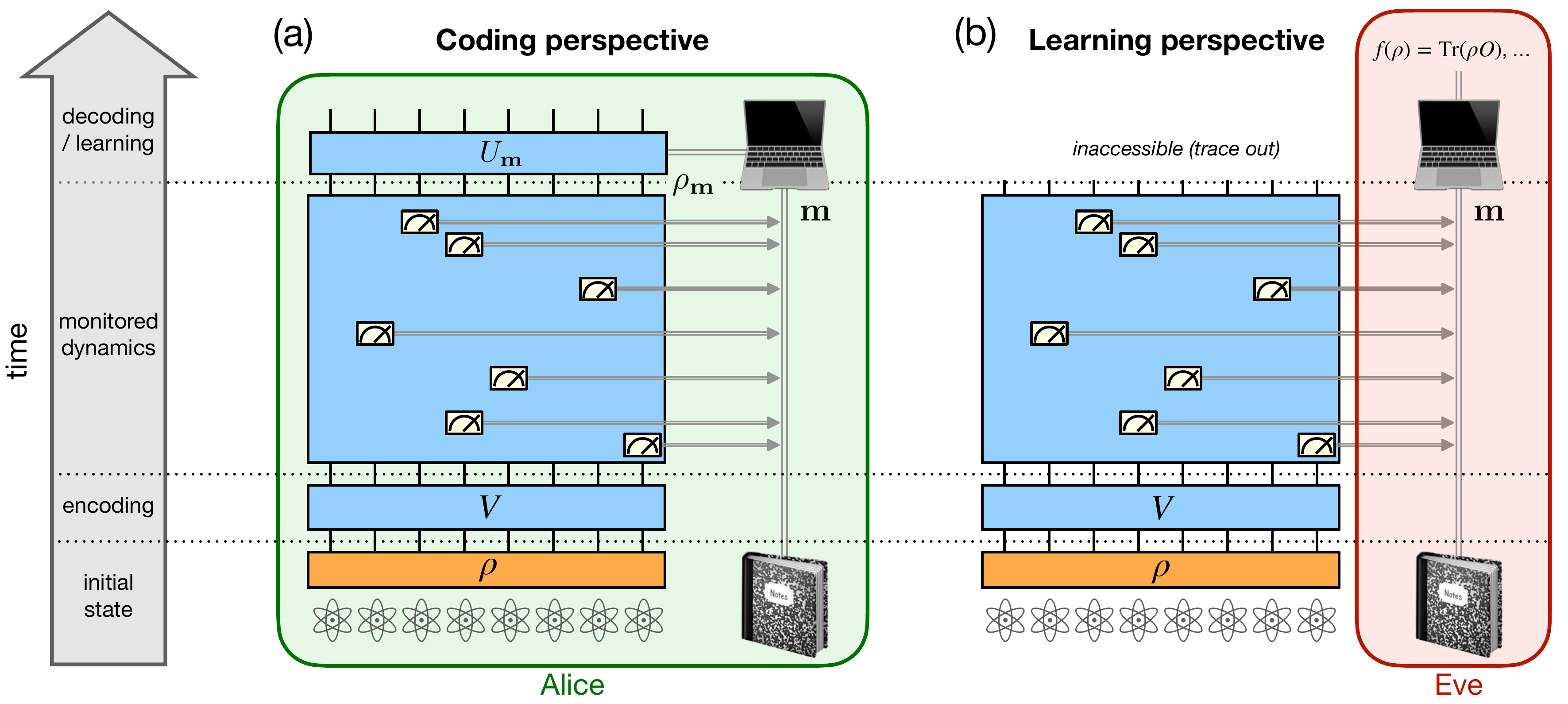}
    \caption{Different perspectives on measurement-induced phases of quantum information.
    (a) {\bf Coding perspective}. A state $\rho$ of a quantum many-body system is subject to monitored dynamics, possibly after an encoding or ``pre-scrambling'' step (the global unitary $V$). The measurement record $\mb m$ is stored in a classical system, here represented by the lab notebook. The experimentalist, Alice, has access to both the quantum and classical systems (green shaded box). After the dynamics, through classical computation conditioned on the measurement record $\mb m$, Alice can in principle find a recovery operation $U_{\mb m}$ and apply it to the quantum system. There is a phase transition in how many qubits from the initial state $\rho$ can be recovered in this way, i.e. in the {\it coding} properties of monitored dynamics.
    (b) {\bf Learning perspective}. An eavesdropper, Eve, has access only to the classical system (red shaded box). She attempts to learn properties of $\rho$ by estimating functions $f(\rho)$ (e.g. expectation values $\Tr(\rho O)$) via classical shadows. There is a phase transition in the sample complexity of these estimation tasks---i.e., in the {\it learnability} of the state $\rho$ from the measured data $\mb m$.
    The two perspectives (a,b) are dual to each other and the transitions coincide.
    }
    \label{fig:idea}
\end{figure*}

\vspace{-6pt}
The coding perspective is primarily understood from the point of view of an experimentalist, Alice, controlling a quantum system, see Fig.~\ref{fig:idea}(a). Over the course of the dynamics, measurements are performed on the system (either by Alice herself or by a particular type of ``environment'' that broadcasts the measurement outcomes). These measurements disturb the initial state of the system. In order to undo this disturbance as much as possible, Alice can perform `recovery' operations on the combined final state of the quantum system and classical measurement apparatus---concretely, she decodes the measurement record $\mb m$ (which is a binary string of measurement outcomes indexed by spacetime locations) to decide on a unitary operation $U_{\mb m}$ to apply to the quantum state. 
As a function of parameters in the monitored dynamics (typically the space-time density of measurements set by $p$), Alice's ability to recover her initial state undergoes a phase transition: on the entangling side, she can successfully recover an extensive amount of quantum information; on the disentangling side, only a subextensive amount\footnote{This setup can equivalently be formulated in terms of Alice's ability to reconstruct a message sent to her by Bob over a noisy quantum channel.}. 

We refer to this point of view as the {\it coding} perspective on the MIPT~\cite{gullans_dynamical_2020,choi_quantum_2020}, due to its close analogy with quantum error correction (QEC)~\cite{shor_scheme_1995,gottesman_stabilizer_1997,dennis_topological_2002}. The measurement record $\mb m$ serves as the ``syndrome''\footnote{More accurately, in this setting the measurements play a dual role---both as the ``errors" that disturb the encoded information, and as the syndromes that allow for its in-principle recovery.} and the conditional unitary $U_{\mb m}$ as the correction / recovery operation; the MIPT arises as a phase transition in the rate of this code, i.e. the ratio of logical qubits to physical qubits, which goes from finite to vanishing. 
In other words, in the entangling phase an {\it extensive amount} of quantum information survives in the combined quantum-classical state of system and measurement record. While its recovery may be practically hard (in terms of classical computation and quantum circuit complexity), its {\it in-principle} presence or absence defines sharp phases. This corresponds to a phase transition in the capacity of the channel that maps the initial quantum state to the combined quantum-classical post-measurement state~\cite{choi_quantum_2020}. This channel capacity also corresponds to the trajectory-averaged entropy of mixed states subject to the monitored evolution, so the coding transition is equally described as a {\it dynamical purification} transition~\cite{gullans_dynamical_2020}. 
Further, the idea of {\it decoding} implicit in this setup has led to groundbreaking developments in our experimental understanding of the MIPT~\cite{gullans_scalable_2020, noel_measurement-induced_2022, hoke_measurement-induced_2023}.

\subsection*{Learning}

In this work, we take a complementary perspective of {\it learning} rather than {\it coding}, i.e. we focus on the information transmitted to the classical measurement record alone, rather than the combined quantum-classical state. This perspective is centered on an ``eavesdropper'', Eve, who does not have access to the quantum many-body system, but wants to learn some properties (e.g. observable expectation values) of its unknown initial state. She may try to do so by collecting classical measurement outcomes and performing suitable computation on them, Fig.~\ref{fig:idea}(b).  

This perspective has  been studied  in the literature in two contexts. The first studies the sensitivity of the distribution of measurement outcomes to changes in the initial state\footnote{These are diagnosed either through the Fisher information of the measurement record~\cite{bao_theory_2020}, or through a  linear cross-entropy diagnostic that compares the measurement record from a quantum experiment to that of a classical simulation of the same circuit with a different input state~\cite{li_cross_2023}.}. In the second context, the monitored dynamics is enriched with a $U(1)$ symmetry, and an eavesdropper attempts to learn the global charge of a system from measurements of local charge densities~\cite{barratt_transitions_2022}. In both cases, the focus is only on the information present in the classical measurement record, as the post-measurement quantum many-body state is considered inaccessible. 
This point of view is complementary to the coding perspective.
An intuitive expectation is that, if coding is successful, then the ``syndrome'' measurements $\mb m$ should reveal no information about the initial state, and thus learning should fail; conversely, if coding fails, information ``leaks'' into the classical measurement record and learning should become possible.

In this work we sharpen this intuition and make it operationally meaningful. 
We view the monitored dynamics as a single, complex ``randomized measurement''~\cite{elben_randomized_2023} performed on the system. From Eve's point of view, this generalized measurement destroys the quantum state and turns it into classical data.
As such, it is impossible for her to recover {\it quantum} information, as in coding (e.g., any entanglement with an outside reference system is destroyed in the process); but it may still be possible to learn a classical description of the state, as in tomography. 

We introduce {\it classical shadows} protocols~\cite{huang_predicting_2020} that Eve may use to learn various properties of the unknown initial state from the outcomes of many shots of these generalized measurements. We then show that the MIPT manifests as a transition in the sample complexity of these tasks, i.e. how many shots of the experiments Eve needs before she can confidently make predictions. The transition in sample complexity may be from polynomial to exponential, between two exponentials, or between two polynomials, depending on the task at hand. Our framework is very general and furnishes a unified language to describe and study learnability transitions in various different contexts (including previous examples from the literature such as  charge learning~\cite{barratt_transitions_2022}). Finally, we argue, and prove in some cases, that these learnability transitions reflect a transition in the {\it informational power}~\cite{dallarno_informational_2011} of monitored dynamics---an intrinsic property independent of the chosen learning protocol.

The balance of this paper is structured as follows. 
In Sec.~\ref{sec:review} we provide a concise, pedagogical review of relevant background topics: generalized measurements, monitored dynamics and classical shadows. Expert readers may safely skip this section.
We then discuss monitored dynamics as a generalized measurement in Sec.~\ref{sec:infopower}.
We review the idea of informational power~\cite{dallarno_informational_2011} of generalized measurements and apply it to monitored dynamics, proving under certain assumptions (and conjecturing more generally) that it undergoes a phase transition at the MIPT. 
In Sec.~\ref{sec:shadows} we introduce {\it ``eavesdropper's shadows''}---classical shadows protocols that an eavesdropper may use to learn the state of the system from measurement records. The consequences of the MIPT on these classical shadows protocols are then analyzed in turn: 
Sec.~\ref{sec:pauli} on the expectation value of Pauli operators, where we additionally investigate the effect of spatial locality on learnability; Sec.~\ref{sec:xeb} on many-body fidelities, which directly connects to recent work on the linear cross-entropy as an order parameter for the transition~\cite{li_cross_2023}; and Sec.~\ref{sec:charge} on learning properties of the charge distribution via $U(1)$-symmetric monitored dynamics~\cite{agrawal_entanglement_2022,barratt_transitions_2022},  
which can also be studied naturally in the formalism of eavesdropper's classical shadows. 
As we do not impose limits on classical computational resources, we find that the learnability transition coincides with the {\it charge-sharpening} transition in Ref.~\cite{agrawal_entanglement_2022}, as expected~\cite{barratt_transitions_2022}.
Finally, in Sec.~\ref{sec:discussion} we summarize our results and point out directions for future work.

\section{Review \label{sec:review}}

In this Section we review essential background topics---generalized measurements (POVMs), monitored dynamics, and classical shadows---for the sake of a self-contained discussion. This part may be skipped by expert readers.

\subsection{Generalized measurements \label{sec:review_povm}}

Generalized measurements in quantum mechanics are described by positive-operator-valued measures (POVMs), sets of operators $\{\effect{\alpha}\}$ (also known as {\it effects}) that obey 
(i) positivity, $\effect{\alpha} \geq 0$ and 
(ii) normalization $\sum_\alpha \effect{\alpha} = \eye$.
These predict the probability of observing each outcome $\alpha$ on a given state $\rho$ by the identification $p_\alpha \equiv \Tr(\rho \effect{\alpha})$, which is guaranteed to be a valid probability distribution by conditions (i-ii). 
A POVM describes measurement outcomes but not the associated post-measurement states of the quantum system. That additional data is contained in the {\it instruments} $\{\kraus{\alpha}\}$ (also known as Kraus operators in the context of quantum channels), that obey $\effect{\alpha} = \kraus{\alpha}^\dagger \kraus{\alpha}$. The state $\rho$ is updated after the measurement as 
\begin{equation}
\rho \mapsto\sum_\alpha p_\alpha \rho_{\alpha},
\qquad \rho_{\alpha} = \kraus{\alpha} \rho \kraus{\alpha}^\dagger/p_\alpha,\end{equation}
where $\rho_{\alpha}$ is the conditional post-measurement state of the system given outcome $\alpha$. 

It may further be helpful to view the whole measurement process as a quantum-classical channel
\begin{equation}
\rho \mapsto \sum_\alpha p_\alpha \rho_{\alpha} \otimes \ketbra{\alpha}_C \label{eq:post_measurement_state}
\end{equation}
where the states $\ketbra{\alpha}_C$ are states of a {\it classical} register, e.g. Alice's lab notebook.
We can always embed such states as orthonormal basis elements\footnote{They need to be orthogonal as they are perfectly distinguishable (classical) states.} of a sufficiently large Hilbert space. 

\subsection{Monitored dynamics \label{sec:review_mipt}}

Monitored dynamics is a type of open-system evolution whose quantum trajectories are labeled by a classical ``measurement record'' $\mb m$. We take $\mb m = (m_{t_1, x_1}, \dots m_{t_M, x_M}) \in \{0,1\}^M$ to be a collection of binary measurement outcomes gathered over different positions and times $(t_i, x_i)$ in the evolution. 
The quantum many-body state $\rho$ evolves into a quantum-classical state 
\begin{equation}
\rho \mapsto \sum_{\mb m} \kraus{\mb m} \rho \kraus{\mb m}^\dagger \otimes \ketbra{\mb m}_C, 
\label{eq:quantumclassical_channel}
\end{equation}
which notably is of the same form as \eqref{eq:post_measurement_state}:
we can in fact view the whole monitored evolution as a single POVM with $2^M$ possible outcomes $\mb m \in \{0,1\}^M$ occurring with probabilities $p_{\mb m} = \Tr(\rho \effect{\mb m})$, $\effect{\mb m} = \kraus{\mb m}^\dagger \kraus{\mb m}$.

Remarkably, it was discovered that the ensemble of quantum trajectories $\rho_{\mb m}\equiv \kraus{\mb m} \rho \kraus{\mb m}^\dagger / p_{\mb m}$ can undergo a sharp phase transition as a function of model parameters, e.g. the density or rate of measurements in the dynamics, from an ``entangling'' to a ``disentangling'' phase~\cite{potter_entanglement_2022,fisher_random_2022,skinner_measurement-induced_2019,li_quantum_2018,li_measurement-driven_2019,choi_quantum_2020,bao_theory_2020,gullans_dynamical_2020,gullans_scalable_2020,ippoliti_entanglement_2021,lavasani_measurement-induced_2021,nahum_measurement_2021,li_statistical_2021,fan_self-organized_2021,feng_measurement-induced_2023,koh_measurement-induced_2023,noel_measurement-induced_2022,hoke_measurement-induced_2023}.
The phenomenology of these phases is very rich and beyond the scope of this review section. Here we focus only on the aspect most relevant to this work, which is {\it dynamical purification}~\cite{gullans_dynamical_2020}, closely related to the coding perspective discussed above: due to the non-unitarity of measurements, an initially-mixed state at late enough times generically becomes nearly pure, $\mathbb{E}_{\mb m}[ \Tr(\rho_{\mb m}^2)] \xrightarrow{t\to\infty} 1$. Here the average is taken over the ensemble of trajectories with Born probability $p_{\mb m} = \Tr(\rho \effect{\mb m})$. 
The time scale over which this happens varies sharply depending on which phase we are in: it is $O(\log(N))$ in the disentangling/pure phase, and $O(\exp(N))$ in the entangling/mixed phase, where $N$ is the size of the system (e.g. number of qubits). 
This reflects the emergence of a quantum code in the entangling phase, which protects some information and prevents it from leaking to the environment for very long times. In particular, taking a dynamic limit with $t,N\to\infty$ with $t = \Theta(N)$ ensures that the average purity becomes an order parameter for the two phases (1 in the disentangling phase, $<1$ in the entangling phase). 

A subtle experimental aspect of this physics is that it is revealed only in nonlinear functions of the trajectories $\{\rho_{\mb m}\}$, such as the aforementioned average purity $\mathbb{E}_{\mb m} \Tr(\rho_{\mb m}^2)$. It does not appear in the average of linear functions, which are fully determined by the average final density matrix $\rho' = \mathbb{E}_{\mb m}[\rho_{\mb m}] = \sum_{\mb m} \kraus{\mb m} \rho \kraus{\mb m}^\dagger$. 
This is the output of {\it dissipative} dynamics and thus generically trivial across the phase diagram. A na\"ive approach based on measuring properties of individual trajectories incurs an exponential sampling overhead, $\sim 2^M$, due to postselection of the $M$ measurement outcomes (producing the same trajectory twice takes of order $2^M$ trials). 

More sophisticated approaches that ameliorate or avoid this prohibitive sampling cost have been proposed~\cite{ippoliti_postselection-free_2021, ippoliti_fractal_2022, lu_spacetime_2021, gullans_scalable_2020} and implemented experimentally~\cite{noel_measurement-induced_2022,hoke_measurement-induced_2023}. 
Ref.~\cite{gullans_scalable_2020} proposed scalable order parameters that can be measured by {\it decoding} the measurement record and applying feedback control to the quantum system, similar to the setup in Fig.~\ref{fig:idea}(a). 
Building on this, other approaches have more recently been proposed that bypass the need for quantum feedback by defining ``hybrid'' quantum-classical order parameters that correlate quantum state read-out with the output of classical simulations~\cite{dehghani_neural-network_2023,garratt_measurements_2023,li_cross_2023,hoke_measurement-induced_2023,garratt_probing_2023}. These approaches generally trade the exponential sample complexity $\sim 2^M$ for the complexity of classical simulation, which may be polynomial or exponential depending on the models. We will show below that our work furnishes a new set of hybrid order parameters for the MIPT, in the form of variances of shadow estimators.  We also note that another approach making use of classical shadows to diagnose the MIPT was proposed recently~\cite{garratt_probing_2023}, with the complementary goal of learning the final, {\it post-measurement} state. The goal is thus conceptually distinct from our learnability perspective, despite employing similar tools---a testament to the generality and wide applicability of the classical shadows framework, reviewed next.

\subsection{Classical shadows \label{sec:review_shadows}}

Classical shadows are a framework for learning properties of quantum states in a relatively sample-efficient manner by making use of randomized measurements~\cite{huang_predicting_2020,elben_randomized_2023,huang_learning_2022,struchalin_experimental_2021,chen_robust_2021,koh_classical_2022,huang_efficient_2021,acharya_shadow_2021,nguyen_optimizing_2022,wan_matchgate_2023,hu_classical_2023,akhtar_scalable_2023,bertoni_shallow_2022,arienzo_closed-form_2022,ippoliti_operator_2023,ippoliti_classical_2023,tran_measuring_2023,shivam_classical_2023,mcginley_shadow_2023}. 
In this work we will make use of classical shadows in order to formulate a  general framework that operationalizes the notion of ``learnability'' in  monitored dynamics. 

The standard shadows protocol~\cite{huang_predicting_2020,elben_randomized_2023} attempts to learn properties of an unknown quantum state $\rho$ by making  measurements in a complete orthonormal basis, e.g. the computational basis after a suitable random rotation $U$. From such a measurement one obtains a snapshot $\ketbra{\mb b}$ of the rotated state $U\rho U^\dagger$, and thus a snapshot $U^\dagger \ketbra{\mb b} U$ of the original state $\rho$ (here $\mb b \in \{0,1\}^N$ is an output bitstring).
Averaging many such snapshots over outcomes of $\mb b$ and random choices of $U$ yields a ``noisy'' version of $\rho$:
\begin{equation}
\mathcal{M}(\rho) = \sum_{\mb b} \int {\rm d}U\, \bra{\mb b}U\rho U^\dagger\ket{\mb b} U^\dagger \ketbra{\mb b} U,
\label{eq:vanilla_shadow_channel}
\end{equation}
where $\bra{\mb b}U\rho U^\dagger\ket{\mb b}$ is the probability of obtaining the measurement outcomes $\mb b$. 
Here ${\rm d}U$ is shorthand for whichever measure on the unitary group we are using (whether continuous or discrete) and $\mathcal{M}$ is known as the {\it shadow channel}. 
Therefore by ``de-noising'' our snapshots we obtain unbiased estimators of the true density matrix:
\begin{equation}
\hat{\rho} = \mathcal{M}^{-1}[U^\dagger \ketbra{\mb b} U].
\label{eq:inverted_snapshot}
\end{equation}
The classical simulation complexity of the protocol is determined by the complexity of obtaining the un-rotated snapshot $U^\dagger \ketbra{\mb b} U$, and of determining and applying the inverse channel $\mathcal{M}^{-1}$ on a classical computer.

The snapshots in Eq.~\eqref{eq:inverted_snapshot} can then be used to predict many properties of $\rho$; e.g. for linear expectation values $\langle O\rangle$, the estimators $\hat{o} = \Tr(\hat{\rho} O)$ average to the correct answer. 
The cost of the learning protocol is quantified by the number of samples that are needed to make an accurate prediction. This is dictated by the {\it shadow norm} $\shdn{O}^2$ (which controls the variance of $\hat{o}$), whose scaling depends on the chosen ensemble of measurements. 
Important results are known for the standard protocols based on local random Pauli (i.e. locally-randomized) and random Clifford (i.e. globally-randomized) measurements. In the former one has $\shdn{P}^2 = 3^k$ where $P$ is a Pauli operator of weight $k$, making the scheme practical for local (few-body) operators. In the latter, $\shdn{O}^2 = \Tr(O^\dagger O)$ for any operator $O$, making the method suitable for low-rank operators such as pure-state projectors. This has the important application of computing many-body fidelity with pure states. The shadows protocol has recently been generalized to {\it shallow shadows}~\cite{akhtar_scalable_2023,bertoni_shallow_2022,arienzo_closed-form_2022,ippoliti_operator_2023} in which the randomization step is effected by finite-depth circuits, and which yields an exponential gain in sampling complexity for the task of learning large, spatially contiguous $k$ body operators.

Classical shadows have been recently extended to the case of generalized (i.e. non-projective) measurements, represented by a POVM $\{\effect{\alpha}\}$~\cite{acharya_shadow_2021,nguyen_optimizing_2022} Given a measured outcome $\alpha$, there are several possibilities for what to use as a ``snapshot'', giving rise to a family of shadow channels which generalize Eq.~\eqref{eq:vanilla_shadow_channel} and take the form
\begin{equation}
\mathcal{M}(\rho) = \sum_\alpha \int{\rm d}U\, \Tr(\rho \effect{\alpha}^U) \eta_\alpha^U.
\label{eq:generalized_shadowchannel}
\end{equation}
Here we use the superscript $U$ to denote that the effect $\effect{\alpha}$ and the snapshot $\eta_\alpha$ both in general depend on random unitary rotations that are part of the protocol, again represented by integration over ${\rm d}U$, but we do not assume a specific form. We will suppress this dependence on random unitaries from our notation in the following.
In Sec.~\ref{sec:shadows} we will see three different choices for $\eta_\alpha$ that are well-motivated by conceptual or practical considerations.

\section{Informational power of monitored dynamics \label{sec:infopower}}

Taking the {\it learning} perspective illustrated in Fig.~\ref{fig:idea}(b), monitored dynamics as a whole is effectively a generalized measurement on the system---a process that maps the quantum state $\rho$ to a probability distribution over outcomes $\mb m$. 
In particular, if $\kraus{\mb m}$ is the evolution operator corresponding to trajectory $\mb m$, then the process is described by a POVM $\{\effect{\mb m} \equiv \kraus{\mb m}^\dagger \kraus{\mb m} \}$ which maps the quantum state $\rho$ to the classical probability distribution $\{ \Tr(\rho\effect{\mb m})\} $. 
The question of learning thus boils down to the ``strength'' of this generalized measurement, or the information content of its outcomes.

To sharpen this notion, let us consider an ensemble of states $\mathcal{E} = \{(p_i,\rho_i)\}$. This is a discrete collection of states $\rho_i$, each one occurring with probability $p_i$, e.g. from some classical stochastic process involved in the state preparation.
We want to know how well a POVM $\Pi = \{\effect{\mb m}\}$ can distinguish the different states $\rho_i$ in the ensemble $\mathcal{E}$. 

If the state $\rho_i$ was drawn, then the outcome $\mb m$ occurs with probability $p_{\mb m|i} \equiv \Tr(\rho_i \effect{\mb m})$. Along with $p_i$ (given as part of the definition of $\mathcal{E}$), this defines the joint distribution
\begin{equation}
    p_{i,\mb m} = p_{\mb m|i}p_i = \Tr(p_i\rho_i \effect{\mb m}) \label{eq:joint_dist}
\end{equation}
and thus also the marginal $p_{\mb m} = \Tr(\rho\effect{\mb m})$, with $\rho = \sum_i p_i \rho_i$ the average state of the ensemble.
With this data, we can define the {\it mutual information} between the POVM $\Pi$ and the state ensemble $\mathcal{E}$ as the mutual information between variables $i$ and $\mb m$ in the joint distribution Eq.~\eqref{eq:joint_dist}:
\begin{align}
    I(\mathcal{E}:\Pi) &= - \sum_{i} p_{i} \ln p_{i} - \sum_{\mb m} p_{\mb m} \ln p_{\mb m} + \sum_{i,\mb m} p_{i,\mb m} \ln p_{i,\mb m} \nonumber\\
    &= \sum_{i,\mb m} p_{i,\mb m} \ln \frac{p_{i,\mb m}}{p_i p_{\mb m}} 
    \label{eq:MI_povm}
\end{align}
where the expression in the second line is in the form of a Kullback-Leibler divergence between the true distribution $p_{i,\mb m}$ and the product of its marginals $p_i p_{\mb m}$, in which the two variables are independent. Informally, this mutual information characterizes how much the measurement outcome $\mb m$ knows about the underlying state $i$.

Finally, maximizing the mutual information $I(\mathcal{E}:\Pi)$ Eq.~\eqref{eq:MI_povm} over possible choices of the ensemble $\mathcal{E}$ yields an intrinsic property of the POVM $\Pi$ known as its {\it informational power}~\cite{dallarno_informational_2011,dallarno_accessible_2014,dallarno_tight_2014}: 
\begin{equation}
W(\Pi) = \max_{\mathcal E} I(\mathcal E: \Pi).
\label{eq:infopower_def}
\end{equation} 

The optimization involved in the definition of informational power, Eq.~\eqref{eq:infopower_def}, may be hard in general.
However, it becomes trivial if we include a ``pre-scrambling'' or ``encoding'' step---meaning the system is rotated by a random unitary $V$ before being measured, as sketched in Fig.~\ref{fig:idea}. 
In that case, we show in Appendix~\ref{app:ip} that 
\begin{equation}
    W(\Pi) = Q(\mathbb{I}/D) - \sum_{\mb m} \pi_{\mb m} Q(\sigma_{\mb m}),
    \label{eq:infopower_result_Q}
\end{equation}
where $D$ is the dimension of the many-body Hilbert space\footnote{In this work we take $D = q^N$, i.e. a system of $N$ $q$-state qudits; we also focus on qubits ($q=2$) when specified.}, $Q(\rho)$ is a function known as the {\it subentropy} (see Appendix~\ref{app:subentropy}), and we have introduced a state ensemble $\mathcal{E}_\Pi = \{(\pi_{\mb m},\sigma_{\mb m})\}$ given by 
\begin{equation}
\left\{
\begin{aligned}
    & \pi_{\mb m} = \Tr(\effect{\mb m}) / D,
    \\
    & \sigma_{\mb m} = \effect{\mb m} / \Tr(\effect{\mb m}).
\end{aligned}
\right.
    \label{eq:canonical_state_ens}
\end{equation}
This state ensemble is dual\footnote{To every state ensemble $\mathcal{E} = \{(p_i,\rho_i)\}$ one can canonically associate the POVM $\Pi_{\mathcal E} = \{ p_i \rho^{-1/2} \rho_i \rho^{-1/2} \}$, with $\rho = \sum_i p_i \rho_i$. This is known as the {\it ``pretty good measurement''}~\cite{hausladen_pretty_1994}. All POVMs $\Pi$ obey $\Pi_{\mathcal{E}_\Pi} = \Pi$, and all state ensembles $\mathcal{E}$ with $\rho = \mathbb{I}/D$ obey $\mathcal{E}_{\Pi_{\mathcal{E}}} = \mathcal E$.}
to our POVM $\Pi = \{\effect{\mb m}\}$.
It is straightforward to check, from the POVM conditions, that this is in fact a state ensemble, i.e. that $\pi_{\mb m}$ is a valid probability distribution and that the $\sigma_{\mb m}$ are states. 
In fact, both objects have intuitive physical interpretations: 
$\pi_{\mb m}$ is the probability of obtaining outcome $\mb m$ when running monitored dynamics on the fully-mixed state $\rho = \mathbb{I}/D$;
$\sigma_{\mb m} \propto \kraus{\mb m}^\dagger (\mathbb{I} / D) K_{\mb m}$ is the output of {\it Heisenberg-picture} monitored evolution $\kraus{\mb m}^\dagger$, also acting on the fully-mixed state. In the most common models of monitored dynamics, made only of unitary gates and projective measurements~\cite{skinner_measurement-induced_2019,li_quantum_2018}, the Schr\"odinger and Heisenber pictures are equivalent (at the ensemble level\footnote{
While individual trajectories are generically not self-adjoint, 
the ensemble (over random unitary operations, locations and outcomes of measurements) is invariant in those models: $\{\kraus{\mb m}\} = \{\kraus{\mb m}^\dagger\}$.}), 
so that we can interpret $\mathcal{E}_\Pi$ as a valid {\it ensemble of trajectories} for a monitored mixed-state evolution. Note that this is {\it not} the ensemble of physical monitored trajectories of the quantum system, $\{(p_{\mb m},\kraus{\mb m}\rho \kraus{\mb m}^\dagger /p_{\mb m})\}$, with $p_{\mb m} = \Tr(\effect{\mb m}\rho)$: such states are inaccessible to Eve. The ensemble $\mathcal{E}_\Pi$ instead emerges as a description of the measurement process, built purely from a classical description of the dynamics (the $\kraus{\mb m}$ operators) available to Eve.
This mapping of the measurement process to an auxiliary ensemble of monitored trajectories plays a key role in this work.

Having established this formalism, we can now interpret the analytical result for the informational power $W(\Pi)$, Eq.~\eqref{eq:infopower_result_Q}.
An exact analytical expression for the subentropy $Q(\rho)$ in terms of the spectrum of $\rho$ is known~\cite{jozsa_lower_1994}, but not particularly illuminating.
However, for stabilizer states (in fact for all states proportional to projectors), one can analytically obtain a more explicit form for the subentropy (see Appendix~\ref{app:ip_clifford}), 
\begin{equation}
    Q(\rho) = 1 - \gamma - \delta H(q^S). \label{eq:subentropy_stab}
\end{equation}
Here $q$ is the local Hilbert space dimension ($D = q^N$), $S$ is the entropy of $\rho$ (in dits), $\gamma = 0.577...$ is Euler-Mascheroni's constant, and $\delta H(x) = H_x - (\ln(x) + \gamma)$ is the deviation of the harmonic sum $H_x = 1 + \frac{1}{2} + \dots + \frac{1}{x}$ from its large-$x$ expansion $\ln(x) + \gamma$. $\delta H$ is non-negative and bounded above by a constant; it vanishes as $\sim \frac{1}{2x}$ for large $x$.

Thus we arrive at the following result for monitored Clifford circuits (where all the trajectories $\sigma_{\mathbf m}$ are stabilizer states):
\begin{equation}
    W(\Pi) = \mathbb{E}_{\mathbf m}[\delta H(q^{S_{\mathbf m}})] - \delta H(q^N),
    \label{eq:infopower_cliff_result}
\end{equation}
with the average over trajectories taken according to the measure $\pi_{\mathbf m}$. 
This explicitly depends on the entropy of monitored trajectories $S_{\mb m} = -\log_q [\Tr(\sigma_{\mb m}^2)] $, meaning that the dynamical purification transition manifests as a transition in the informational power.
In the entangling phase, the trajectories $\sigma_{\mb m}$ remain highly mixed, with $S_{\mb m}\propto N$; for large $N$, employing the expansion $\delta H(x)\sim 1/(2x)$, we obtain
\begin{equation}
    W(\Pi) \simeq \frac{1}{2}\mathbb{E}_{\mathbf m}[\Tr(\sigma_{\mb m}^2)] - \frac{1}{2D} 
    \xrightarrow{D\to\infty} 0,
    \label{eq:infopower_vs_purity}
\end{equation}
so that the informational power, which quantifies learnability from the measurement record, goes to zero in the entangling phase. 
In the disentangling phase, the trajectories purify and the entropies $S_{\mb m}$ quickly decay towards zero (reaching $O(1)$ values at logarithmic depth) and thus $\delta H(q^{S_{\mb m}})$ remains finite.

To summarize, we have shown that the informational power of pre-scrambled Clifford monitored dynamics of depth $t = {\rm poly}(N)$ for large systems $N\gg 1$ obeys
\begin{equation}
    W(\Pi) = 
    \left\{ 
    \begin{aligned}
    & \sim q^{-sN} & \text{ (entangling phase)},\\
    & \textsf{const.}>0 & \text{ (disentangling phase)},
    \end{aligned} \right.
    \label{eq:infopower_summary}
\end{equation}
with $s \in [0,1]$ the order parameter of the purification phase transition (entropy density). 
We conjecture that the same transition holds for generic (non-Clifford) monitored dynamics. A suggestive result to this effect is that a Renyi-2 version of the mutual information $I(\mathcal{E}:\Pi)$ can be computed exactly and depends only on the average purity of the system, thus manifestly displaying the purification transition (see Appendix~\ref{app:ip_renyi2}). While this is not a valid mutual information, in randomized settings it is often a good proxy for the qualitative behavior of the true mutual information.

We note that the informational power is related to the Fisher information diagnostic in Ref.~\cite{bao_theory_2020} which measures the susceptibility of the measurement outcome distribution to small changes in the initial state. Like the informational power, the Fisher information aims to quantify how much information about the quantum state flows into the measurement record, thus the two approaches are closely related. 
A technical difference is that the Fisher information depends on the initial state and the choice of perturbation, while the informational power is an intrinsic property of the monitored dynamics.
More importantly, the Fisher information diagnostic in Ref.~\cite{bao_theory_2020} generically requires the collection of exponentially many samples (so that probability distributions can be estimated with some accuracy), while in our work we will show that the 
transition in informational power is reflected in the complexity of classical shadows, and can be determined from few samples---the complexity bottleneck in our scheme lies instead in the classical simulation of the quantum system (needed to carry out classical shadow estimation). 

Finally, we note that the informational power is equal to the channel capacity of the quantum-to-classical channel mapping the state $\rho$ to a classical probability distribution over measurement records $\mb m$~\cite{dallarno_informational_2011}: $\rho \mapsto \sum_{\mathbf m} \Tr(\effect{\mb m} \rho) \ketbra{\mathbf m}_C$, where $\{\ket{\mathbf m}_C\}$ is an orthonormal basis of classical states of the measurement device, as in Eq.~\eqref{eq:quantumclassical_channel}. Therefore $W(\Pi)$ directly quantifies the flow of information from Alice's unknown state to Eve's classical data, sketched in Fig.~\ref{fig:idea}(b). The MIPT arises as a sharp transition in this flow of information. In the rest of this work we examine how this transition affects Eve's ability to learn properties of the quantum state $\rho$.

\section{Eavesdropper's shadows \label{sec:shadows}}

The informational power transition discussed above suggests a general characterization of the MIPT as a learnability phase transition.
To assign an operational meaning to the transition, one needs to consider concrete protocols that Eve might employ to learn features of Alice's unknown initial state $\rho$ from the eavesdropped classical data $\mathbf m$.
Classical shadows, reviewed in Sec.~\ref{sec:review_shadows}, have emerged as a general and powerful framework for addressing this type of problem. Here we apply them to the generalized measurement associated with monitored dynamics. We will refer to these protocols as {\it ``eavesdropper's shadows''}.

\subsection{Setup}

We consider an experimentalist, Alice, who controls a quantum many-body system, and an eavesdropper, Eve, who wants to learn properties of Alice's system without having access to it.
Alice prepares an initial state $\rho$ and runs some model of monitored dynamics on it (e.g. a brickwork circuit with single-qubit projective measurements~\cite{skinner_measurement-induced_2019,li_quantum_2018}). She iterates this process many times, with the same state $\rho$ but a different realization of the dynamics each time.
Eve only has access to the following, {\it purely classical} data, for each run of the experiment:
\begin{enumerate}
\item[(i)] complete classical description of the monitored dynamics (e.g. circuit architecture, gates, locations and basis of measurements);
\item[(ii)] mid-circuit measurement record $\mb m$.
\end{enumerate}
Eve aims to learn as much as she can about the initial state $\rho$ from as few runs of the experiment as possible. This setup in sketched in Fig.~\ref{fig:idea}(b).

Eve's task can be readily cast in the framework of classical shadows with generalized measurements~\cite{acharya_shadow_2021,nguyen_optimizing_2022}.
From Eve's point of view, this setup is equivalent to a generalized measurement of $\rho$, namely the POVM $\Pi = \{ \effect{\mb m} = \kraus{\mb m}^\dagger \kraus{\mb m}\}$, with $\kraus{\mb m}$ being the Kraus operator for quantum trajectory $\mb m$ of the monitored dynamics.
While we have left it implicit to lighten our notation, $\effect{\mb m}$ also depends on random unitary gates that vary with each realization; therefore, it is a type of randomized measurement~\cite{elben_randomized_2023}. 
The problem of learning about $\rho$ from outcomes $\mb m$ of the randomized measurements is thus formally analogous to the standard classical shadows protocol~\cite{huang_predicting_2020,elben_randomized_2023}, Sec.~\ref{sec:review_shadows}.

\subsection{Protocol}

In the standard protocol, based on a projective POVM $\{U^\dagger \ketbra{\mb b} U\}$, the choice of a post-measurement ``snapshot'' state is automatic---upon getting outcome $\mb b$, the best guess for the pre-measurement state is just the POVM element itself, $\sigma_{U, \mb b} \equiv U^\dagger \ketbra{\mb b} U$. 
In the general case, with a non-projective POVM $\{\effect{\mb m}\}$, the choice is less obvious. 
In fact, there are several valid choices motivated on practical or conceptual grounds, as we will see below; all of them recover the ``canonical'' choice in the limit of the POVM becoming projective.
Denoting a choice of snapshot state by $\eta_{\mb m}$, we have a measure-and-prepare channel
\begin{equation}
    \mathcal{M}(\rho) 
    = \sum_{\mathbf m} \Tr(\effect{\mb m} \rho) \eta_{\mathbf m}
\end{equation}
which can in principle be used for classical shadow estimation, along the same steps outlined in Sec.~\ref{sec:review_shadows}.

A natural choice for $\eta_{\mb m}$ is based on a general mapping between POVMs and state ensembles, Eq.~\eqref{eq:canonical_state_ens}: we propose setting $\eta_{\mb m} = \sigma_{\mb m}$.
This choice closely mirrors the standard protocol, Sec.~\ref{sec:review_shadows}, upon replacing $U \mapsto \kraus{\mb m}$ (random monitored dynamics instead of random unitary rotation) and $\ketbra{\mb b} \mapsto \mathbb{I}/D$.
The latter is a uniform average over all possible outcomes $\ket{\mb b}$, representing Eve's complete ignorance about the final state of the system.
Beyond this heuristic reasoning, we note that this choice can be formally motivated as the {\it Petz recovery map}~\cite{petz_sufficient_1986,barnum_reversing_2002,wilde_recoverability_2015,penington_replica_2022} from the space of measurement records to the space of quantum states, see App.~\ref{app:shadows}.

With this prescription, the shadow channel reads
\begin{align}
    \mathcal{M}(\rho) &= \sum_{\mb m} \Tr(\rho \effect{\mb m}) \sigma_{\mb m} \nonumber \\
    &= D \sum_{\mb m} \pi_{\mb m} \Tr(\rho \sigma_{\mb m})   \sigma_{\mb m}  \nonumber \\
    &= D \Tr_2[(\mathbb{I}\otimes \rho) \sigma^{(2)}],
    \label{eq:shadow_channel_petz}
\end{align}
where ${\rm Tr}_2$ is the partial trace over the second replica, and $\sigma^{(2)}$ is the {\it second moment operator} of the state ensemble $\mathcal{E}_\Pi = \{ (\pi_{\mb m}, \sigma_{\mb m}) \}$ dual to our POVM $\Pi = \{\effect{\mb m}\}$ [see Eq.~\eqref{eq:canonical_state_ens}]:
\begin{equation}
    \sigma^{(2)} = \sum_{\mb m} \pi_{\mb m} \sigma_{\mb m}^{\otimes 2}. 
    \label{eq:second_moment_operator}
\end{equation}
For most typical models of monitored dynamics (featuring only unitary evolution and projective measurements), the states $\sigma_{\mb m}$ are quantum trajectories of a valid monitored evolution $\kraus{\mb m}^\dagger$ acting on the fully-mixed state; thus the second-moment operator $\sigma^{(2)}$ is directly sensitive to the MIPT. 
For instance the order parameter of {\it dynamical purification phases}~\cite{gullans_dynamical_2020,gullans_scalable_2020} (Sec.~\ref{sec:review_mipt}), the trajectory-averaged purity $\avgP$, can be obtained as an expectation value on the two-replica state $\sigma^{(2)}$:
\begin{equation}
    \avgP = \sum_{\mb m} \pi_{\mb m} \Tr(\sigma_{\mb m}^2)
    = \Tr(\sigma^{(2)} \hat \tau), \label{eq:avgP_from_2ndmoment}
\end{equation}
with $\hat \tau$ the replica SWAP operator.
Thus $\sigma^{(2)}$ undergoes a sharp change at the MIPT, and by extension so does the shadow channel $\mathcal{M}$, Eq.~\eqref{eq:shadow_channel_petz}.
We will investigate the consequences of this sharp change in terms of {\it learnability transitions} in the rest of the manuscript.

\subsection{Alternative prescriptions}

Before proceeding, we note that other choices for the ``snapshot'' $\eta_{\mathbf m}$ are possible and well-motivated. 
In particular, two choices have been considered in the literature in the context of classical shadows with generalized measurements~\cite{acharya_shadow_2021,nguyen_optimizing_2022}. We discuss them in detail in Appendix~\ref{app:shadows}, and briefly summarize the results below:
\begin{itemize}
    \item {\it Least squares}~\cite{nguyen_optimizing_2022}: set $\eta_{\mb m} = \effect{\mb m}$. This is {\it not} a state (due to trace normalization), and the resulting ``shadow channel'' $\mathcal{M}(\rho) = \sum_{\mb m} \Tr(\rho \effect{\mb m}) \effect{\mb m}$ is thus not a channel. Classical shadows work regardless. 
    This choice minimizes the two-norm distance between the observed ($\Tr(\rho\effect{\mb m})$) and predicted ($\Tr(\hat{\rho} \effect{\mb m})$, with $\hat\rho$ the classical shadow of $\rho$) measurement outcome distributions.
    The shadow channel takes a form analogous to Eq.~\eqref{eq:shadow_channel_petz} [see Eq.~\eqref{eq:meas_channel_ls}], but with a modified second-moment operator
    \begin{equation}
        \tilde{\sigma}^{(2)} = \sum_{\mb m} \tilde{\pi}_{\mb m} \sigma_{\mb m}^{\otimes 2}
        \label{eq:secondmoment_ls}
    \end{equation}
    where the probabilities are $\tilde{\pi}_{\mb m} = \pi_{\mb m}^2 / \sum_{\mb m'} \pi_{\mb m'}^2$.
    This also features a MIPT, albeit with a different universality class due to a different re-weighting of the trajectories~\cite{bao_theory_2020, vasseur_entanglement_2019, jian_measurement-induced_2020, bao_symmetry_2021,li_cross_2023}.
    \item {\it Maximum fidelity}~\cite{acharya_shadow_2021}: set $\eta_{\mb m} = \ketbra{\psi_{\mb m}}$ where $\ket{\psi_{\mb m}}$ is the leading eigenvector of $\effect{\mb m}$ (we neglect degeneracies). 
    This choice maximizes the fidelity $\bra{\phi} \mathcal{M}(\ketbra{\phi}) \ket{\phi}$ between the input and output of $\mathcal{M}$, on average over Haar-random input states $\ket{\phi}$.
    Again the shadow channel takes a form analogous to Eq.~\eqref{eq:shadow_channel_petz} [see Eq.~\eqref{eq:shadow_channel_maxfid}], but with $\sigma^{(2)}$ replaced by the state
    \begin{equation}
        \sigma^{(\infty,1)} = \sum_{\mb m} \pi_{\mb m} \ketbra{\psi_{\mb m}} \otimes \sigma_{\mb m}.
        \label{eq:secondmoment_maxfid}
    \end{equation}
    The expectation of the replica SWAP $\hat{\tau}$ on this state yields the average of $q^{-S_\infty}$, meaning this is also sensitive to the MIPT (with the same universality and critical point as Eq.~\eqref{eq:shadow_channel_petz} in this case\footnote{All Renyi entropies with indices $n>1$ are bounded within a multiplicative constant of each other, so $S_2$ and $S_\infty$ have the same critical properties.}).
\end{itemize}

We note that all three prescriptions reduce to the one from Ref.~\cite{huang_predicting_2020} for the case of projective measurements, but they differ for generalized measurements.
In the following we will use the prescription $\eta_{\mb m} = \sigma_{\mb m}$ unless otherwise specified; the qualitative conclusions would be unchanged with either prescription, since as we saw each of them is sensitive to the MIPT.

\begin{figure*}
    \centering
    \includegraphics[width=\textwidth]{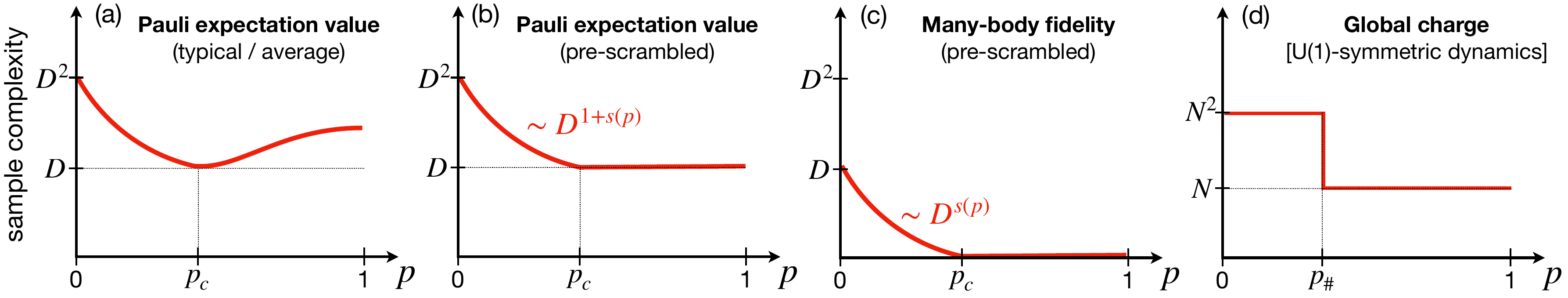}
    \caption{Schematic summary of main results for the sample complexity of various estimation tasks via ``eavesdropper's shadows'', as a function of the measurement rate $p$. $N$ is the system size, $D = q^N$ is the Hilbert space dimension, and $s(p)$ is the entropy density (order parameter of the entangling phase). 
    (a) Estimation of Pauli expectation values, Sec.~\ref{sec:pauli_locality}. Different operators have different complexity; the curve shows the qualitative behavior of average or typical elements of the $N$-qudit Pauli group. The MIPT ($p = p_c$) emerges as a point of minimum complexity.
    (b) Estimation of Pauli expectation values with a pre-scrambling step, Sec.~\ref{sec:prescrambling}. All non-identity Paulis have the same complexity, $\sim D^{1+s(p)}$, which changes non-analytically at the MIPT.
    (c) Estimation of many-body fidelity (with a pre-scrambling step), Sec.~\ref{sec:xeb}. The complexity transitions from constant to exponential in $N$ at the MIPT.
    (d) Estimation of global charge from dynamics with $U(1)$ symmetry, Sec.~\ref{sec:charge}. The complexity transitions from $\sim N$ to $\sim N^2/t$ at the charge-sharpening transition, $p = p_\#$.
    \label{fig:results_summary} 
    }
\end{figure*}

To summarize, we have examined different strategies that Eve might use to learn properties of Alice's unknown initial state $\rho$ from the measurement record $\mb m$ via classical shadows.
We have found that the shadow channel is determined by a {\it moment operator} for an emergent ensemble of monitored quantum trajectories of a mixed-state dynamics, Eq.~\eqref{eq:canonical_state_ens}. 
As the second moment of trajectories is sensitive to the MIPT, we expect this to lead to a qualitative change in the performance of classical shadows.
In the following sections we work out the consequences of this observation on a range of different manifestations of the measurement-induced phase transition. Our results are schematically summarized in Fig.~\ref{fig:results_summary}.

\section{Learning Pauli expectation values \label{sec:pauli}}

In this section we begin to unravel the consequences of the connection between ``eavesdropper's shadows'' and dynamical purification of mixed states introduced in Sec.~\ref{sec:shadows}. We start by focusing on Eve's ability to learn Pauli expectation values on the unknown initial state $\rho$, namely to estimate $\langle P \rangle = \Tr(\rho P)$ ($P$ is a Pauli operator\footnote{We use generalized Pauli operators generated by the `clock' and `shift' operators when $q>2$.}) to constant additive error $\epsilon$. 

\subsection{Shadow norm and entanglement \label{sec:shdn_entanglement}}

In general, the sample complexity of learning the expectation of an operator $O$ on a state $\rho$ is given by its squared {\it shadow norm} $\shdn{O}^2$, see Sec.~\ref{sec:review_shadows}.
This is given formally by\footnote{Note that there is, implicit in this notation, an average over random unitary gates $\{u\}$ which enter the monitored dynamics. An average $\int {\rm d}\{u\}$ is implicitly present concurrently with each trajectory average $\sum_{\mb m} \pi_{\mb m}$; we drop it to lighten the notation.}
\begin{equation}
    \shdn{O}^2 = D \Tr( \rho \otimes \mathcal{M}^{-1}(O) \otimes \mathcal{M}^{-1}(O)\, {\sigma^{(3)}}),\label{eq:shdn_general}
\end{equation}
where ${\sigma^{(3)}} = \sum_{\mathbf m} \pi_{\mathbf m} \sigma_{\mathbf m}^{\otimes 3}$ is the third moment operator of the trajectory ensemble $\mathcal{E} = \{(\pi_{\mb m}, \sigma_{\mb m})\}$.
If the ensemble is Pauli-invariant~\cite{hu_classical_2023,bu_classical_2022} and $O$ is a Pauli operator, Eq.~\eqref{eq:shdn_general} simplifies to 
\begin{equation}
    \shdn{O}^2 = \frac{1}{D} \Tr(O\mathcal{M}^{-1}(O)), \label{eq:shdn_simplified}
\end{equation}
which is independent of $\rho$.
Furthermore, if the POVM is invariant under multiplication by single-site Pauli operators~\cite{bu_classical_2022} (as is the case in typical models of monitored dynamics, made with random Haar or Clifford gates), the Pauli operators are eigenmodes of the channel: $\mathcal{M}(P) = \lambda_P P$. Thus, by Eq.~\eqref{eq:shdn_simplified}, their shadow norm is $\shdn{P}^2 = \lambda_P^{-1}$. 

We now derive a relationship between the eigenvalues $\lambda_P$ (controlling the shadow norms of Pauli operators) and the entanglement structure of the ensemble of trajectories $\mathcal{E}$. 
We have, using operator-to-state notation [$(O|$ for super-bras, $|O)$ for super-kets, with inner product $(A|B) = \Tr(A^\dagger B)$],
\begin{equation}
    \lambda_P = \frac{(P|\mathcal{M}|P)}{(P|P)} = \Tr[P^{\otimes 2} \sigma^{(2)}].
\end{equation}
At the same time, the averaged purity of a subsystem $A$ in the ensemble of monitored trajectories $\mathcal{E}$ is 
\begin{equation}
    \avgP_A = \Tr[\hat{\tau}_A \sigma^{(2)}]
\end{equation}
with $\hat{\tau}_A$ the replica SWAP operator acting only on subsystem $A$.
We can expand $\hat \tau$ in the Pauli basis\footnote{Expanding $\hat\tau$ in the two-replica Pauli basis as $\hat{\tau} = \sum_{\alpha\beta} c_{\alpha\beta} P_\alpha \otimes P_\beta$, the coefficients $c_{\alpha \beta}$ are given by $c_{\alpha \beta} = \Tr[\hat{\tau}(P_\alpha \otimes P_\beta)] / D^2 = \Tr(P_\alpha P_\beta)/D^2 = \delta_{\alpha \beta}/D$.} as 
\begin{equation}
    \hat{\tau}_A = \frac{1}{D_A} \sum_{P:\, {\sf supp}(P)\subseteq A} P^{\otimes 2},
\end{equation}
where $D_A$ is the Hilbert space dimension for subsystem $A$ and $\textsf{supp}(P)$ denotes the support of $P$, i.e. the subsystem where $P$ is non-identity.
A relationship between the entanglement feature $\{\avgP_A\}$ and the eigenvalues $\{\lambda_A\}$ of channel $\mathcal{M}$ follows: 
\begin{equation}
    D_A \avgP_A = \sum_{P:\, {\sf supp}(P)\subseteq A} \lambda_P = \sum_{B\subseteq A} (q^2-1)^{|B|} \lambda_B, \label{eq:purity_vs_lambdas}
\end{equation}
where the first sum is over Pauli operators $P$ supported inside $A$, while the second is over subsystems $B$ contained inside $A$\footnote{We use $\lambda_A$ ($A$ being a subsystem) and $\lambda_P$ ($P$ being a Pauli operator) interchangeably, with the understanding that $\lambda_P = \lambda_{{\sf supp}(P)}$.}. The second equality holds because there are $(q^2-1)^{|B|}$ distinct Pauli operators with support $B$.
The inverse of Eq.~\eqref{eq:purity_vs_lambdas} yields
\begin{align}
    \lambda_A 
    & = (1-q^2)^{-|A|} \sum_{B\subseteq A} \avgP_B (-q)^{|B|},
    \label{eq:lambda_vs_purities}
\end{align}
which is a well-known relationship between entanglement and shadow norm in the theory of classical shadows~\cite{hu_classical_2023,akhtar_scalable_2023,bu_classical_2022,ippoliti_operator_2023,ippoliti_classical_2023}; we see that it straightforwardly extends to our setting of shadows with generalized measurements, when taking $\{\avgP_A\}$ to be the entanglement feature of monitored trajectories $\sigma_{\mb m}$. 

Eq.~\eqref{eq:purity_vs_lambdas} and \eqref{eq:lambda_vs_purities} connect entanglement properties of the trajectories with shadow norms of Pauli operators. This is interesting as it suggests a sharp change in the performance of classical shadows at the dynamical purification transition, consistent with our prior analysis in Sec.~\ref{sec:infopower} and \ref{sec:shadows}.
The connection between entanglement and shadow norms in Eq.~\eqref{eq:lambda_vs_purities} is not straightforward, as it is a sum of exponentially many terms with alternating signs.
Nonetheless, a simple exact statement can be made about the {\it harmonic mean} of $\shdn{P}^2$ for all Pauli operators supported inside a subsystem $A$: rewriting Eq.~\eqref{eq:purity_vs_lambdas}, we have
\begin{equation}
    \left( \frac{1}{D_A^2} \sum_{P:\, {\sf supp}(P)\subseteq A}  \shdn{P}^{-2} \right)^{-1} = D_A / \avgP_A \sim D_A^{1+s},
    \label{eq:harm_mean}
\end{equation}
where $s\in[0,1]$ is an entropy density defined by the scaling of average purity $\avgP_A \sim q^{-sN_A} = D_A^{-s}$.
This implies a sharp change of the shadow norm distribution at the purification transition, which separates the pure phase ($s = 0$) from the mixed phase ($s>0$). 
The harmonic mean of squared shadow norms scales exponentially in subsystem size $N_A$ on both sides of the transition, as $D_A^{1+s} =  q^{[1+s(p)]N_A}$, but the coefficient in the exponential changes non-analytically at the critical point $p = p_c$ (where $s(p) \simeq \Theta(p_c-p) |p-p_c|^\nu$, with $\Theta$ the Heaviside theta and $\nu\simeq 1.3$ a critical exponent~\cite{gullans_dynamical_2020}). Notably, the harmonic mean Eq.~\eqref{eq:harm_mean} is a lower bound to both the {\it average} shadow norm (arithmetic mean) and the {\it typical} shadow norm (geometric mean), implying that both must diverge as $\Omega(2^{(1+s)N_A})$ in the entangling phase and as $\Omega(2^{N_A})$ in the disentangling phase.

\subsection{Optimal learning at the MIPT \label{sec:pauli_locality}}

\begin{figure}
    \centering
    \includegraphics[width=\columnwidth]{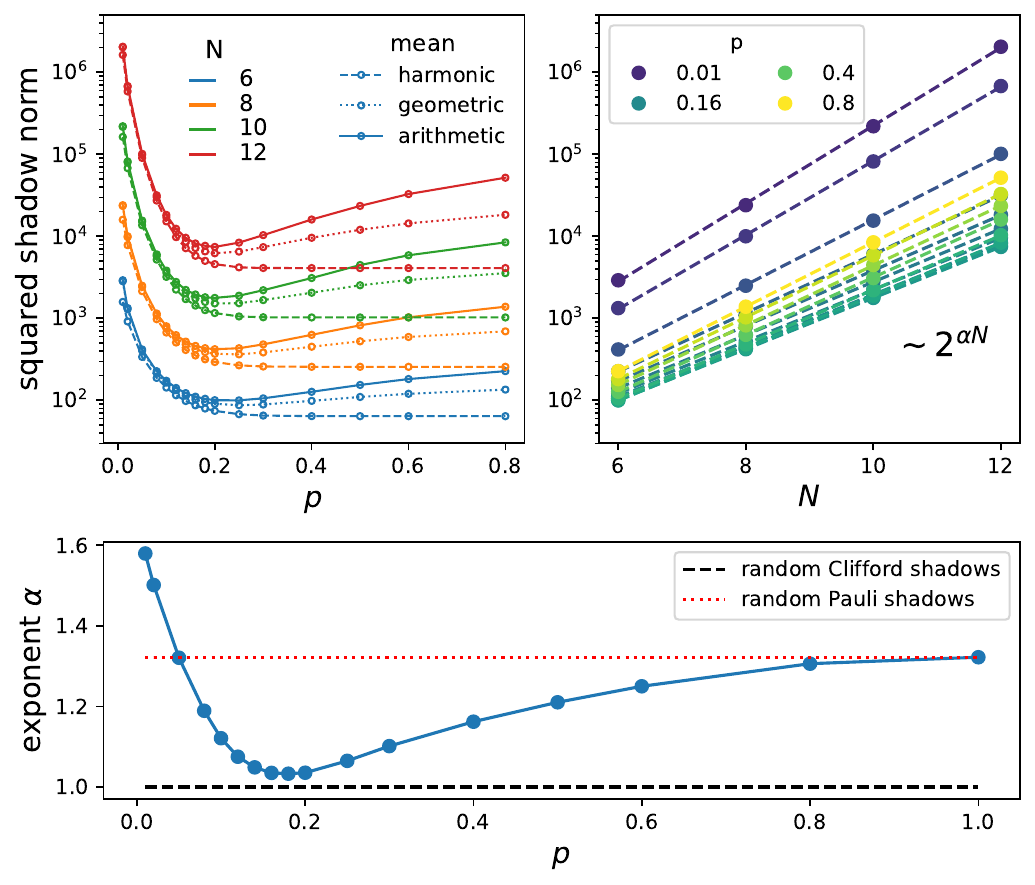}
    \caption{
    (a) Squared shadow norms of Pauli operators under eavesdropper's shadows as a function of measurement rate $p$, for different system sizes $N$. Data is from exact numerical simulations of quantum trajectory density matrices, averaged over between 2000 and 40000 random circuit realizations depending on size. The harmonic mean (dotted lines) is monotonically-decreasing in $p$, being completely determined by the system's purity; the arithmetic mean (solid) and geometric mean (dashed) show a minimum at the MIPT. 
    (b) Same data (for the arithmetic mean) plotted as a function of $N$ displays clear exponential scaling $\sim D^\alpha = 2^{\alpha N}$. Dashed lines are fits.
    (c) Fit coefficient $\alpha$ vs $p$ shows a minimum at the MIPT.
    }
    \label{fig:shdn}
\end{figure}

To gain more insight on the structure of the shadow norm distribution, beyond the exact harmonic-mean result of Eq.~\eqref{eq:harm_mean} and the bounds it implies, we turn to numerical simulations.
We perform exact numerical simulations of mixed-state monitored dynamics (a standard model made of brickwork layers of Haar-random gates and single-qubit $Z$ measurements with probability $p\in[0,1]$) on up to $N = 12$ qubits. Note we are limited to this size by the fact that we simulate the full density matrix dynamics (equivalent to a pure state of $2N$ qubits).
We obtain the entanglement feature $\{\avgP_A\}$ averaged over many realizations of the dynamics; from the entanglement feature we obtain the full set of shadow norms $\{\lambda_A^{-1}\}$ via Eq.~\eqref{eq:lambda_vs_purities}. 
Results are shown in Fig.~\ref{fig:shdn}.
The harmonic mean of shadow norms, as predicted, is a function only of the averaged purity, and is thus monotonically decreasing in the measurement rate $p$. However the arithmetic mean $D^{-2} \sum_P \shdn{P}^2$ and geometric mean $\exp(D^{-2} \sum_P \log \shdn{P}^2)$ alike exhibit non-monotonic behavior, with a minimum near the purification transition $p \approx 0.16$. 

This is explained by the fact that, deep in the pure phase, one recovers {\it random Pauli shadows} (i.e. shadows with random {\it local, single-qubit} Pauli measurements~\cite{huang_predicting_2020}). This is exactly true at $p = 1$, and we expect it to be a fairly accurate approximation throughout $p \geq 0.5$, where the circuit is non-percolating\footnote{For $1/2 < p < 1$, a more precise analogy is with classical shadows with locally-entangled measurements~\cite{ippoliti_classical_2023}, as the measurement basis typically breaks up into a tensor product of finite-sized bases, one for each non-percolating cluster of the circuit.}. In this regime, even with complete access to the system, {\it spatial locality} makes learning large Pauli operators very inefficient.
On the other hand, in the mixed phase, as seen earlier, the informational power of Eve's measurements goes to zero, making any learning inefficient.
The MIPT is a ``sweet spot'' between these two obstructions: the informational power is still finite (at polynomial depth), while the restriction of locality is alleviated. 

Approaching the transition from the area-law side, we expect eavesdropper's shadows to perform similarly to {\it shallow shadows}~\cite{akhtar_scalable_2023,bertoni_shallow_2022,arienzo_closed-form_2022,ippoliti_operator_2023} with finite depth: i.e., Eve manages to eventually read out all the information, but this takes an amount of time that grows as the transition is approached, giving information more time to spread, analogous to increasing circuit depth in shallow shadows.
At the transition, this ``effective depth'' should diverge; it is tempting to speculate a relationship with shallow shadows at $\log(N)$ depth, which were shown to give Pauli shadow norms $\sim q^N$ for large Pauli operators (consistent with the scaling of both harmonic and arithmetic-mean $\shdn{P}^2$ at the MIPT observed here).

Thus there are two conceptually-independent reasons why Eve's task of learning the expectation values of general Pauli operators via classical shadows may be hard: spatial locality at large $p$ (pure/disentangling phase) and a fundamental lack of information at small $p$ (entangling/mixed/coding phase).
Numerical results indicate that an optimum is reached near the purification transition, $p_c \approx 0.16$, where the performance is close to random Clifford shadows ($\shdn{P}^2 \sim D$ for all traceless Paulis). 
This is a novel characterization of the MIPT, as the optimum rate $p$ for learning Pauli expectation values from the measurement record.

\subsection{Complexity transition \label{sec:prescrambling}}

For the rest of this work, we will focus on the presence or absence of {\it any} information in the measurement record, regardless of its locality structure. 
For this reason, it is advantageous to ``pre-scramble'' the input state $\rho$ with a random global Clifford operation $V$, as in Sec.~\ref{sec:infopower}, which eliminates issues with spatial locality. 
This dramatically simplifies the picture, as the channel $\mathcal{M}$ now depends on a single property of the trajectory ensemble---its average global purity $\avgP$.
We have
\begin{align}
    \mathcal{M}(\rho) 
    & = D \int{\rm d}V\, \Tr_2[(\mathbb{I}\otimes \rho)\, V^{\otimes 2} \sigma^{(2)} (V^\dagger)^{\otimes 2}] \nonumber \\
    & = \frac{(D-\avgP)\Tr(\rho)\mathbb{I} + (D\avgP-1)\rho}{D^2-1} \label{eq:M_expression_prescr}
\end{align}
This reproduces the familiar random-Clifford result $(\mathbb{I}+\rho)/(D+1)$ for $\avgP = 1$. 
For $\avgP = 1/D$ (no measurements), it correctly gives $\mathbb{I}/D$: there is no information at all about $\rho$, the measurement channel is a global erasure and is not invertible. Intermediate values of $\avgP$ interpolate between these two extremes. In particular the channel is invertible whenever $\avgP > 1/D$.  

It follows that all traceless operators are eigenmodes of $\mathcal{M}$ with eigenvalue
\begin{equation}
    \lambda = \frac{D\avgP - 1}{D^2-1} = \frac{D^{1-s}-1}{D^2-1} \sim D^{-(1+s)} \label{eq:prescr_lambda}
\end{equation}
where $\sim$ denotes asymptotic scaling at large $D$. 
Thus in particular, all Pauli operators $P \neq I$ have shadow norm 
\begin{equation}
    \shdn{P}^2 \sim D^{1+s} = q^{(1+s)N} \label{eq:prescr_shdn}
\end{equation}
which indeed changes sharply at the purification transition, as the entropy density goes from $s = 0$ (pure phase) to $s>0$ (mixed phase).

\subsection{Information extracted per measurement \label{sec:pauli_infopermeas}}

The result in Eq.~\eqref{eq:prescr_shdn} implies that, to learn the expectation of $P$ on the unknown state $\rho$, Eve needs many more samples in the mixed phase than she does in the pure phase---a factor of $\sim q^{sN}$ more.
In other words, the amount of information about $\rho$ leaking into the measurement record $\mb m$ is suppressed {\it exponentially} in the mixed phase. This characterization of the mixed or coding phase~\cite{choi_quantum_2020, gullans_dynamical_2020,gullans_scalable_2020} is complementary to the exponentially-long lifetime (also $\sim q^{sN}$) of information in the system.

In the present framework, however, we can make an even more precise statement about the way in which information on $\rho$ leaks into the measurement record.
A key result in dynamical purification~\cite{gullans_dynamical_2020,li_statistical_2021} is that, as a function of circuit depth $t$, the entropy density $s$ in the mixed phase decreases as 
\begin{equation}
    s(t) \sim s_0 - \frac{1}{N}\log_q(t)
    \label{eq:purif}
\end{equation} 
(at times $1 \ll t \ll q^{s_0N}$), where $s_0$ is the entropy density ``plateau'' that characterizes the mixed phase.
The ansatz in Eq.~\eqref{eq:purif}, plugged into Eq.~\eqref{eq:prescr_shdn}, gives squared shadow norm
\begin{equation}
    \shdn{P}^2 \sim \frac{1}{t} D^{1+s_0} \label{eq:shdn_1overt}
\end{equation} 
for any traceless Pauli $P$.
This quantifies the total number of circuit runs needed to learn $\langle P \rangle$ up to constant error.
We can translate this number of circuits into a total number of measurements, $M_{\rm tot}$:
with $M \sim pNt$ measurements per circuit, we have 
\begin{equation}
M_{\rm tot} \sim M\times \shdn{P}^2 \sim N q^{(1+s_0)N} \label{eq:total_n_measurements}
\end{equation}
which is $t$-independent\footnote{Note this conclusion depends crucially on the coefficient of $\log(t)$ being exactly 1, which was argued e.g. in Refs.~\cite{gullans_dynamical_2020,li_statistical_2021}.}.

The fact that $M_{\rm tot}$, the total number of measurements needed, is approximately $t$-independent gives it an invariant meaning: there is, effectively, a fixed amount of information learned by Eve {\it per measurement}; this amount is $\sim q^{-(1+s_0)N}$ bits. The factor of $q^{-N}$ comes from Haar-random encoding of the initial state (pre-scrambling), while the factor of $q^{-s_0N}$ is the additional encoding coming from the dynamics in the mixed phase.
This gives a sharp, operational meaning to the idea that measurements fail to read out information about the encoded quantum state in the mixed phase.
This sample complexity further saturates the scaling of the informational power in the entangling phase, Sec.~\ref{sec:infopower}, showing that eavesdropper's shadows are (near-)optimal for this task.

\section{Learning many-body fidelity \label{sec:xeb}}

Another observable of interest in many applications, e.g. benchmarking, is the fidelity with a pure many-body state $F = \bra{\psi} \rho \ket{\psi}$. The complexity of learning $F$ via classical shadows is given by the shadow norm of the rank-1 projector $\ketbra{\psi}$. Notably, this shadow norm is $O(1)$ in random Clifford shadows~\cite{huang_learning_2022}, which makes these observables interesting as practical targets. 
Ref.~\cite{li_cross_2023} recently proposed a diagnostic for measurement-induced phases based on a linear cross-entropy function which intuitively captures the (in)distinguishability between measurement records drawn from monitored dynamics acting on different initial states; here we show that this diagnostic can in fact be readily framed in terms of fidelity estimation via eavesdropper's shadows.

\subsection{Review of linear XEB diagnostic \label{sec:xeb_review}}

The linear cross-entropy diagnostic proposed in Ref.~\cite{li_cross_2023} reads
\begin{equation}
    {\sf XEB} = \left \langle \frac{p(\mathbf{m}|\rho_0) } {\sum_{\mathbf m'} p(\mathbf{m}'|\rho_0)^2} \right \rangle_{\mathbf m\sim p(\mathbf m|\rho)}  \label{eq:xeb}
\end{equation}
where $p(\mathbf m|\rho)$ is the probability of drawing measurement record $\mathbf m$ in an experiment on the initial state $\rho$. 
The idea is that $\rho_0$ is a ``simple'' (e.g. stabilizer) initial state whose probabilities $p(\mathbf{m}|\rho_0)$ are computed classically, whereas $\rho$ is a generic initial state, and $\mathbf{m}\sim p(\mathbf m|\rho)$ denotes samples drawn from an experiment on quantum hardware initialized in state $\rho$. 
We again assume the monitored circuit is prefaced by a global random Clifford operation, as in Sec.~\ref{sec:prescrambling} and in Ref.~\cite{li_cross_2023}.

In the notation of our work, we have $p(\mb m| \rho) = \Tr(\effect{\mb m}\rho)$.
Thus in the limit of a large number of experimental samples, the linear-XEB diagnostic reads
\begin{align}
    {\sf XEB} 
    & = \frac{\sum_{\mathbf m} p(\mathbf m|\rho) p(\mathbf m|\rho_0)}{\sum_{\mathbf m} p(\mathbf m|\rho_0)^2} 
    = \frac{\sum_{\mathbf m} \pi_{\mathbf m}^2 \Tr(\sigma_{\mathbf m}^{\otimes 2} \rho\otimes \rho_0)} {\sum_{\mathbf m} \pi_{\mathbf m}^2 \Tr(\sigma_{\mathbf m}^{\otimes 2} \rho_0^{\otimes 2}) } \nonumber \\
    & = \Tr[(\rho \otimes \rho_0) \tilde{\sigma}^{(2)}] / \Tr[\rho_0^{\otimes 2} \tilde{\sigma}^{(2)}]
    \label{eq:xeb_ournotation}
\end{align}
where $\tilde{\sigma}^{(2)}$ is the modified second-moment operator from Eq.~\eqref{eq:secondmoment_ls}, 
defined according to probabilities $\tilde{\pi}_{\mb m} \propto \pi_{\mb m}^2$, that also appears in the least-squares shadows prescription. As we discussed in Sec.~\ref{sec:shadows} (see also App.~\ref{app:shadows}), the ensemble of trajectories $\mathcal{\tilde{E}} = \{(\tilde{\pi}_{\mb m}, \sigma_{\mb m})\}$ is known to also display an entanglement phase transition, albeit with different universality~\cite{bao_theory_2020}.
Thus the ${\sf XEB}$ quantity in Eq.~\eqref{eq:xeb_ournotation} is sensitive to a MIPT. 

Ref.~\cite{li_cross_2023} in particular showed that, when including a pre-scrambling stage (as in Sec.~\ref{sec:infopower} and \ref{sec:prescrambling}), in the mixed phase we have ${\sf XEB}\simeq 1$ independent of $\rho$, while in the pure phase ${\sf XEB}$ becomes sensitive to $\rho$ and in particular approaches a finite value $<1$ if $\rho$ differs significantly from $\rho_0$. Aside from the specific details of the protocol, this is suggestive of a phase transition in {\it learnability of the initial state}: in the pure phase we can successfully tell if $\rho$ and $\rho_0$ are different, in the mixed phase we fail to do so. 
In the following we clarify this connection to a learnability phase transition and frame this result in the language of shadow estimation.

\subsection{Fidelity from a modified linear-XEB \label{sec:mod_xeb}}

To sharpen the connection between the {\sf XEB} diagnostic of Eq.~\eqref{eq:xeb}~\cite{li_cross_2023} and learnability, let us start by introducing a slight variation of the quantity, based on how much a new measurement record $\mathbf m$ from the experiment updates Eve's belief about the unknown initial state $\rho$ of the system.

We define the quantity
\begin{equation}
    {\sf XEB}' = \left\langle p(\rho_0|\mathbf{m}) \right \rangle_{\mathbf m \sim p(\mathbf m|\rho)},
\end{equation}
which differs from Eq.~\eqref{eq:xeb} by the order of conditioning ($p(\rho_0|\mathbf m)$ instead of $p(\mathbf m| \rho_0)$ in the numerator of Eq.~\eqref{eq:xeb}) and by the absence of a normalization factor, which becomes unnecessary in this case. 
This quantity has an intuitive interpretation as Eve's updated belief about the initial state, given her new information about a measurement record $\mathbf m$ eavesdropped from Alice. 

To suitably define the conditional probability $p(\rho_0|\mb{m})$, we start with Eve's joint probability distribution over initial (pure) states of the system $\rho$ and measurement records $\mathbf m$:
\begin{equation}
p(\rho, \mathbf m)\,{\rm d}\rho = \Tr(\rho \effect{\mb m})\,{\rm d}\rho,  \label{eq:joint_prob_dist}   
\end{equation}
where ${\rm d}\rho$ is a measure over quantum states that reflects Eve's prior beliefs about Alice's initial state $\rho$ (e.g. a uniform measure representing complete ignorance); we require $\int {\rm d}\rho\, \rho = \mathbb{I}/D$.
From this joint distribution we can obtain the marginal
\begin{equation}
    p(\mb m) = \int{\rm d}\rho\, p(\rho,\mathbf m) = \Tr(\effect{\mb m}/D) = \pi_{\mathbf m}, \label{eq:marginal}
\end{equation}
and thus the conditional probability via Bayes' rule:
\begin{equation}
    p(\rho |\mathbf{m})\, {\rm d}\rho
    = \frac{p(\rho, \mathbf{m})\, {\rm d}\rho}{p(\mathbf m)}
    = D\Tr(\rho \sigma_{\mathbf m})\, {\rm d}\rho.
    \label{eq:conditional_prob}
\end{equation}
In the limit of a large number of experimental shots, we thus obtain
\begin{align}
    {\sf XEB}' & = \sum_{\mathbf m} p(\mathbf m|\rho) p(\rho_0|\mathbf m) 
    = D \sum_{\mathbf m} \Tr(\rho \effect{\mathbf{m}}) \Tr(\rho_0 \sigma_{\mathbf{m}}) \nonumber \\
    & = D \Tr(\rho_0 \mathcal{M}(\rho)),
    \label{eq:xebprime_vsM}
\end{align}
which is explicitly a function of the shadow channel $\mathcal{M}$ in Eq.~\eqref{eq:shadow_channel_petz}, and thus sensitive to the standard MIPT (`standard' meaning with correct trajectory weights $\pi_{\mb m}$).

Finally, with pre-scrambling, we use Eq.~\eqref{eq:M_expression_prescr} to obtain
\begin{align}
    {\sf XEB}' 
    & = \frac{D - \avgP + (D\avgP - 1)F}{D - 1/D} 
    \simeq 1 + \avgP F
    \label{eq:xebprime_avg}
\end{align}
where $F = \Tr(\rho \rho_0)$ is the fidelity between the true quantum state $\rho$ and the classical guess $\rho_0$ (taken to be pure).
The $\simeq$ denotes the asymptotic scaling at large $D$, to leading order. 
Thus in the mixed phase, where $\avgP \sim D^{-s} \ll 1$, ${\sf XEB}'$ is exponentially close to 1 regardless of the fidelity $F$ between the unknown state $\rho$ and the guess $\rho_0$, whereas in the pure phase it approaches a constant value that is informative about $F$:
\begin{equation}
        \overline{{\sf XEB}'}  \xrightarrow{D\to\infty} \left\{ 
    \begin{aligned} & 1 & (\text{mixed phase}) \\ 
    & 1+{\rm const.}\times F & (\text{pure phase}) \end{aligned} \right.
    \label{eq:xebprime_transition}
\end{equation}

To practically assess the learnability of $F$, we must also consider the fluctuations of $p(\rho_0|\mathbf{m})$ across experimental shots $\mathbf{m}$. We address this question in App.~\ref{app:xeb}, where we show that 
\begin{equation}
    \delta {\sf XEB}' \sim D^{-s/2} 
    \label{eq:xeb_uncertainty}
\end{equation}
to leading order in large $D$. 
Learning the value of the fidelity $F$ to additive error $\epsilon$ requires learning the value of ${\sf XEB}' \simeq 1 + \avgP F$ to additive error $\avgP \epsilon \sim D^{-s} \epsilon$; this requires a number of repetitions 
\begin{equation}
    M \sim \frac{(\delta {\sf XEB}')^2}{(\avgP \epsilon)^2} \sim D^s \epsilon^{-2},
    \label{eq:xeb_sample_complexity}
\end{equation} 
which undergoes a sharp change form constant to exponential at the MIPT.

\subsection{Shadow estimation \label{sec:xeb_shadows}}

The same phase transition in sample complexity as Eq.~\eqref{eq:xeb_sample_complexity} can be straightforwardly obtained from shadow estimation of the fidelity $F = \bra{\psi} \rho \ket{\psi}$ with a many-body state $\ket{\psi}$. 
The sample complexity of learning $F$ is quantified by the squared shadow norm of the operator $O = \ketbra{\psi}$:
\begin{equation}
    \shdn{\ketbra{\psi}}^2 = D \Tr(\rho \otimes \mathcal{M}^{-1}(\ketbra{\psi})^{\otimes 2}\, \sigma^{(3)}).
    \label{eq:shdn_psipsi}
\end{equation}
with $\sigma^{(3)}$ the third moment of the ensemble of trajectories:
\begin{equation}
    \sigma^{(3)} = \sum_{\mb m} \pi_{\mb m} \sigma_{\mb m}^{\otimes 3}.
    \label{eq:third_moment}
\end{equation}

With pre-scrambling, Eq.~\eqref{eq:shdn_psipsi} can be computed analytically; see App.~\ref{app:xeb} for details.
In all, we obtain to leading order in large $D$
\begin{align}
    \shdn{\ketbra{\psi}}^2 
    & = \avgP^{-1} + F \avgP^{-2} \avgPthree + O(1/D) \nonumber \\
    & \sim D^s , \label{eq:shdn_xeb_leading}
\end{align}
where $\avgPthree = \mathbb{E}_{\mb m} \Tr(\sigma_{\mb m}^3)$ relates to the third Renyi entropy of the ensemble of trajectories, and the second line uses the fact that $\avgP \geq \avgPthree$.
This gives the same sample complexity ($\sim D^s$) as Eq.~\eqref{eq:xeb_sample_complexity}.

To summarize, we have shown that the cross-entropy diagnostic of Ref.~\cite{li_cross_2023} (Sec.~\ref{sec:xeb_review}) can be reinterpreted, with a small modification, as a protocol to learn the fidelity of unknown initial state $\rho$ with another state $\rho_0 = \ketbra{\psi}$ (Sec.~\ref{sec:mod_xeb}); this task is easy (requires a constant number of experimental samples) in the disentangling phase, and hard (requires an exponential number of samples $\sim q^{sN}$) in the entangling phase.
In turn, the task can be straightforwardly phrased in terms of estimation of the fidelity under eavesdropper's shadows (Sec.~\ref{sec:xeb_shadows}); its complexity, given by the squared shadow norm of the projector $\rho_0 = \ketbra{\psi}$, exhibits a transition from constant to exponential at the MIPT.

\section{Learning charge in $U(1)$-symmetric dynamics \label{sec:charge}}

A setting where {learnability transitions} in monitored dynamics are already well-established is that of systems with a $U(1)$ symmetry~\cite{agrawal_entanglement_2022,barratt_transitions_2022,barratt_field_2022}. There, one may try to learn the global charge $Q$ of the system from measurements of the local charge density $q_x$. The success of this learning task depends on the decoder (i.e. classical prediction algorithm) used; however, granting Eve arbitrary classical computational resources, the learnability transition~\cite{barratt_transitions_2022} was found to coincide with the {\it charge-sharpening} transition, discussed below~\cite{agrawal_entanglement_2022,barratt_field_2022}.
Here we recover the same result in the framework of eavesdropper's shadows, where it takes the form of a transition in the shadow norm of the global charge operator $\shdn{\hat Q}$.

\subsection{Setup \label{sec:u1_setup}}

We consider a system of $N$ qubits with a $U(1)$ symmetry generated by the charge operator 
\begin{equation}    
\hat{Q} = \frac{1}{2} \sum_i Z_i = \sum_{Q=-N/2}^{N/2} Q \hat{\Pi}_Q
\end{equation}
where $\hat{\Pi}_Q$ are orthogonal projectors on the charge sectors, of rank $\binom{N}{Q+N/2}$. 
The system evolves under a combination of $U(1)$-symmetric unitary gates and measurements of the local charge density $Z_i$. 

Ref.~\cite{agrawal_entanglement_2022} identified a {\it charge sharpening} transition in this class of models.
Charge sharpening is the loss of charge fluctuations over the course of symmetric monitored dynamics; it is analogous to the dynamical purification transition, but restricted to a state's {\it number entropy}\footnote{Writing a symmetric state $\rho$ as a direct sum of states in each charge block $\rho = \bigoplus_Q p_Q \rho_Q$, the number entropy is $S_n = -\sum_Q p_Q \ln p_Q$.}. 
Purification of the number entropy corresponds to projecting a state into a single charge sector. The time scale for this process (sharpening time, $t_\#$) undergoes a transition from $\sim \log(N)$ to $\sim N$. 
The transition was located at a critical measurement rate $p = p_\# < p_c$, where $p_c$ is the entanglement (or purification) critical point. In other words, one has a transition between a {\it fuzzy phase} ($p < p_\#$) and a {\it sharp phase} ($p > p_\#$) within the entangling phase, while the disentangling phase ($p > p_c$) is always sharp. 

The charge sharpening transition $p = p_\#$ was found to correspond to a transition in learnability of the total charge in the initial state $\rho$, at least in the limit where Eve has access to complete information about the circuit and unlimited classical computational resources (which is the setting we consider).
This result can be recovered straightforwardly in our ``eavesdropper's shadows''.
The formalism of Sec.~\ref{sec:shadows} carries over to this case; however, the shadow channel $\mathcal{M}$, Eq.~\eqref{eq:shadow_channel_petz}, is {\it not} invertible. 
To see this, let us note that the states $\sigma_{\mathbf m}$ are diagonal in the charge: they are produced by starting from the fully-mixed state $\mathbb{I}/D$ and acting with $Z$ measurements and $U(1)$-symmetric gates, neither of which can create coherences between charge sectors. Then, defining for each operator $O$ the {\it charge-diagonal} component $O_\Delta = \int\frac{{\rm d}\phi}{2\pi} e^{-i\phi\hat{Q}} O e^{i\phi \hat{Q}} = \sum_Q {\hat \Pi}_Q O {\hat \Pi}_Q$, we have
\begin{align}
    \mathcal{M}(O_\Delta) 
    & = D \sum_{\mathbf m} \pi_{\mathbf m} \Tr(\sigma_{\mathbf m}O_\Delta) \sigma_{\mathbf m}  \nonumber \\
    & = D \sum_{\mathbf m} \pi_{\mathbf m} \Tr(\sigma_{\mathbf m, \Delta} O) \sigma_{\mathbf m} = \mathcal{M}(O)
    \label{eq:M_chargediagonal}
\end{align}
(we have used cyclicity of the trace and the fact that $\sigma_{\mathbf m, \Delta} = \sigma_{\mathbf m}$ for all $\mathbf m$).
It follows that all operators that are off-diagonal in the charge are in the kernel of $\mathcal{M}$:
\begin{equation}
    \mathcal{M}(O - O_\Delta) = 0.
\end{equation}
This means that such operators are {\it not learnable}, with any number of samples, under this classical shadows protocol. However, when restricted to the subspace of charge-diagonal operators, $\mathcal M$ may become invertible and it may be possible to successfully learn all charge-diagonal operators.
Here we focus on the problem of learning total charge $Q$, which by definition is charge-diagonal and so in principle learnable via eavesdropper's shadows.
The sample complexity of this task is determined by the squared shadow norm $\shdn{\hat Q}^2$, which we study next.

\subsection{Shadow norm of the charge operator}

The shadow norm in general is given by 
\begin{equation}
    \shdn{\hat Q}^2 = D \Tr(\rho \otimes \mathcal{M}^{-1} (\hat Q)^{\otimes 2} \sigma^{(3)}),
    \label{eq:Q_shdn}
\end{equation}
which depends on the initial state $\rho$. 
Different choices for the initial state are possible---e.g. Ref.~\cite{barratt_transitions_2022} considers initial states that are in either of two charge sectors $Q_0$, $Q_1$. 
Here for simplicity and generality we take an average over all possible initial states (according to any 1-design distribution, such that $\int {\rm d}\rho\, \rho = \mathbb{I}/D$).
This yields the state-averaged shadow norm:
\begin{equation}
    \avgshadownorm{\hat Q}^2 
    = \Tr(\mathcal{M}^{-1}(\hat Q)^{\otimes 2} \sigma^{(2)})
    = \frac{1}{D} \Tr[\hat{Q} \mathcal{M}^{-1}(\hat Q)]
    \label{eq:Q_avgshdn}
\end{equation}
where the first equality comes from the fact that $\Tr_1(\sigma^{(3)}) = \sigma^{(2)}$ and the second from the definition of $\mathcal{M}$, Eq.~\eqref{eq:shadow_channel_petz}.

It is now helpful to introduce super-ket/bra notation and write Eq.~\eqref{eq:Q_avgshdn} as $\avgshadownorm{\hat Q}^2 = (\hat{Q}| \mathcal{M}^{-1}|\hat{Q}) / D$.
For all invertible Hermitian forms $H$ and all nonzero complex vectors $v$ we have, by convexity, $v^\dagger H^{-1} v / (v^\dagger v) \geq [v^\dagger H v / (v^\dagger v)]^{-1}$.
Therefore the following bound holds:
\begin{equation}
    \avgshadownorm{\hat{Q}}^2 \geq \frac{(\hat{Q}|\hat{Q})^2}{D(\hat{Q}|\mathcal{M}|\hat{Q})}.
\end{equation}
The numerator is computed straightforwardly: 
\begin{equation}
(\hat{Q}|\hat{Q}) = \Tr(\hat{Q}^2) =\frac{1}{4} \sum_{i,j} \Tr(Z_i Z_j) = \frac{N}{4} D.
\end{equation}
For the denominator, we note
\begin{align}
    \frac{(\hat{Q}|\mathcal{M}|\hat{Q})}{D}
    & = \sum_{\mathbf m} \pi_{\mathbf m} \Tr(\hat{Q} \sigma_{\mb m})^2
    = \mathbb{E}_{\mathbf m}[\langle \hat{Q}\rangle_{\mathbf m}^2].
    \label{eq:u1_shdn_bound}
\end{align}
This is the {\it variance across trajectories} of the charge expectation value\footnote{Since $\mathbb{E}_{\mb m} \sigma_{\mb m} = \mathbb{I}/D$, we have $\mathbb{E}_{\mathbf m}[\langle \hat{Q} \rangle_{\mathbf m}] = \Tr(\hat Q)/D = 0$ and thus $\mathbb{E}_{\mathbf m}[ \langle \hat{Q} \rangle_{\mathbf m}^2] = {\rm var}_{\mathbf m} \langle \hat{Q}\rangle_{\mathbf m} $.}. 
This is directly related to the order parameter of the charge sharpening transition used in Ref.~\cite{agrawal_entanglement_2022}, the {\it trajectory-averaged charge fluctuation} $\delta Q = \mathbb{E}_{\mathbf m} [\langle \hat{Q}^2 \rangle_{\mathbf m} - \langle \hat{Q} \rangle_{\mathbf m}^2]$: 
in particular, we have 
\begin{equation}
{\rm var}_{\mathbf m}[\langle \hat{Q} \rangle_{\mathbf m}] = \frac{N}{4} - \delta Q.
\end{equation}
This follows from the fact that the first term in $\delta Q$ is taken in the fully-mixed state and gives $\Tr(\hat{Q}^2)/D = N/4$. 

We can thus recast the bound Eq.~\eqref{eq:u1_shdn_bound} in terms of the charge-sharpening order parameter:
\begin{equation}
        \avgshadownorm{\hat{Q}}^2 
        \geq \frac{\delta Q_0}{1 - \delta Q(t)/\delta Q_0}
        \label{eq:shdn_charge_result}
\end{equation}
where $\delta Q_0 = N/4 = \Tr(\hat{Q}^2)$ is the quantum fluctuation of charge in a completely-mixed state. 
Below we work out the consequences of this bound in each phase.

{\bf Sharp phase.} 
We have $\delta Q(t) \sim \delta Q_0 e^{-ct}$ with $c>0$ a constant. At sufficiently large constant depth $t \sim \frac{1}{c} \log(1/\epsilon)$ the charge sharpens to within tolerance $\epsilon$, and we have $\shdn{\hat{Q}}^2 \geq (1-\epsilon)N/4$: it takes $\Omega(N)$ experiments, or a total of $\Omega(N^2)$ measurements, to learn the charge of the initial state within error $\epsilon$. This bound is expected from a simple central limit theorem argument~\cite{barratt_transitions_2022} and would apply even upon measuring all qubits immediately\footnote{The random initial state we consider need not have definite charge. It has mean charge $0$ and fluctuations $O(\sqrt{N})$, so lowering the uncertainty to $O(1)$ takes of order $N$ experiments. Note that this is unlike Ref.~\cite{barratt_transitions_2022}, where the initial state is promised to be of definite charge and so a single shot may suffice.}, so the bound in this phase is trivial. It is easy to see that, if each outcome $\mb m$ uniquely specifies a value $Q$ of the charge (as is the case after sharpening), then $O(N)$ experiments are also sufficient.

{\bf Fuzzy phase.} 
We have~\cite{agrawal_entanglement_2022} $\delta Q(t) \sim \delta Q_0 e^{-ct/N}$---i.e., the sharpening time $t_\#$ diverges linearly in system size $N$. 
This gives
\begin{equation}
    \avgshadownorm{\hat{Q}}^2 \geq \frac{N/4}{1-e^{-ct/N}} \simeq \frac{N^2}{4ct}
    \label{eq:shdn_fuzzy}
\end{equation}
where the approximation holds at times $t\ll N$. 
At finite $t$, this proves that $\sim N^2/t$ samples are needed, for a total of $M_{\rm tot} \sim (N^2/t) \times (Nt) \sim N^3$ local measurements. 
This result is parametrically larger than in the sharp phase, scaling as $\sim N^3$ rather than $\sim N^2$. Furthermore it is analogous to our previous result on learning Pauli operators, Sec.~\ref{sec:pauli_infopermeas}, in that an invariant unit of information extracted per measurement emerges\footnote{This holds as long as $t$ is not much larger than the sharpening time $t_\# \sim N$ so that the scaling ansatz for $\delta Q(t)$ applies.}. In the fuzzy phase, this unit is suppressed by a factor of $N$ relative to the sharp phase.

To summarize, we have analyzed the state-averaged shadow norm of the charge operator $\avgshadownorm{\hat{Q}}^2$, which quantifies the number of experimental repetitions needed to learn the charge $\langle \hat{Q}\rangle$ of an unknown initial state from the measurement record and knowledge of the circuit. 
We have shown that this behaves differently in the two measurement-induced phases: it is $O(N)$ throughout the sharp phase, whereas it is bounded below by $\sim N^2/t$ in the fuzzy phase. 
The total number of measurements (there are $\sim Nt$ measurements per experimental repetition) thus transitions from $O(N^2)$ (sharp phase) to $\Omega(N^3)$ (fuzzy phase).
Thus, in the same way in which the entanglement phase transition was shown to be a transition in learnability of generic properties such as Pauli expectation values (Sec.~\ref{sec:pauli}) and many-body fidelities (Sec.~\ref{sec:xeb}), the charge-sharpening transition emerges as a transition in learnability of the charge $\hat{Q}$ in symmetric monitored dynamics.

\section{Discussion \label{sec:discussion}}

\subsection{Summary}

We have presented a {\it learnability} perspective on measurement-induced phases of quantum information, based on the ability of an eavesdropper to learn properties of an unknown quantum state of the system from classical mid-circuit measurement data. 
The learnability perspective is complementary to more well-established perspectives on the MIPT based on the entanglement properties of post-measurement states of the quantum many-body system, or to the {\it coding} perspective which focuses on recovering quantum information from the combined quantum-classical state of the many-body system and measurement device. 
In our setting, the eavesdropper, Eve, collects measurement records $\mb m$ from multiple repetitions of the experiment. In combination with a complete classical description of the dynamics and unconstrained classical computing resources, Eve can try to learn properties of the system's unknown initial state $\rho$---{\it without} access to the final, post-measurement state. 

We have shown that the MIPT generically coincides with a complexity phase transition for these tasks, i.e., a transition in the number of measurement outcomes needed to predict properties of $\rho$ to a fixed accuracy. 
This is underpinned by a transition in the {\it informational power} of the POVM associated to monitored dynamics. This is a an invariant measure of the information flow from the quantum state to the classical measurement record (technically a channel capacity) which sharply changes from finite to exponentially small at the MIPT.

We give an operational meaning to this transition through the framework of classical shadows, which furnishes a unified language to describe learnability transitions in various different contexts. To this end, we introduced a family of classical shadows protocols that the eavesdropper may use to concretely predict properties of $\rho$, and analytically showed that they carry signatures of the MIPT; namely the shadow channel used in the estimation process depends on the second moment of an associated ensemble of monitored quantum trajectories, which can undergo a MIPT. 
We have then unpacked the consequences of this result on several estimation tasks of interest: 

{\it Pauli expectation values.}
We have found the critical point $p = p_c$ to be (on average) optimal for the estimation of Pauli operators---the critical point balances the negative effects of spatial locality in the disentangling phase (which makes learning large Pauli operators more difficult) against the overall lack of information in the entangling phase. 
Washing out the effects of locality with a pre-scrambling step (i.e. a sufficiently-deep random unitary circuit preceding the monitored dynamics), we analytically derived the sample complexity of Pauli estimation for any traceless Pauli to be $\sim q^{(1+s)N}$, with $q$ the local Hilbert space dimension, $N$ the size of the system, and $s \in [0,1]$ the order parameter of the entangling phase (entropy density). The MIPT thus manifests as a non-analyticity in the coefficient of the exponential in this case.

{\it Many-body fidelity.}
We analytically derived the sample complexity of fidelity estimation (with pre-scrambling) to be $\sim q^{sN}$, transitioning from constant to exponential at the MIPT. Furthermore, we found a close connection between shadow estimation of the fidelity and a previously-proposed order parameter for the MIPT based on a linear cross-entropy diagnostic~\cite{li_cross_2023}.

{\it Charge.} We considered models of monitored dynamics with a $U(1)$ symmetry and derived the sample complexity of learning the global charge expectation $\langle \hat{Q} \rangle$ on the initial state. We have found a transition, this time between distinct power-laws in $N$, at the charge-sharpening transition~\cite{agrawal_entanglement_2022}. Thus we have recast previous results on charge learnability transitions in the unified language of shadow estimation, on the same footing as the other learnability transitions identified above.
A striking result, derived both for the entanglement and charge-sharpening transitions, is the emergence of an invariant amount of information extracted {\it per measurement} by the eavesdropper.

\subsection{Experimental implications}

In this work we have focused uniquely on the {\it sample complexity} of the various learning tasks: how many repetitions of the quantum experiment are necessary for learning. We have intentionally neglected the issue of {\it classical} computational complexity. This is key to any practical considerations. 

Several recent works have, in various ways, introduced ``hybrid'' quantum-classical order parameters for the MIPT that trade experimental sample complexity (associated to a `postselection overhead' of obtaining post-measurement properties) for classical computational complexity~\cite{gullans_scalable_2020,dehghani_neural-network_2023,garratt_measurements_2023,li_cross_2023,hoke_measurement-induced_2023,garratt_probing_2023}. This is generally advantageous as 
(i) classical resources are cheaper and more available than quantum ones,
(ii) the required classical simulation may in fact be efficient (e.g. in Clifford or matchgate circuits, etc), and
(iii) even when that is not the case, the exponential barrier for classical simulation (typically $\sim \exp(N)$) may be more favorable than that of quantum sampling (typically $\sim \exp(NT)$, $T$ being the duration of the dynamics).

Our perspective in this work automatically leads to a family of these quantum-classical order parameters, namely the {\it variances} of shadow estimators for various properties of $\rho$. Such variances can be estimated from a small number of experimental datapoints in both phases. For example, learning a Pauli expectation value $\langle P \rangle$ in the mixed phase is hard due to the large variance ${\rm var}(\hat{p}) \sim D^{1+s}$ of the shadow estimator $\hat{p}$; our estimate of $\langle P\rangle$ after $M$ iterations of the experiment, $\frac{1}{M} \sum_{i=1}^M \hat{p}_i$, carries an uncertainty $\sim \sqrt{{\rm var}(\hat p)/M}$, so that $M \sim D^{1+s}$ experiments are needed in order to make the error small. However, the variance ${\rm var}(\hat{p})$ itself is easily estimated from $M = O(1)$ iterations of the experiment, and its scaling with system size $N$ ($\sim q^{[1+s]N}$) serves as an order parameter of the MIPT.
We note that all samples generated by the quantum experiment are valid and can be used for shadow estimation. The practical complexity reduces entirely to {\it classical computation}, as various steps of shadow estimation (computation of the snapshots, the shadow channel, the inverted snapshots, and the observable estimators) may require exponential classical resources.

\subsection{Outlook}

Our work opens some interesting directions for future work. First of all, we have chosen to focus only on the {\it sample complexity} of learning, neglecting the classical computational complexity. It would be interesting, especially with an eye to practical applications, to revisit our results with restrictions on classical computation resources: how much can Eve learn, e.g., with only polynomial-time classical algorithms?
The methods used in this work generically require exponential-time computation; are there other more efficient methods that can still capture the learnability transition? 

For the problem of learning the global charge from $U(1)$-symmetric dynamics~\cite{barratt_transitions_2022}, it was found that polynomial-time decoders still give rise to learnability transitions, albeit at a larger measurement rate $p$; with increasing computational resources, one eventually recovers the ``intrinsic'' transition at $p = p_\#$ (the charge-sharpening transition~\cite{agrawal_entanglement_2022}).
Does a similar picture hold for the entanglement transition? We have found that, with unlimited classical computation, the MIPT $p = p_c$ yields a learnability transition; would the transition move to some larger measurement rate $p > p_c$ upon restricting classical computational resources?
The percolation threshold (e.g. $p=1/2$ in one-dimensional brickwork circuits~\cite{skinner_measurement-induced_2019}), above which the monitored dynamics breaks down into finite-sized space-time regions, may be an upper bound for such a transition in polynomial-time learnability. We leave this question as an interesting direction for follow-up research.

In this work we have focused on {\it non-Gaussian} systems. However, Gaussian systems can exhibit potentially richer types of measurement-induced criticality and different entanglement phase diagrams (e.g. without a stable volume-law phase)~\cite{cao_entanglement_2019,buchhold_effective_2021,fava_nonlinear_2023}. It would be interesting to understand how such behaviors map onto the performance of eavesdropper's shadows, relative to standard {\it `matchgate shadows'} for fermionic systems~\cite{wan_matchgate_2023}.

Another open question is the precise nature of eavesdropper's shadows across the phase diagram, and especially at the critical point. We have found that, without pre-scrambling, the MIPT appears as an optimum in the learnability of typical or average Pauli operators on the system, see Sec.~\ref{sec:pauli_locality} and Fig.~\ref{fig:shdn}. This is due to the combination of two effects: the lack of informational power in the entangling phase, and the role of locality in the disentangling phase (the latter makes large Pauli operators particularly hard to learn). 
We have conjectured that in the entangling phase, eavesdropper's shadows function similarly to random Clifford shadows~\cite{huang_predicting_2020}, up to an overall inflation of sample complexity by the exponential factor $q^{sN}$ (due to the suppressed informational power); whereas in the disentangling phase, they function similarly to {\it shallow shadows}~\cite{akhtar_scalable_2023,bertoni_shallow_2022,arienzo_closed-form_2022,ippoliti_operator_2023} of variable depth (recovering the zero-depth limit, i.e. random Pauli measurements, at $p=1$).
This leaves the question of eavesdropper's shadows at the MIPT. The logarithmic scaling of entanglement at the critical point is expected to yield a distinctive fingerprint on the shadow norm distribution (via Eq.~\eqref{eq:lambda_vs_purities}), perhaps similar to classical shadows based on tree tensor networks. 
Testing and quantifying these conjectures, and finding potential applications, are exciting directions for future research. 

{\it Note added:} Shortly after completion of our manuscript we became aware of related work on classical shadows from monitored dynamics~\cite{akhtar_measurement-induced_2023}.

\acknowledgments 
We thank Ehud Altman, Bryan Clark, Xiaozhou Feng, Sam Garratt, Sarang Gopalakrishnan and Tibor Rakovszky for helpful discussions. 
We are especially grateful to Yaodong Li for insightful discussions in the early stages of this project.
M.~I. was partly supported by the Gordon and Betty Moore Foundation's EPiQS Initiative through Grant GBMF8686.
V.~K. acknowledges support from the US Department of Energy, Office of Science, Basic Energy Sciences, under Early Career Award Nos. DE-SC0021111, from the Alfred P. Sloan Foundation through a Sloan Research Fellowship and from the Packard Foundation through a Packard Fellowship in Science and Engineering. 
Numerical simulations were carried out on Stanford Research Computing Center's Sherlock cluster. 
This project originated at the KITP program {\it ``Quantum Many-Body Dynamics and Noisy Intermediate-Scale Quantum Systems''}; KITP is supported by the National Science Foundation under Grant No. NSF PHY-1748958.

\appendix

\section{Informational power \label{app:ip}}

In this Appendix we collect various technical results related to the computation of the informational power,  Sec.~\ref{sec:infopower}.

\subsection{Derivation of Eq.~\eqref{eq:infopower_result_Q}}

Here we derive the analytical expression for the pre-scrambled informational power in terms of the subentropy, Eq.~\eqref{eq:infopower_result_Q}. 

With pre-scrambling by a random many-body unitary $V$, our POVM is given by $\Pi = \{V^\dagger \effect{\mb m} V \}$, with effects indexed by the pair $(\mathbf{m}, V)$ which plays the role of a generalized ``outcome''. The pair $(\mb{m}, V)$ occurs with probability\footnote{Note this is a probability density over the continuous variable $V$ with the Haar measure ${\rm d}V$. The POVM normalization condition reads $\sum_{\mathbf m} \int {\rm d}V\, (V^\dagger \effect{\mb m} V ) = \sum_{\mathbf m} \Tr(\effect{\mb m}) \mathbb{I}/D = \mathbb{I}$.} $\Tr(V\rho V^\dagger \effect{\mb m})$. 
Due to pre-scrambling, {\it all} pure-state ensembles $\mathcal{E} = \{(p_i, \ketbra{\psi_i})\}$ that unravel the same density matrix $\rho$ have the same mutual information with $\Pi$~\cite{jozsa_lower_1994}:
indeed, we have
\begin{align}
    I(\mathcal{E}: \Pi)
    & = \sum_i p_i \sum_{\mathbf m} \int {\rm d}V\, 
    \bra{\psi_i} V^\dagger \effect{\mb m} V \ket{\psi_i} \times \nonumber \\
    & \qquad \times \ln \frac{\bra{\psi_i} V^\dagger \effect{\mb m} V \ket{\psi_i} }{ \Tr(\effect{\mb m}\rho)};
\end{align}
then, introducing a Haar-random state $\ket{\phi} \equiv V \ket{\psi_i}$ and replacing integration over the unitary group (Haar measure ${\rm d}V$) with integration over the Hilbert space (Haar measure ${\rm d}\phi$), we obtain
\begin{align}
    I(\mathcal{E}: \Pi)
    & = \sum_{\mb{m}} \pi_{\mathbf m} \int {\rm d}\phi\, 
    f( \bra{\phi} D\sigma_{\mathbf m} \ket{\phi}) \nonumber \\
    & \qquad - \sum_{\mb m} \pi_{\mb m} \int {\rm d}V\, f[\Tr(D\sigma_{\mb m}V\rho V^\dagger)]
    \label{eq:MI_integral}
\end{align}
where $\sigma_{\mb m}$ is as in Eq.~\eqref{eq:canonical_state_ens} and we have introduced the shorthand $f(x) = x\ln(x)$.
This shows that all pure-state ensembles $\mathcal{E}$ that unravel the same $\rho$ have the same mutual information $I(\mathcal{E}:\Pi)$, as claimed.

We can now proceed to maximize the mutual information. It can be shown that the second line of Eq.~\eqref{eq:MI_integral} is $\leq 0$ (by convexity of $f(x)$), thus it can be maximized by setting $\rho = \mathbb{I}/D$. Since the optimal ensemble $\mathcal{E}$ is always made of pure states\footnote{
The mutual information is non-decreasing under replacement of a mixed state $\rho_i$ in the ensemble with a pure decomposition, $(p_i, \rho_i) \mapsto \{( p_{ij}, \ketbra{\psi_{ij}}) \}$ such that $\sum_j p_{ij} \ketbra{\psi_{ij}} = p_i \rho_i$. This follows from convexity of $f(x) = x\ln(x)$. Thus the optimal ensemble can always be written in terms of pure states only.}, 
it follows that the optimization is trivial---any 1-design ensemble of pure states (e.g. the computational basis) maximizes the mutual information. 
The informational power, following Eq.~\eqref{eq:MI_integral}, is thus given by
\begin{align}
    W(\Pi) & = \sum_{\mathbf m} \pi_{\mathbf m} \mathcal{G}(\sigma_{\mathbf m}), \label{eq:infopower_prescrambled} \\
    \mathcal{G}(\sigma) & = \int{\rm d}\phi\, \bra{\phi} D\sigma\ket{\phi} \ln \bra{\phi} D\sigma\ket{\phi}. \label{eq:G_of_rho}
\end{align}

The Haar integral $\mathcal{G}$ has been worked out for general $\sigma$ in Ref.~\cite{jozsa_lower_1994}, and gives
\begin{equation}
    \mathcal{G}(\sigma) = Q(\mathbb{I}/D) - Q(\sigma),
    \label{eq:infopower_vs_subentropy}
\end{equation}
where $Q(\sigma)$ is the subentropy, see below. This yields Eq.~\eqref{eq:infopower_result_Q}.

\subsection{Subentropy \label{app:subentropy}}

Here we provide some more details about the subentropy $Q(\rho)$~\cite{jozsa_lower_1994}.

Like the von Neumann entropy $S(\rho)$, the subentropy $Q(\rho)$ is solely a function of the spectrum of $\rho$, $\{\lambda_j\}_{j = 1}^D$:
\begin{equation}
Q(\rho) = -\sum_{j=1}^D \frac{\lambda_j \ln \lambda_j}{\prod_{k\neq j} (1-\lambda_k/\lambda_j) }.
\label{eq:subentropy_explicit}
\end{equation}
If the spectrum is degenerate with $\lambda_i = \lambda_j$, the formula can be regularized by taking a limit $\lambda_i \to \lambda_j$; $Q(\rho)$ is finite and well-defined.

Another connection between entropy and subentropy is that they bound (above and below, respectively) the ``accessible information'' of a state ensemble $\mathcal{E}$, defined as
\begin{equation}
A(\mathcal{E})\equiv \max_\Pi I(\mathcal{E}:\Pi).
\label{eq:accessible_info}
\end{equation}
Note the duality with the informational power of a POVM, cf Eq.~\eqref{eq:infopower_def}. 
As shown in Ref.~\cite{jozsa_lower_1994}, one has $Q(\rho) \leq A(\mathcal{E}) \leq S(\rho)$, where $\rho = \sum_i p_i \rho_i$ is the average density matrix of the state ensemble.
The ensemble $\mathcal{E}$ attaining the lower bound $A(\mathcal{E}) = Q(\rho)$ for a given $\rho$ is known as the {\it Scrooge ensemble} and has recently emerged as a candidate universal distribution for post-measurement states of subsystems in chaotic dynamics~\cite{cotler_emergent_2023,choi_preparing_2023,ippoliti_solvable_2022}.

An important difference with the entropy is that the subentropy cannot be extensive. In fact it is bounded above by a constant: we have $Q(\rho) \leq Q(\mathbb{I}/D) = 1-\gamma - \delta H(D) < 1-\gamma < 0.424$. 
(Here $\gamma = 0.577\dots$ is Euler-Mascheroni's constant and $\delta H(x) = \sum_{j=1}^x 1/j - \ln(x) -\gamma$, as in the main text).

\subsection{Informational power of Clifford monitored dynamics \label{app:ip_clifford}}

Here we derive Eq.~\eqref{eq:subentropy_stab}, which underlies the result for the informational power of Clifford monitored circuits, Eq.~\eqref{eq:infopower_cliff_result}.

We consider the Haar integral $\mathcal{G}(\rho)$ from Eq.~\eqref{eq:G_of_rho} for the case in which $\rho$ is proportional to a projector, $\rho = \Pi/r$ where $\Pi^2 = \Pi$ and $r$ is the rank of $\Pi$. 
This includes the case of stabilizer states. 
First we use a standard replica trick to write
\begin{equation}
    \mathcal{G}(\rho) = \partial_n \int {\rm d}\phi\, \bra{\phi} D\rho \ket{\phi}^{1+n} \big|_{n = 0}.
\end{equation}
We then evaluate the integral for integer values of $n$ by using the form of the $(n+1)$-th moment of the Haar measure on a $D$-dimensional Hilbert space~\cite{cotler_emergent_2023,ippoliti_solvable_2022}:
\begin{equation}
    \rho_{{\rm Haar},D}^{(n+1)} \equiv \int{\rm d}\phi \ketbra{\phi}^{\otimes n+1} = \frac{(D-1)!}{(D+n)!} \sum_{\sigma \in S_{n+1}} \hat{\sigma} \label{eq:rho_haar_D}
\end{equation}
where $\sigma$ is a permutation of $n+1$ elements and $\hat{\sigma}$ is the associated replica permutation operator.
We obtain
\begin{equation}
    \mathcal{G}(\rho) = \partial_n \left[ \frac{(D-1)!}{(D+n)!} \sum_{\sigma \in S_{n+1}} \Tr((D\rho)^{\otimes n+1} \hat{\sigma}) \right]_{n = 0}.
\end{equation}

Using the fact that $\rho = \Pi/r$ ($r$ being the rank of the projector $\Pi$) we have $\Tr(\Pi^{\otimes n+1} \hat{\sigma}) = \prod_{i=1}^{|\sigma|} \Tr(\Pi^{n_i}) = r^{|\sigma|}$, with $|\sigma|$ the number of cycles in the permutation $\sigma$, and $n_i$ the length of each cycle. 
Thus
\begin{equation}
    \mathcal{G}(\rho) = \partial_n \left[ \frac{(D-1)!}{(D+n)!} \left(\frac{D}{r} \right)^{n+1} \sum_{\sigma \in S_{n+1}} r^{|\sigma|} \right]_{n = 0}.
\end{equation}
The summation over $\sigma$ can be done exactly by noting that, upon taking the trace of Eq.~\eqref{eq:rho_haar_D} and replacing $D\mapsto r$, one has
\begin{equation}
    \frac{(r-1)!}{(r+n)!} \sum_{\sigma \in S_{n+1}} r^{|\sigma|} =  \Tr( \rho_{{\rm Haar},r}^{(n+1)})
    =  1.
\end{equation}
It follows that
\begin{equation}
\mathcal{G}(\rho) = \partial_n \left[ \left(\frac{D}{r} \right)^n \frac{D!}{(D+n)!} \frac{(r+n)!}{r!}  \right]_{n = 0},
\end{equation}
where the Hilbert space replicas are gone and we can now take the derivative (i.e. replica limit). 

Analytically continuing the factorial to the $\Gamma$ function, we have $(x+n)! = x! + x!(H_x-\gamma) n + O(n^2)$, with $H_x = \sum_{j=1}^x 1/j$ the harmonic sum. 
We conclude
\begin{equation}
\mathcal{G}(\rho) = \ln(D/r) + H_r - H_D = \delta H(r) - \delta H(D),
\end{equation}
with $\delta H(x) = H_x - (\ln(x)+\gamma)$ as in the main text.
Finally, writing the rank $r$ of the state in terms of the entropy as $q^S$ we obtain Eq.~\eqref{eq:subentropy_stab}.

\subsection{Renyi-2 informational power of general monitored dynamics \label{app:ip_renyi2}}

Here we introduce a Renyi-2 version of the informational power which is computable for general (non-stabilizer) states. Note that the Renyi-2 version of the mutual information on which this construction is based is {\it not} a valid mutual information (e.g. does not obey positivity); nonetheless in randomized settings it often behaves in a qualitatively similar way to the true mutual information, so our result here is suggestive of the presence of an informational power transition in general (non-stabilizer) monitored dynamics.

We define the Renyi-2 informational power $W_2$ by maximizing (over state ensembles $\mathcal{E}$) the Renyi-2 ``mutual information''
\begin{align}
    I_2(\mathcal{E}, \Pi) 
    & = S_2(p_i) + S_2(p_\alpha) - S_2(p_{i,\alpha}) \nonumber \\
    & = \ln \frac{\sum_{i,\alpha} p_{i,\alpha}^2}{\sum_{i,\alpha} p_i^2 p_\alpha^2}
\end{align}
Going through the same manipulations as in Eq.~\eqref{eq:MI_integral}, we have
\begin{align}
    I_2(\mathcal{E}, \Pi) 
    & = \ln \frac{\sum_i p_i^2 \sum_{\mathbf m} \int {\rm d}\phi\, \bra{\phi}\effect{\mb m}\ket{\phi}^2 }{ \sum_i p_i^2 \sum_{\mathbf m} \Tr(\effect{\mb m})^2/D^2}
\end{align}
Defining modified probabilities $\tilde{p}_i = p_i^2 / \sum_j p_j^2$ and $\tilde{\pi}_{\mathbf m} = \pi_{\mathbf m}^2 / \sum_{\mathbf m'} \pi_{\mathbf m '}^2 $
we arrive at
\begin{align}
    I_2(\mathcal{E}, \Pi) 
    & = \ln \sum_i \tilde{p}_i \sum_{\mathbf m} D^2 \tilde{\pi}_{\mathbf m} \int{\rm d}\phi\, \bra{\phi}\sigma_{\mathbf m}\ket{\phi}^2 
\end{align}
The integrand is independent of $i$, which again shows that the mutual information is independent of $\mathcal{E}$ owing to pre-scrambling. 
We get
\begin{align}
    W_2(\Pi) 
    & = \ln \sum_{\mathbf m} \tilde{\pi}_{\mathbf m} \frac{D}{D+1} \left[1 + \Tr(\sigma_{\mathbf m}^2) \right] \nonumber \\
    & = \ln \frac{1 + \avgP}{1 + 1/D}
\end{align}
where $\avgP$ is the average purity of the trajectories $\sigma_{\mb m}$, averaged over the modified distribution $\tilde{\pi}_{\mathbf m}$. 

It follows that the above-defined ``Renyi-2 informational power'' exhibits a MIPT: 
in the mixed phase, with $\avgP = D^{-s}$ and $s>0$, we have $W_2(\Pi)\to 0$ in the large-system limit;
in the pure phase, with $\avgP = q^{-S}$ and $S$ finite, we have $W_2(\Pi) \to \ln(1+q^{-S}) > 0$. 
Note that, due to the modified measure over trajectories, the transition is in a different universality class from the standard one~\cite{bao_theory_2020,li_cross_2023}.

\section{Alternative constructions for eavesdropper's shadows \label{app:shadows}}

Here we complete the discussion in Sec.~\ref{sec:shadows} by providing details on multiple options for classical shadows protocols based on generalized measurements, and how they relate to the MIPT. 
We start from the prescription discussed in the main text, and how it can be formally interpreted as {\it Petz recovery} of the quantum state $\rho$ from the measurement record $\mb m$. 
We then review two existing approaches for shadows based on generalized measurements---one based on least squares, one on maximum fidelity. We obtain the associated shadow channels and discuss how they are sensitive to the MIPT.

\subsection{Petz recovery}

We begin with the prescription followed in Sec.~\ref{sec:shadows}, i.e. $\eta_{\mb m} = \sigma_{\mb m}$.
To complement the heuristic justification based on time-reversed monitored dynamics, we show that the same prescription arises as the {\it Petz recovery map}~\cite{petz_sufficient_1986,barnum_reversing_2002,wilde_recoverability_2015,penington_replica_2022} relative to the channel $\mathcal{N}(\rho) = \sum_{\mb m} \Tr(\effect{\mb m} \rho) \ketbra{\mb m}$ (mapping quantum states to classical measurement records) and the reference state $\rho_0 = \mathbb{I}/D$. The Petz map for a noise channel $\mathcal{N}$ and reference state $\rho_0$ is defined in general as 
\begin{equation}
    \mathcal{R}_{\rm Petz}^{\rho_0, \mathcal{N}}(\bullet ) = \rho_0^{1/2} \mathcal{N}^\dagger \left[ \mathcal{N}(\rho_0)^{-1/2} \bullet \mathcal{N}(\rho_0)^{-1/2} \right] \rho_0^{1/2}. 
    \label{eq:petz_general}
\end{equation}
The Petz map tries to undo the action of the channel $\mathcal{N}$, which plays the role of noise; it manifestly succeeds for the reference state $\rho_0$, as seen by plugging in $\bullet = \mathcal{N}(\rho_0)$ in Eq.~\eqref{eq:petz_general}. 
It also succeeds for any other states whose relative entropy with $\rho_0$ is non-decreasing under the action of $\mathcal{N}$~\cite{petz_sufficient_1986,wilde_recoverability_2015}.
These properties have made it a useful tool from formal quantum information to applications in error correction~\cite{ng_simple_2010} and gravity~\cite{cotler_entanglement_2019,penington_replica_2022}.
For the task at hand, $\mathcal{R}_{\rm Petz}$ maps classical states (i.e. probability distributions over measurement records) to quantum states, which is indeed our goal: given some eavesdropped measurement record $\mb m$, predict the quantum state $\rho$ it came from. 

Using the fact that $\mathcal{N}(\rho_0) = \sum_{\mb m} \pi_{\mb m} \ketbra{\mb m}$ and $\mathcal{N}^\dagger[\ketbra{\mb m}] = \effect{\mb m}$,
explicit calculation of the Petz recovery on a classical state $p^{\rm exp} = \sum_{\mb m} p^{\rm exp}_{\mb m} \ketbra{\mb m}$ yields
\begin{equation}
    \mathcal{R}^{\rho_0, \mathcal{N}}_{\rm Petz} (p^{\rm exp}) 
    = \sum_{\mb m} p^{\rm exp}_{\mb m} \frac{\effect{\mb m}}{D\pi_{\mb m}}
    = \sum_{\mb m} p^{\rm exp}_{\mb m} \sigma_{\mb m}.
\end{equation}
In conclusion, the Petz recovery prescription says that, given an experimental outcome $\mb m$ (representable as a delta-function distribution $p^{\rm exp}_{\mb m'} = \delta_{\mb m, \mb m'}$), Eve should prepare the state $\sigma_{\mb m}$.

\subsection{Least squares}

Ref.~\cite{nguyen_optimizing_2022} proposes using $\eta_{\mathbf m} = \effect{\mb m}$ based on a least-squares criterion. Namely, given an experimentally-observed measurement record distribution $p^{\rm exp}_{\mb m}$, we can try to reconstruct the unknown state $\rho$ by minimizing the cost function 
\begin{equation}
    \mathcal{L}(\rho) = \sum_{\mb m} \left[ p^{\rm exp}_{\mb m} - \Tr(\effect{\mb m}\rho) \right]^2
    \label{eq:cost_function_ls}
\end{equation}
i.e. the two-norm of the distance between observed distribution $p^{\rm exp}_{\mb m}$ and predicted distribution $\Tr(\effect{\mb m}\rho)$. 
At this stage it is convenient to introduce some extra notation: we use $\dket{\mb{m}} = \ketbra{\mb m}$ to denote classical states of the measurement record, and $|A)$ to denote quantum operators as states in a doubled Hilbert space.
Defining again the quantum-to-classical channel $\mathcal{N} = \sum_{\mb m} \dket{\mb{m}} (\effect{\mb m}|$ already encountered in the discussion of the Petz recovery above, we can write the predicted distribution for a given $\rho$ as $\Tr(\rho\effect{\mb m}) = \dbra{\mb m} \mathcal{N} |\rho)$.
The cost function thus reads
\begin{equation}
    \mathcal{L}(\rho) = \left\| \dket{p^{\rm exp}} - \mathcal{N} |\rho) \right\|^2,
    \label{eq:cost_function_ls2}
\end{equation}
with $\dket{p^{\rm exp}} = \sum_{\mb m} p^{\rm exp}_{\mb m} \dket{\mb m}$ for short,
and its optimization reduces to usual least squares, with the well-known result
\begin{equation}
    |\hat{\rho}) = (\mathcal{N}^\dagger \circ \mathcal{N})^{-1} \mathcal{N}^\dagger \dket{p^{\rm exp}}.
    \label{eq:ls_solution}
\end{equation}

In a nutshell, this prescription says that for every run of the experiment, giving some outcome $\mb m$, we should construct a ``snapshot'' $\mathcal{N}^\dagger \dket{\mb m} = |\effect{\mb m})$ and then an ``inverted snapshot'' $\mathcal{M}^{-1}(\effect{\mb m})$, where the ``measurement channel'' $\mathcal{M}$ is given by
\begin{equation}
    \mathcal{M} = \mathcal{N}^\dagger \circ \mathcal{N} = \sum_{\mb m} |\effect{\mb m})(\effect{\mb m}|.
    \label{eq:meas_channel_ls_implicit}
\end{equation}

The operator $\effect{\mb m}$ is evidently not a state, due to its trace normalization: we have $\effect{\mb m} = D \pi_{\mb m} \sigma_{\mb m}$ in the notation introduced in Sec.~\ref{sec:review_povm}.
It follows that $\mathcal M$ is not a channel (it is not trace preserving). The shadows protocol works regardless, as the application of the inverse shadow channel $\mathcal{M}^{-1}$ takes care of the incorrect normalization. 

Finally, it is helpful to rewrite the shadow channel in terms of a state ensemble dual to our POVM $\{\effect{\mb m}\}$, in analogy with Eq.~\eqref{eq:shadow_channel_petz}. We have
\begin{align}
    \mathcal{M}(\rho) 
    & = \sum_{\mathbf m} \Tr(\rho \effect{\mb m})\effect{\mb m} 
    = D^2 \sum_{\mathbf m} \pi^2_{\mathbf m} \Tr(\rho\sigma_{\mathbf m})\sigma_{\mathbf m} \nonumber \\
    & = D^2 \left(\sum_{\mb m} \pi_{\mb m}^2\right) \Tr[ (\mathbb{I}\otimes \rho) \tilde{\sigma}^{(2)}] ,
    \label{eq:meas_channel_ls}
\end{align}
where $\tilde{\sigma}^{(2)}$ is the {second moment operator} of the ensemble $\mathcal{\tilde{E}} = \{ (\tilde{\pi}_{\mb m}, \sigma_{\mb m})\}$ defined by the usual states $\sigma_{\mb m} = \effect{\mb m} / \Tr(\effect{\mb m})$ [cf Eq.~\eqref{eq:canonical_state_ens}] but with a modified probability distribution $\tilde{\pi}_{\mb m} =  \pi_{\mb m}^2 / \sum_{\mb m'} \pi_{\mb m'}^2$:
\begin{equation}
    \tilde{\sigma}^{(2)} = \sum_{\mb m}\tilde{\pi}_{\mb m} \sigma_{\mb m}^{\otimes 2}.
\end{equation}
This trajectory ensemble also features a MIPT, but due to the modified weights it has a different universality~\cite{bao_theory_2020} and generally occurs at a different (though empirically very close~\cite{li_cross_2023}) measurement rate. 

\subsection{Maximum fidelity}

Finally, Ref.~\cite{acharya_shadow_2021} proposes using the pure state corresponding to the leading eigenvalue\footnote{We neglect degeneracies at this stage.} in $\effect{\mathbf m}$: $\ketbra{\psi_{\mathbf m}} = \lim_{n\to\infty} \effect{\mb m}^n / \Tr(\effect{\mb m}^n)$. 
This choice maximizes the the Haar-averaged fidelity between input and output states of the shadow channel, $F = \int {\rm d}\phi\, \bra{\phi} \mathcal{M}(\ketbra{\phi}) \ket{\phi}$: we have
\begin{align}
    F 
    & = D \sum_{\mb m} \pi_{\mb m} \int {\rm d}\phi \bra{\phi} \sigma_{\mb m} \ket{\phi}
    \bra{\phi} \eta_{\mb m} \ket{\phi} \nonumber \\
    & = \sum_{\mb m} \pi_{\mb m} \frac{1 + \Tr(\sigma_{\mb m} \eta_{\mb m})}{D+1}.
\end{align}
This is maximized by taking $\eta_{\mb m} = \ketbra{\psi_{\mb m}}$, the projector on the leading eigenvector of $\sigma_{\mb m}$, as claimed.
We note that this construction is closely related to a previous proposal for a notion of ``quantumness'' of a Hilbert space~\cite{fuchs_squeezing_2003,fuchs_quantumness_2004}, based on the ability of a classical eavesdropper to read and resend the information without being detected.

The resulting shadow channel is given by
\begin{equation}
    \mathcal{M}(\rho) = D \Tr[ (\mathbb{I}\otimes \rho) \sigma^{(\infty,1)}] \label{eq:shadow_channel_maxfid}
\end{equation}
where we defined a generalized ``moment operator''
\begin{equation}
    \sigma^{(\infty,1)} = \sum_{\mb m} \pi_{\mb m} \ketbra{\psi_{\mb m}} \otimes \sigma_{\mb m} \label{eq:sigma_infinity}
\end{equation}
for the ensemble of trajectories. This operator is sensitive to the MIPT; e.g., the expectation value of the replica SWAP operator $\hat{\tau}$ yields
\begin{equation}
    \Tr(\sigma^{(\infty,1)} \hat{\tau} ) = \sum_{\mb m} \pi_{\mb m} \bra{\psi_{\mb m}} \sigma_{\mb m} \ket{\psi_{\mb m}} = \mathbb{E}_{\mb m} [e^{-S_{\infty, \mb m}}], \label{eq:renyi_infty_ev}
\end{equation}
the ``annealed average'' over trajectories of the Renyi-$\infty$ entropy. The measure over trajectories in this case is the conventional one, $\pi_{\mb m}$.

\section{XEB calculations \label{app:xeb}}

Here we derive two results relating to the sample complexity of fidelity estimation from Sec.~\ref{sec:xeb}---Eq.~\eqref{eq:xeb_uncertainty} and \eqref{eq:shdn_xeb_leading}---both of which involve a third-moment expectation value.

\subsection{Computation of third moment quantity \label{app:third_moment}}

We start by obtaining an auxiliary result: evaluating the thrid-moment quantity
\begin{equation}
\Gamma \equiv \Tr(\rho \otimes \rho_0^{\otimes 2}\, \sigma^{(3)})
\end{equation}
with $\rho_0 = \ketbra{\psi}$ and $\sigma^{(3)}$ the third-moment operator given in Eq.~\eqref{eq:third_moment}.
Due to pre-scrambling, the latter is expressed as a sum of three-replica permutations, 
\begin{align}
    \sigma^{(3)} & = \sum_{\nu \in S_3} c_\nu \hat{\nu} \nonumber \\
    & = c_e \hat{e} + c_\tau (\hat{\tau}_{1,2} + \hat{\tau}_{2,3} + \hat{\tau}_{3,1}) + c_\chi (\hat{\chi}_+ + \hat{\chi}_-), \label{eq:perm_expansion}
\end{align}
where $e$ is the identity permutation, $\tau_{i,j}$ is the transposition of elements $i$, $j$, and $\chi_\pm$ are the cyclical permutations. 
By plugging Eq.~\eqref{eq:perm_expansion} into the definition of $\Gamma$, we get
\begin{equation}
    \Gamma = c_e + c_\tau + 2(c_\tau + c_\chi) F
\end{equation}
where $F = \Tr(\rho \rho_0)$ is the fidelity between the true state $\rho$ and the guess $\rho_0 = \ketbra{\psi}$.

By making use of Weingarten functions~\cite{kostenberger_weingarten_2021}
\begin{equation}
\left\{  \begin{aligned}
{\sf Wg}(e) & = (D^2-2)/g(D) \\
{\sf Wg}(\tau_{ij}) & = -D/g(D) \\
{\sf Wg}(\chi_\pm) & = 2/g(D)
\end{aligned} \right.
\end{equation}
 with $g(D) = D(D^2-1)(D^2-4)$, we get
\begin{eqnarray}
\begin{pmatrix} c_e \\ c_\tau \\ c_\chi 
\end{pmatrix}
= \frac{1}{g(D)}
\begin{pmatrix}
    D^2-2 & -3D & 4 \\
    -D & D^2+2 & -2D \\
    2 & -3D & D^2
\end{pmatrix}
\begin{pmatrix}
    1 \\ \avgP \\ \avgPthree
\end{pmatrix}
\end{eqnarray}
where $\avgPthree = \Tr(\chi_\pm \sigma^{(3)}) = \mathbb{E}_{\mathbf m} \Tr(\sigma_{\mathbf m}^3)$ is related to the third Renyi entropy of the trajectories. 
Therefore we have
\begin{align}
    c_e + c_\tau 
    & = \frac{D^2 -D-2 + \avgP (D^2-3D+2) + \avgPthree (4-2D)}{D(D^2-1)(D^2-4)} \nonumber \\
    & \simeq D^{-3}(1 + \avgP) 
\end{align}
and
\begin{align}
    c_\tau + c_\chi
    & = \frac{2-D + \avgP (D^2-3D+2) + \avgPthree (D^2-2D)}{D(D^2-1)(D^2-4)} \nonumber \\
    & \simeq D^{-3}(\avgP + \avgPthree).
\end{align}
Here $\simeq$ denotes leading order in $D$. 
With this we conclude
\begin{equation}
    \Gamma \simeq D^{-3}[ 1+ \avgP + 2F(\avgP + \avgPthree)].
    \label{eq:gamma_result}
\end{equation}

\subsection{Statistical fluctuations of modified linear-XEB \label{app:xeb_std}}

Here we derive Eq.~\eqref{eq:xeb_uncertainty}.
The standard deviation $\delta {\sf XEB}'$ is given by
\begin{equation}
    (\delta {\sf XEB}')^2 = \langle p(\rho_0|\mathbf{m})^2 \rangle_{\mathbf m\sim p(\mathbf m|\rho)} - \langle p(\rho_0|\mathbf{m}) \rangle_{\mathbf m\sim p(\mathbf m|\rho)}^2.
\end{equation}
The second term is simply $({\sf XEB}')^2$, already computed in Eq.~\eqref{eq:xebprime_avg}.
Focusing on the first, we have
\begin{equation}
    \langle p(\rho_0|\mathbf{m})^2 \rangle_{\mathbf m\sim p(\mathbf m|\rho)}
    = D^3 \Tr(\rho\otimes \rho_0 \otimes \rho_0\ \sigma^{(3)}) \label{eq:xeb_3rdmoment}
\end{equation}
which is proportional to the $\Gamma$ quantity in Eq.~\eqref{eq:gamma_result}.
It follows that
\begin{align}    
(\delta {\sf XEB}')^2
    & = \avgP + 2F \avgPthree - (F\avgP)^2.
\end{align}
In the pure phase, both $\avgP$ and $\avgPthree$ are constant.
In the mixed phase they both vanish asymptotically, with $\avgP \sim D^{-s}$ and $\avgPthree \leq \avgP$. Eq.~\eqref{eq:xeb_uncertainty} follows.

\subsection{Shadow norm computation}

Here we derive Eq.~\eqref{eq:shdn_xeb_leading}.
The inverse channel $\mathcal{M}^{-1}$, derivable from Eq.~\eqref{eq:M_expression_prescr}, reads $\mathcal{M}^{-1}(\rho) = \lambda^{-1} \rho-c\mathbb{I}$ with $\lambda = (D\avgP-1)/(D^2-1)$ and $c = (D - \avgP) / (D \avgP -1)$. This yields
\begin{align}
    \shdn{\rho_0}^2 
    & = D \left\{ 
    c^2 \Tr(\rho\, \sigma^{(1)}) - 2c \lambda^{-1} \Tr(\rho\otimes \rho_0\,  \sigma^{(2)}) \right. \nonumber \\
    & \left. \qquad + \lambda^{-2} \Tr(\rho\otimes\rho_0^{\otimes 2}\, \sigma^{(3)}) \right\}
\end{align}
where we have used the fact that moment operators obey $\Tr_1({\sigma^{(k)}}) = {\sigma^{(k-1)}}$ (tracing over one of the $k$ replicas yields the moment operator on $k-1$ replicas).
The first two terms are straightforwardly evaluated by noting that, due to pre-scrambling, ${\sigma^{(1)}} = \mathbb{I}/D$ and $\sigma^{(2)} = [(1-\avgP/D)e + (\avgP-1/D)\chi] / (D^2-1)$ ($e,\chi$ are the two replica permutations, identity and swap). 
The last term, involving the third moment, is again given by the $\Gamma$ quantity evaluated in Eq.~\eqref{eq:gamma_result}. Explicit evaluation yields Eq.~\eqref{eq:shdn_xeb_leading}.

\bibliography{shadows_MIPT}

\begin{thebibliography}{77}%
\makeatletter
\providecommand \@ifxundefined [1]{%
 \@ifx{#1\undefined}
}%
\providecommand \@ifnum [1]{%
 \ifnum #1\expandafter \@firstoftwo
 \else \expandafter \@secondoftwo
 \fi
}%
\providecommand \@ifx [1]{%
 \ifx #1\expandafter \@firstoftwo
 \else \expandafter \@secondoftwo
 \fi
}%
\providecommand \natexlab [1]{#1}%
\providecommand \enquote  [1]{``#1''}%
\providecommand \bibnamefont  [1]{#1}%
\providecommand \bibfnamefont [1]{#1}%
\providecommand \citenamefont [1]{#1}%
\providecommand \href@noop [0]{\@secondoftwo}%
\providecommand \href [0]{\begingroup \@sanitize@url \@href}%
\providecommand \@href[1]{\@@startlink{#1}\@@href}%
\providecommand \@@href[1]{\endgroup#1\@@endlink}%
\providecommand \@sanitize@url [0]{\catcode `\\12\catcode `\$12\catcode
  `\&12\catcode `\#12\catcode `\^12\catcode `\_12\catcode `\%12\relax}%
\providecommand \@@startlink[1]{}%
\providecommand \@@endlink[0]{}%
\providecommand \url  [0]{\begingroup\@sanitize@url \@url }%
\providecommand \@url [1]{\endgroup\@href {#1}{\urlprefix }}%
\providecommand \urlprefix  [0]{URL }%
\providecommand \Eprint [0]{\href }%
\providecommand \doibase [0]{http://dx.doi.org/}%
\providecommand \selectlanguage [0]{\@gobble}%
\providecommand \bibinfo  [0]{\@secondoftwo}%
\providecommand \bibfield  [0]{\@secondoftwo}%
\providecommand \translation [1]{[#1]}%
\providecommand \BibitemOpen [0]{}%
\providecommand \bibitemStop [0]{}%
\providecommand \bibitemNoStop [0]{.\EOS\space}%
\providecommand \EOS [0]{\spacefactor3000\relax}%
\providecommand \BibitemShut  [1]{\csname bibitem#1\endcsname}%
\let\auto@bib@innerbib\@empty
\bibitem [{\citenamefont {Skinner}\ \emph {et~al.}(2019)\citenamefont
  {Skinner}, \citenamefont {Ruhman},\ and\ \citenamefont
  {Nahum}}]{skinner_measurement-induced_2019}%
  \BibitemOpen
  \bibfield  {author} {\bibinfo {author} {\bibfnamefont {Brian}\ \bibnamefont
  {Skinner}}, \bibinfo {author} {\bibfnamefont {Jonathan}\ \bibnamefont
  {Ruhman}}, \ and\ \bibinfo {author} {\bibfnamefont {Adam}\ \bibnamefont
  {Nahum}},\ }\bibfield  {title} {\enquote {\bibinfo {title}
  {Measurement-{Induced} {Phase} {Transitions} in the {Dynamics} of
  {Entanglement}},}\ }\href {\doibase 10.1103/PhysRevX.9.031009} {\bibfield
  {journal} {\bibinfo  {journal} {Physical Review X}\ }\textbf {\bibinfo
  {volume} {9}},\ \bibinfo {pages} {031009} (\bibinfo {year}
  {2019})}\BibitemShut {NoStop}%
\bibitem [{\citenamefont {Li}\ \emph {et~al.}(2018)\citenamefont {Li},
  \citenamefont {Chen},\ and\ \citenamefont {Fisher}}]{li_quantum_2018}%
  \BibitemOpen
  \bibfield  {author} {\bibinfo {author} {\bibfnamefont {Yaodong}\ \bibnamefont
  {Li}}, \bibinfo {author} {\bibfnamefont {Xiao}\ \bibnamefont {Chen}}, \ and\
  \bibinfo {author} {\bibfnamefont {Matthew P.~A.}\ \bibnamefont {Fisher}},\
  }\bibfield  {title} {\enquote {\bibinfo {title} {Quantum {Zeno} effect and
  the many-body entanglement transition},}\ }\href {\doibase
  10.1103/PhysRevB.98.205136} {\bibfield  {journal} {\bibinfo  {journal}
  {Physical Review B}\ }\textbf {\bibinfo {volume} {98}},\ \bibinfo {pages}
  {205136} (\bibinfo {year} {2018})}\BibitemShut {NoStop}%
\bibitem [{\citenamefont {Li}\ \emph {et~al.}(2019)\citenamefont {Li},
  \citenamefont {Chen},\ and\ \citenamefont
  {Fisher}}]{li_measurement-driven_2019}%
  \BibitemOpen
  \bibfield  {author} {\bibinfo {author} {\bibfnamefont {Yaodong}\ \bibnamefont
  {Li}}, \bibinfo {author} {\bibfnamefont {Xiao}\ \bibnamefont {Chen}}, \ and\
  \bibinfo {author} {\bibfnamefont {Matthew P.~A.}\ \bibnamefont {Fisher}},\
  }\bibfield  {title} {\enquote {\bibinfo {title} {Measurement-driven
  entanglement transition in hybrid quantum circuits},}\ }\href {\doibase
  10.1103/PhysRevB.100.134306} {\bibfield  {journal} {\bibinfo  {journal}
  {Physical Review B}\ }\textbf {\bibinfo {volume} {100}},\ \bibinfo {pages}
  {134306} (\bibinfo {year} {2019})}\BibitemShut {NoStop}%
\bibitem [{\citenamefont {Potter}\ and\ \citenamefont
  {Vasseur}(2022)}]{potter_entanglement_2022}%
  \BibitemOpen
  \bibfield  {author} {\bibinfo {author} {\bibfnamefont {Andrew~C.}\
  \bibnamefont {Potter}}\ and\ \bibinfo {author} {\bibfnamefont {Romain}\
  \bibnamefont {Vasseur}},\ }\bibfield  {title} {\enquote {\bibinfo {title}
  {Entanglement {Dynamics} in {Hybrid} {Quantum} {Circuits}},}\ }in\ \href
  {\doibase 10.1007/978-3-031-03998-0_9} {\emph {\bibinfo {booktitle}
  {Entanglement in {Spin} {Chains}: {From} {Theory} to {Quantum} {Technology}
  {Applications}}}},\ \bibinfo {series and number} {Quantum {Science} and
  {Technology}},\ \bibinfo {editor} {edited by\ \bibinfo {editor}
  {\bibfnamefont {Abolfazl}\ \bibnamefont {Bayat}}, \bibinfo {editor}
  {\bibfnamefont {Sougato}\ \bibnamefont {Bose}}, \ and\ \bibinfo {editor}
  {\bibfnamefont {Henrik}\ \bibnamefont {Johannesson}}}\ (\bibinfo {address}
  {Cham},\ \bibinfo {year} {2022})\ pp.\ \bibinfo {pages}
  {211--249}\BibitemShut {NoStop}%
\bibitem [{\citenamefont {Fisher}\ \emph {et~al.}(2022)\citenamefont {Fisher},
  \citenamefont {Khemani}, \citenamefont {Nahum},\ and\ \citenamefont
  {Vijay}}]{fisher_random_2022}%
  \BibitemOpen
  \bibfield  {author} {\bibinfo {author} {\bibfnamefont {Matthew P.~A.}\
  \bibnamefont {Fisher}}, \bibinfo {author} {\bibfnamefont {Vedika}\
  \bibnamefont {Khemani}}, \bibinfo {author} {\bibfnamefont {Adam}\
  \bibnamefont {Nahum}}, \ and\ \bibinfo {author} {\bibfnamefont {Sagar}\
  \bibnamefont {Vijay}},\ }\bibfield  {title} {\enquote {\bibinfo {title}
  {Random {Quantum} {Circuits}},}\ }\href {\doibase 10.48550/arXiv.2207.14280}
  {\bibfield  {journal} {\bibinfo  {journal} {arXiv e-prints}\ } (\bibinfo
  {year} {2022}),\ 10.48550/arXiv.2207.14280}\BibitemShut {NoStop}%
\bibitem [{\citenamefont {Huang}\ \emph {et~al.}(2020)\citenamefont {Huang},
  \citenamefont {Kueng},\ and\ \citenamefont
  {Preskill}}]{huang_predicting_2020}%
  \BibitemOpen
  \bibfield  {author} {\bibinfo {author} {\bibfnamefont {Hsin-Yuan}\
  \bibnamefont {Huang}}, \bibinfo {author} {\bibfnamefont {Richard}\
  \bibnamefont {Kueng}}, \ and\ \bibinfo {author} {\bibfnamefont {John}\
  \bibnamefont {Preskill}},\ }\bibfield  {title} {\enquote {\bibinfo {title}
  {Predicting many properties of a quantum system from very few
  measurements},}\ }\href {\doibase 10.1038/s41567-020-0932-7} {\bibfield
  {journal} {\bibinfo  {journal} {Nature Physics}\ }\textbf {\bibinfo {volume}
  {16}},\ \bibinfo {pages} {1050--1057} (\bibinfo {year} {2020})}\BibitemShut
  {NoStop}%
\bibitem [{\citenamefont {Elben}\ \emph {et~al.}(2023)\citenamefont {Elben},
  \citenamefont {Flammia}, \citenamefont {Huang}, \citenamefont {Kueng},
  \citenamefont {Preskill}, \citenamefont {Vermersch},\ and\ \citenamefont
  {Zoller}}]{elben_randomized_2023}%
  \BibitemOpen
  \bibfield  {author} {\bibinfo {author} {\bibfnamefont {Andreas}\ \bibnamefont
  {Elben}}, \bibinfo {author} {\bibfnamefont {Steven~T.}\ \bibnamefont
  {Flammia}}, \bibinfo {author} {\bibfnamefont {Hsin-Yuan}\ \bibnamefont
  {Huang}}, \bibinfo {author} {\bibfnamefont {Richard}\ \bibnamefont {Kueng}},
  \bibinfo {author} {\bibfnamefont {John}\ \bibnamefont {Preskill}}, \bibinfo
  {author} {\bibfnamefont {Benoit}\ \bibnamefont {Vermersch}}, \ and\ \bibinfo
  {author} {\bibfnamefont {Peter}\ \bibnamefont {Zoller}},\ }\bibfield  {title}
  {\enquote {\bibinfo {title} {The randomized measurement toolbox},}\ }\href
  {\doibase 10.1038/s42254-022-00535-2} {\bibfield  {journal} {\bibinfo
  {journal} {Nature Reviews Physics}\ }\textbf {\bibinfo {volume} {5}},\
  \bibinfo {pages} {9--24} (\bibinfo {year} {2023})}\BibitemShut {NoStop}%
\bibitem [{\citenamefont {Cao}\ \emph {et~al.}(2019)\citenamefont {Cao},
  \citenamefont {Tilloy},\ and\ \citenamefont
  {De~Luca}}]{cao_entanglement_2019}%
  \BibitemOpen
  \bibfield  {author} {\bibinfo {author} {\bibfnamefont {Xiangyu}\ \bibnamefont
  {Cao}}, \bibinfo {author} {\bibfnamefont {Antoine}\ \bibnamefont {Tilloy}}, \
  and\ \bibinfo {author} {\bibfnamefont {Andrea}\ \bibnamefont {De~Luca}},\
  }\bibfield  {title} {\enquote {\bibinfo {title} {Entanglement in a fermion
  chain under continuous monitoring},}\ }\href {\doibase
  10.21468/SciPostPhys.7.2.024} {\bibfield  {journal} {\bibinfo  {journal}
  {SciPost Physics}\ }\textbf {\bibinfo {volume} {7}},\ \bibinfo {pages} {024}
  (\bibinfo {year} {2019})}\BibitemShut {NoStop}%
\bibitem [{\citenamefont {Fidkowski}\ \emph {et~al.}(2021)\citenamefont
  {Fidkowski}, \citenamefont {Haah},\ and\ \citenamefont
  {Hastings}}]{fidkowski_how_2021}%
  \BibitemOpen
  \bibfield  {author} {\bibinfo {author} {\bibfnamefont {Lukasz}\ \bibnamefont
  {Fidkowski}}, \bibinfo {author} {\bibfnamefont {Jeongwan}\ \bibnamefont
  {Haah}}, \ and\ \bibinfo {author} {\bibfnamefont {Matthew~B.}\ \bibnamefont
  {Hastings}},\ }\bibfield  {title} {\enquote {\bibinfo {title} {How
  {Dynamical} {Quantum} {Memories} {Forget}},}\ }\href {\doibase
  10.22331/q-2021-01-17-382} {\bibfield  {journal} {\bibinfo  {journal}
  {Quantum}\ }\textbf {\bibinfo {volume} {5}},\ \bibinfo {pages} {382}
  (\bibinfo {year} {2021})}\BibitemShut {NoStop}%
\bibitem [{\citenamefont {Fava}\ \emph {et~al.}(2023)\citenamefont {Fava},
  \citenamefont {Piroli}, \citenamefont {Swann}, \citenamefont {Bernard},\ and\
  \citenamefont {Nahum}}]{fava_nonlinear_2023}%
  \BibitemOpen
  \bibfield  {author} {\bibinfo {author} {\bibfnamefont {Michele}\ \bibnamefont
  {Fava}}, \bibinfo {author} {\bibfnamefont {Lorenzo}\ \bibnamefont {Piroli}},
  \bibinfo {author} {\bibfnamefont {Tobias}\ \bibnamefont {Swann}}, \bibinfo
  {author} {\bibfnamefont {Denis}\ \bibnamefont {Bernard}}, \ and\ \bibinfo
  {author} {\bibfnamefont {Adam}\ \bibnamefont {Nahum}},\ }\bibfield  {title}
  {\enquote {\bibinfo {title} {Nonlinear {Sigma} {Models} for {Monitored}
  {Dynamics} of {Free} {Fermions}},}\ }\href {\doibase
  10.1103/PhysRevX.13.041045} {\bibfield  {journal} {\bibinfo  {journal}
  {Physical Review X}\ }\textbf {\bibinfo {volume} {13}},\ \bibinfo {pages}
  {041045} (\bibinfo {year} {2023})}\BibitemShut {NoStop}%
\bibitem [{\citenamefont {Bao}\ \emph {et~al.}(2021)\citenamefont {Bao},
  \citenamefont {Choi},\ and\ \citenamefont {Altman}}]{bao_symmetry_2021}%
  \BibitemOpen
  \bibfield  {author} {\bibinfo {author} {\bibfnamefont {Yimu}\ \bibnamefont
  {Bao}}, \bibinfo {author} {\bibfnamefont {Soonwon}\ \bibnamefont {Choi}}, \
  and\ \bibinfo {author} {\bibfnamefont {Ehud}\ \bibnamefont {Altman}},\
  }\bibfield  {title} {\enquote {\bibinfo {title} {Symmetry enriched phases of
  quantum circuits},}\ }\href {\doibase 10.1016/j.aop.2021.168618} {\bibfield
  {journal} {\bibinfo  {journal} {Annals of Physics}\ }\bibinfo {series}
  {Special issue on {Philip} {W}. {Anderson}},\ \textbf {\bibinfo {volume}
  {435}},\ \bibinfo {pages} {168618} (\bibinfo {year} {2021})}\BibitemShut
  {NoStop}%
\bibitem [{\citenamefont {Agrawal}\ \emph {et~al.}(2022)\citenamefont
  {Agrawal}, \citenamefont {Zabalo}, \citenamefont {Chen}, \citenamefont
  {Wilson}, \citenamefont {Potter}, \citenamefont {Pixley}, \citenamefont
  {Gopalakrishnan},\ and\ \citenamefont {Vasseur}}]{agrawal_entanglement_2022}%
  \BibitemOpen
  \bibfield  {author} {\bibinfo {author} {\bibfnamefont {Utkarsh}\ \bibnamefont
  {Agrawal}}, \bibinfo {author} {\bibfnamefont {Aidan}\ \bibnamefont {Zabalo}},
  \bibinfo {author} {\bibfnamefont {Kun}\ \bibnamefont {Chen}}, \bibinfo
  {author} {\bibfnamefont {Justin~H.}\ \bibnamefont {Wilson}}, \bibinfo
  {author} {\bibfnamefont {Andrew~C.}\ \bibnamefont {Potter}}, \bibinfo
  {author} {\bibfnamefont {J.~H.}\ \bibnamefont {Pixley}}, \bibinfo {author}
  {\bibfnamefont {Sarang}\ \bibnamefont {Gopalakrishnan}}, \ and\ \bibinfo
  {author} {\bibfnamefont {Romain}\ \bibnamefont {Vasseur}},\ }\bibfield
  {title} {\enquote {\bibinfo {title} {Entanglement and {Charge}-{Sharpening}
  {Transitions} in {U}(1) {Symmetric} {Monitored} {Quantum} {Circuits}},}\
  }\href {\doibase 10.1103/PhysRevX.12.041002} {\bibfield  {journal} {\bibinfo
  {journal} {Physical Review X}\ }\textbf {\bibinfo {volume} {12}},\ \bibinfo
  {pages} {041002} (\bibinfo {year} {2022})}\BibitemShut {NoStop}%
\bibitem [{\citenamefont {Majidy}\ \emph {et~al.}(2023)\citenamefont {Majidy},
  \citenamefont {Agrawal}, \citenamefont {Gopalakrishnan}, \citenamefont
  {Potter}, \citenamefont {Vasseur},\ and\ \citenamefont
  {Halpern}}]{majidy_critical_2023}%
  \BibitemOpen
  \bibfield  {author} {\bibinfo {author} {\bibfnamefont {Shayan}\ \bibnamefont
  {Majidy}}, \bibinfo {author} {\bibfnamefont {Utkarsh}\ \bibnamefont
  {Agrawal}}, \bibinfo {author} {\bibfnamefont {Sarang}\ \bibnamefont
  {Gopalakrishnan}}, \bibinfo {author} {\bibfnamefont {Andrew~C.}\ \bibnamefont
  {Potter}}, \bibinfo {author} {\bibfnamefont {Romain}\ \bibnamefont
  {Vasseur}}, \ and\ \bibinfo {author} {\bibfnamefont {Nicole~Yunger}\
  \bibnamefont {Halpern}},\ }\bibfield  {title} {\enquote {\bibinfo {title}
  {Critical phase and spin sharpening in {SU}(2)-symmetric monitored quantum
  circuits},}\ }\href {\doibase 10.1103/PhysRevB.108.054307} {\bibfield
  {journal} {\bibinfo  {journal} {Physical Review B}\ }\textbf {\bibinfo
  {volume} {108}},\ \bibinfo {pages} {054307} (\bibinfo {year}
  {2023})}\BibitemShut {NoStop}%
\bibitem [{\citenamefont {Choi}\ \emph {et~al.}(2020)\citenamefont {Choi},
  \citenamefont {Bao}, \citenamefont {Qi},\ and\ \citenamefont
  {Altman}}]{choi_quantum_2020}%
  \BibitemOpen
  \bibfield  {author} {\bibinfo {author} {\bibfnamefont {Soonwon}\ \bibnamefont
  {Choi}}, \bibinfo {author} {\bibfnamefont {Yimu}\ \bibnamefont {Bao}},
  \bibinfo {author} {\bibfnamefont {Xiao-Liang}\ \bibnamefont {Qi}}, \ and\
  \bibinfo {author} {\bibfnamefont {Ehud}\ \bibnamefont {Altman}},\ }\bibfield
  {title} {\enquote {\bibinfo {title} {Quantum {Error} {Correction} in
  {Scrambling} {Dynamics} and {Measurement}-{Induced} {Phase} {Transition}},}\
  }\href {\doibase 10.1103/PhysRevLett.125.030505} {\bibfield  {journal}
  {\bibinfo  {journal} {Physical Review Letters}\ }\textbf {\bibinfo {volume}
  {125}},\ \bibinfo {pages} {030505} (\bibinfo {year} {2020})}\BibitemShut
  {NoStop}%
\bibitem [{\citenamefont {Gullans}\ and\ \citenamefont
  {Huse}(2020{\natexlab{a}})}]{gullans_dynamical_2020}%
  \BibitemOpen
  \bibfield  {author} {\bibinfo {author} {\bibfnamefont {Michael~J.}\
  \bibnamefont {Gullans}}\ and\ \bibinfo {author} {\bibfnamefont {David~A.}\
  \bibnamefont {Huse}},\ }\bibfield  {title} {\enquote {\bibinfo {title}
  {Dynamical {Purification} {Phase} {Transition} {Induced} by {Quantum}
  {Measurements}},}\ }\href {\doibase 10.1103/PhysRevX.10.041020} {\bibfield
  {journal} {\bibinfo  {journal} {Physical Review X}\ }\textbf {\bibinfo
  {volume} {10}},\ \bibinfo {pages} {041020} (\bibinfo {year}
  {2020}{\natexlab{a}})}\BibitemShut {NoStop}%
\bibitem [{\citenamefont {Shor}(1995)}]{shor_scheme_1995}%
  \BibitemOpen
  \bibfield  {author} {\bibinfo {author} {\bibfnamefont {Peter~W.}\
  \bibnamefont {Shor}},\ }\bibfield  {title} {\enquote {\bibinfo {title}
  {Scheme for reducing decoherence in quantum computer memory},}\ }\href
  {\doibase 10.1103/PhysRevA.52.R2493} {\bibfield  {journal} {\bibinfo
  {journal} {Physical Review A}\ }\textbf {\bibinfo {volume} {52}},\ \bibinfo
  {pages} {R2493--R2496} (\bibinfo {year} {1995})}\BibitemShut {NoStop}%
\bibitem [{\citenamefont {Gottesman}(1997)}]{gottesman_stabilizer_1997}%
  \BibitemOpen
  \bibfield  {author} {\bibinfo {author} {\bibfnamefont {Daniel}\ \bibnamefont
  {Gottesman}},\ }\bibfield  {title} {\enquote {\bibinfo {title} {Stabilizer
  {Codes} and {Quantum} {Error} {Correction}},}\ }\href {\doibase
  10.48550/arXiv.quant-ph/9705052} {\  (\bibinfo {year} {1997}),\
  10.48550/arXiv.quant-ph/9705052}\BibitemShut {NoStop}%
\bibitem [{\citenamefont {Dennis}\ \emph {et~al.}(2002)\citenamefont {Dennis},
  \citenamefont {Kitaev}, \citenamefont {Landahl},\ and\ \citenamefont
  {Preskill}}]{dennis_topological_2002}%
  \BibitemOpen
  \bibfield  {author} {\bibinfo {author} {\bibfnamefont {Eric}\ \bibnamefont
  {Dennis}}, \bibinfo {author} {\bibfnamefont {Alexei}\ \bibnamefont {Kitaev}},
  \bibinfo {author} {\bibfnamefont {Andrew}\ \bibnamefont {Landahl}}, \ and\
  \bibinfo {author} {\bibfnamefont {John}\ \bibnamefont {Preskill}},\
  }\bibfield  {title} {\enquote {\bibinfo {title} {Topological quantum
  memory},}\ }\href {\doibase 10.1063/1.1499754} {\bibfield  {journal}
  {\bibinfo  {journal} {Journal of Mathematical Physics}\ }\textbf {\bibinfo
  {volume} {43}},\ \bibinfo {pages} {4452--4505} (\bibinfo {year}
  {2002})}\BibitemShut {NoStop}%
\bibitem [{\citenamefont {Gullans}\ and\ \citenamefont
  {Huse}(2020{\natexlab{b}})}]{gullans_scalable_2020}%
  \BibitemOpen
  \bibfield  {author} {\bibinfo {author} {\bibfnamefont {Michael~J.}\
  \bibnamefont {Gullans}}\ and\ \bibinfo {author} {\bibfnamefont {David~A.}\
  \bibnamefont {Huse}},\ }\bibfield  {title} {\enquote {\bibinfo {title}
  {Scalable {Probes} of {Measurement}-{Induced} {Criticality}},}\ }\href
  {\doibase 10.1103/PhysRevLett.125.070606} {\bibfield  {journal} {\bibinfo
  {journal} {Physical Review Letters}\ }\textbf {\bibinfo {volume} {125}},\
  \bibinfo {pages} {070606} (\bibinfo {year} {2020}{\natexlab{b}})}\BibitemShut
  {NoStop}%
\bibitem [{\citenamefont {Noel}\ \emph {et~al.}(2022)\citenamefont {Noel},
  \citenamefont {Niroula}, \citenamefont {Zhu}, \citenamefont {Risinger},
  \citenamefont {Egan}, \citenamefont {Biswas}, \citenamefont {Cetina},
  \citenamefont {Gorshkov} \emph {et~al.}}]{noel_measurement-induced_2022}%
  \BibitemOpen
  \bibfield  {author} {\bibinfo {author} {\bibfnamefont {Crystal}\ \bibnamefont
  {Noel}}, \bibinfo {author} {\bibfnamefont {Pradeep}\ \bibnamefont {Niroula}},
  \bibinfo {author} {\bibfnamefont {Daiwei}\ \bibnamefont {Zhu}}, \bibinfo
  {author} {\bibfnamefont {Andrew}\ \bibnamefont {Risinger}}, \bibinfo {author}
  {\bibfnamefont {Laird}\ \bibnamefont {Egan}}, \bibinfo {author}
  {\bibfnamefont {Debopriyo}\ \bibnamefont {Biswas}}, \bibinfo {author}
  {\bibfnamefont {Marko}\ \bibnamefont {Cetina}}, \bibinfo {author}
  {\bibfnamefont {Alexey~V.}\ \bibnamefont {Gorshkov}},  \emph {et~al.},\
  }\bibfield  {title} {\enquote {\bibinfo {title} {Measurement-induced quantum
  phases realized in a trapped-ion quantum computer},}\ }\href {\doibase
  10.1038/s41567-022-01619-7} {\bibfield  {journal} {\bibinfo  {journal}
  {Nature Physics}\ }\textbf {\bibinfo {volume} {18}},\ \bibinfo {pages}
  {760--764} (\bibinfo {year} {2022})}\BibitemShut {NoStop}%
\bibitem [{\citenamefont {Hoke}\ \emph {et~al.}(2023)\citenamefont {Hoke},
  \citenamefont {Ippoliti}, \citenamefont {Rosenberg}, \citenamefont {Abanin},
  \citenamefont {Acharya}, \citenamefont {Andersen}, \citenamefont {Ansmann},
  \citenamefont {Arute} \emph {et~al.}}]{hoke_measurement-induced_2023}%
  \BibitemOpen
  \bibfield  {author} {\bibinfo {author} {\bibfnamefont {J.~C.}\ \bibnamefont
  {Hoke}}, \bibinfo {author} {\bibfnamefont {M.}~\bibnamefont {Ippoliti}},
  \bibinfo {author} {\bibfnamefont {E.}~\bibnamefont {Rosenberg}}, \bibinfo
  {author} {\bibfnamefont {D.}~\bibnamefont {Abanin}}, \bibinfo {author}
  {\bibfnamefont {R.}~\bibnamefont {Acharya}}, \bibinfo {author} {\bibfnamefont
  {T.~I.}\ \bibnamefont {Andersen}}, \bibinfo {author} {\bibfnamefont
  {M.}~\bibnamefont {Ansmann}}, \bibinfo {author} {\bibfnamefont
  {F.}~\bibnamefont {Arute}},  \emph {et~al.},\ }\bibfield  {title} {\enquote
  {\bibinfo {title} {Measurement-induced entanglement and teleportation on a
  noisy quantum processor},}\ }\href {\doibase 10.1038/s41586-023-06505-7}
  {\bibfield  {journal} {\bibinfo  {journal} {Nature}\ }\textbf {\bibinfo
  {volume} {622}},\ \bibinfo {pages} {481--486} (\bibinfo {year}
  {2023})}\BibitemShut {NoStop}%
\bibitem [{\citenamefont {Bao}\ \emph {et~al.}(2020)\citenamefont {Bao},
  \citenamefont {Choi},\ and\ \citenamefont {Altman}}]{bao_theory_2020}%
  \BibitemOpen
  \bibfield  {author} {\bibinfo {author} {\bibfnamefont {Yimu}\ \bibnamefont
  {Bao}}, \bibinfo {author} {\bibfnamefont {Soonwon}\ \bibnamefont {Choi}}, \
  and\ \bibinfo {author} {\bibfnamefont {Ehud}\ \bibnamefont {Altman}},\
  }\bibfield  {title} {\enquote {\bibinfo {title} {Theory of the phase
  transition in random unitary circuits with measurements},}\ }\href {\doibase
  10.1103/PhysRevB.101.104301} {\bibfield  {journal} {\bibinfo  {journal}
  {Physical Review B}\ }\textbf {\bibinfo {volume} {101}},\ \bibinfo {pages}
  {104301} (\bibinfo {year} {2020})}\BibitemShut {NoStop}%
\bibitem [{\citenamefont {Li}\ \emph {et~al.}(2023)\citenamefont {Li},
  \citenamefont {Zou}, \citenamefont {Glorioso}, \citenamefont {Altman},\ and\
  \citenamefont {Fisher}}]{li_cross_2023}%
  \BibitemOpen
  \bibfield  {author} {\bibinfo {author} {\bibfnamefont {Yaodong}\ \bibnamefont
  {Li}}, \bibinfo {author} {\bibfnamefont {Yijian}\ \bibnamefont {Zou}},
  \bibinfo {author} {\bibfnamefont {Paolo}\ \bibnamefont {Glorioso}}, \bibinfo
  {author} {\bibfnamefont {Ehud}\ \bibnamefont {Altman}}, \ and\ \bibinfo
  {author} {\bibfnamefont {Matthew P.~A.}\ \bibnamefont {Fisher}},\ }\bibfield
  {title} {\enquote {\bibinfo {title} {Cross {Entropy} {Benchmark} for
  {Measurement}-{Induced} {Phase} {Transitions}},}\ }\href {\doibase
  10.1103/PhysRevLett.130.220404} {\bibfield  {journal} {\bibinfo  {journal}
  {Physical Review Letters}\ }\textbf {\bibinfo {volume} {130}},\ \bibinfo
  {pages} {220404} (\bibinfo {year} {2023})}\BibitemShut {NoStop}%
\bibitem [{\citenamefont {Barratt}\ \emph
  {et~al.}(2022{\natexlab{a}})\citenamefont {Barratt}, \citenamefont {Agrawal},
  \citenamefont {Potter}, \citenamefont {Gopalakrishnan},\ and\ \citenamefont
  {Vasseur}}]{barratt_transitions_2022}%
  \BibitemOpen
  \bibfield  {author} {\bibinfo {author} {\bibfnamefont {Fergus}\ \bibnamefont
  {Barratt}}, \bibinfo {author} {\bibfnamefont {Utkarsh}\ \bibnamefont
  {Agrawal}}, \bibinfo {author} {\bibfnamefont {Andrew~C.}\ \bibnamefont
  {Potter}}, \bibinfo {author} {\bibfnamefont {Sarang}\ \bibnamefont
  {Gopalakrishnan}}, \ and\ \bibinfo {author} {\bibfnamefont {Romain}\
  \bibnamefont {Vasseur}},\ }\bibfield  {title} {\enquote {\bibinfo {title}
  {Transitions in the {Learnability} of {Global} {Charges} from {Local}
  {Measurements}},}\ }\href {\doibase 10.1103/PhysRevLett.129.200602}
  {\bibfield  {journal} {\bibinfo  {journal} {Physical Review Letters}\
  }\textbf {\bibinfo {volume} {129}},\ \bibinfo {pages} {200602} (\bibinfo
  {year} {2022}{\natexlab{a}})}\BibitemShut {NoStop}%
\bibitem [{\citenamefont {Dall'Arno}\ \emph {et~al.}(2011)\citenamefont
  {Dall'Arno}, \citenamefont {D'Ariano},\ and\ \citenamefont
  {Sacchi}}]{dallarno_informational_2011}%
  \BibitemOpen
  \bibfield  {author} {\bibinfo {author} {\bibfnamefont {Michele}\ \bibnamefont
  {Dall'Arno}}, \bibinfo {author} {\bibfnamefont {Giacomo~Mauro}\ \bibnamefont
  {D'Ariano}}, \ and\ \bibinfo {author} {\bibfnamefont {Massimiliano~F.}\
  \bibnamefont {Sacchi}},\ }\bibfield  {title} {\enquote {\bibinfo {title}
  {Informational power of quantum measurements},}\ }\href {\doibase
  10.1103/PhysRevA.83.062304} {\bibfield  {journal} {\bibinfo  {journal}
  {Physical Review A}\ }\textbf {\bibinfo {volume} {83}},\ \bibinfo {pages}
  {062304} (\bibinfo {year} {2011})}\BibitemShut {NoStop}%
\bibitem [{\citenamefont {Ippoliti}\ \emph {et~al.}(2021)\citenamefont
  {Ippoliti}, \citenamefont {Gullans}, \citenamefont {Gopalakrishnan},
  \citenamefont {Huse},\ and\ \citenamefont
  {Khemani}}]{ippoliti_entanglement_2021}%
  \BibitemOpen
  \bibfield  {author} {\bibinfo {author} {\bibfnamefont {Matteo}\ \bibnamefont
  {Ippoliti}}, \bibinfo {author} {\bibfnamefont {Michael~J.}\ \bibnamefont
  {Gullans}}, \bibinfo {author} {\bibfnamefont {Sarang}\ \bibnamefont
  {Gopalakrishnan}}, \bibinfo {author} {\bibfnamefont {David~A.}\ \bibnamefont
  {Huse}}, \ and\ \bibinfo {author} {\bibfnamefont {Vedika}\ \bibnamefont
  {Khemani}},\ }\bibfield  {title} {\enquote {\bibinfo {title} {Entanglement
  {Phase} {Transitions} in {Measurement}-{Only} {Dynamics}},}\ }\href {\doibase
  10.1103/PhysRevX.11.011030} {\bibfield  {journal} {\bibinfo  {journal}
  {Physical Review X}\ }\textbf {\bibinfo {volume} {11}},\ \bibinfo {pages}
  {011030} (\bibinfo {year} {2021})}\BibitemShut {NoStop}%
\bibitem [{\citenamefont {Lavasani}\ \emph {et~al.}(2021)\citenamefont
  {Lavasani}, \citenamefont {Alavirad},\ and\ \citenamefont
  {Barkeshli}}]{lavasani_measurement-induced_2021}%
  \BibitemOpen
  \bibfield  {author} {\bibinfo {author} {\bibfnamefont {Ali}\ \bibnamefont
  {Lavasani}}, \bibinfo {author} {\bibfnamefont {Yahya}\ \bibnamefont
  {Alavirad}}, \ and\ \bibinfo {author} {\bibfnamefont {Maissam}\ \bibnamefont
  {Barkeshli}},\ }\bibfield  {title} {\enquote {\bibinfo {title}
  {Measurement-induced topological entanglement transitions in symmetric random
  quantum circuits},}\ }\href {\doibase 10.1038/s41567-020-01112-z} {\bibfield
  {journal} {\bibinfo  {journal} {Nature Physics}\ }\textbf {\bibinfo {volume}
  {17}},\ \bibinfo {pages} {342--347} (\bibinfo {year} {2021})}\BibitemShut
  {NoStop}%
\bibitem [{\citenamefont {Nahum}\ \emph {et~al.}(2021)\citenamefont {Nahum},
  \citenamefont {Roy}, \citenamefont {Skinner},\ and\ \citenamefont
  {Ruhman}}]{nahum_measurement_2021}%
  \BibitemOpen
  \bibfield  {author} {\bibinfo {author} {\bibfnamefont {Adam}\ \bibnamefont
  {Nahum}}, \bibinfo {author} {\bibfnamefont {Sthitadhi}\ \bibnamefont {Roy}},
  \bibinfo {author} {\bibfnamefont {Brian}\ \bibnamefont {Skinner}}, \ and\
  \bibinfo {author} {\bibfnamefont {Jonathan}\ \bibnamefont {Ruhman}},\
  }\bibfield  {title} {\enquote {\bibinfo {title} {Measurement and
  {Entanglement} {Phase} {Transitions} in {All}-{To}-{All} {Quantum}
  {Circuits}, on {Quantum} {Trees}, and in {Landau}-{Ginsburg} {Theory}},}\
  }\href {\doibase 10.1103/PRXQuantum.2.010352} {\bibfield  {journal} {\bibinfo
   {journal} {PRX Quantum}\ }\textbf {\bibinfo {volume} {2}},\ \bibinfo {pages}
  {010352} (\bibinfo {year} {2021})}\BibitemShut {NoStop}%
\bibitem [{\citenamefont {Li}\ and\ \citenamefont
  {Fisher}(2021)}]{li_statistical_2021}%
  \BibitemOpen
  \bibfield  {author} {\bibinfo {author} {\bibfnamefont {Yaodong}\ \bibnamefont
  {Li}}\ and\ \bibinfo {author} {\bibfnamefont {Matthew P.~A.}\ \bibnamefont
  {Fisher}},\ }\bibfield  {title} {\enquote {\bibinfo {title} {Statistical
  mechanics of quantum error correcting codes},}\ }\href {\doibase
  10.1103/PhysRevB.103.104306} {\bibfield  {journal} {\bibinfo  {journal}
  {Physical Review B}\ }\textbf {\bibinfo {volume} {103}},\ \bibinfo {pages}
  {104306} (\bibinfo {year} {2021})}\BibitemShut {NoStop}%
\bibitem [{\citenamefont {Fan}\ \emph {et~al.}(2021)\citenamefont {Fan},
  \citenamefont {Vijay}, \citenamefont {Vishwanath},\ and\ \citenamefont
  {You}}]{fan_self-organized_2021}%
  \BibitemOpen
  \bibfield  {author} {\bibinfo {author} {\bibfnamefont {Ruihua}\ \bibnamefont
  {Fan}}, \bibinfo {author} {\bibfnamefont {Sagar}\ \bibnamefont {Vijay}},
  \bibinfo {author} {\bibfnamefont {Ashvin}\ \bibnamefont {Vishwanath}}, \ and\
  \bibinfo {author} {\bibfnamefont {Yi-Zhuang}\ \bibnamefont {You}},\
  }\bibfield  {title} {\enquote {\bibinfo {title} {Self-organized error
  correction in random unitary circuits with measurement},}\ }\href {\doibase
  10.1103/PhysRevB.103.174309} {\bibfield  {journal} {\bibinfo  {journal}
  {Phys. Rev. B}\ }\textbf {\bibinfo {volume} {103}},\ \bibinfo {pages}
  {174309} (\bibinfo {year} {2021})}\BibitemShut {NoStop}%
\bibitem [{\citenamefont {Feng}\ \emph {et~al.}(2023)\citenamefont {Feng},
  \citenamefont {Skinner},\ and\ \citenamefont
  {Nahum}}]{feng_measurement-induced_2023}%
  \BibitemOpen
  \bibfield  {author} {\bibinfo {author} {\bibfnamefont {Xiaozhou}\
  \bibnamefont {Feng}}, \bibinfo {author} {\bibfnamefont {Brian}\ \bibnamefont
  {Skinner}}, \ and\ \bibinfo {author} {\bibfnamefont {Adam}\ \bibnamefont
  {Nahum}},\ }\bibfield  {title} {\enquote {\bibinfo {title}
  {Measurement-{Induced} {Phase} {Transitions} on {Dynamical} {Quantum}
  {Trees}},}\ }\href {\doibase 10.1103/PRXQuantum.4.030333} {\bibfield
  {journal} {\bibinfo  {journal} {PRX Quantum}\ }\textbf {\bibinfo {volume}
  {4}},\ \bibinfo {pages} {030333} (\bibinfo {year} {2023})}\BibitemShut
  {NoStop}%
\bibitem [{\citenamefont {Koh}\ \emph {et~al.}(2023)\citenamefont {Koh},
  \citenamefont {Sun}, \citenamefont {Motta},\ and\ \citenamefont
  {Minnich}}]{koh_measurement-induced_2023}%
  \BibitemOpen
  \bibfield  {author} {\bibinfo {author} {\bibfnamefont {Jin~Ming}\
  \bibnamefont {Koh}}, \bibinfo {author} {\bibfnamefont {Shi-Ning}\
  \bibnamefont {Sun}}, \bibinfo {author} {\bibfnamefont {Mario}\ \bibnamefont
  {Motta}}, \ and\ \bibinfo {author} {\bibfnamefont {Austin~J.}\ \bibnamefont
  {Minnich}},\ }\bibfield  {title} {\enquote {\bibinfo {title}
  {Measurement-induced entanglement phase transition on a superconducting
  quantum processor with mid-circuit readout},}\ }\href {\doibase
  10.1038/s41567-023-02076-6} {\bibfield  {journal} {\bibinfo  {journal}
  {Nature Physics}\ }\textbf {\bibinfo {volume} {19}},\ \bibinfo {pages}
  {1314--1319} (\bibinfo {year} {2023})}\BibitemShut {NoStop}%
\bibitem [{\citenamefont {Ippoliti}\ and\ \citenamefont
  {Khemani}(2021)}]{ippoliti_postselection-free_2021}%
  \BibitemOpen
  \bibfield  {author} {\bibinfo {author} {\bibfnamefont {Matteo}\ \bibnamefont
  {Ippoliti}}\ and\ \bibinfo {author} {\bibfnamefont {Vedika}\ \bibnamefont
  {Khemani}},\ }\bibfield  {title} {\enquote {\bibinfo {title}
  {Postselection-{Free} {Entanglement} {Dynamics} via {Spacetime} {Duality}},}\
  }\href {\doibase 10.1103/PhysRevLett.126.060501} {\bibfield  {journal}
  {\bibinfo  {journal} {Physical Review Letters}\ }\textbf {\bibinfo {volume}
  {126}},\ \bibinfo {pages} {060501} (\bibinfo {year} {2021})}\BibitemShut
  {NoStop}%
\bibitem [{\citenamefont {Ippoliti}\ \emph {et~al.}(2022)\citenamefont
  {Ippoliti}, \citenamefont {Rakovszky},\ and\ \citenamefont
  {Khemani}}]{ippoliti_fractal_2022}%
  \BibitemOpen
  \bibfield  {author} {\bibinfo {author} {\bibfnamefont {Matteo}\ \bibnamefont
  {Ippoliti}}, \bibinfo {author} {\bibfnamefont {Tibor}\ \bibnamefont
  {Rakovszky}}, \ and\ \bibinfo {author} {\bibfnamefont {Vedika}\ \bibnamefont
  {Khemani}},\ }\bibfield  {title} {\enquote {\bibinfo {title} {Fractal,
  {Logarithmic}, and {Volume}-{Law} {Entangled} {Nonthermal} {Steady} {States}
  via {Spacetime} {Duality}},}\ }\href {\doibase 10.1103/PhysRevX.12.011045}
  {\bibfield  {journal} {\bibinfo  {journal} {Physical Review X}\ }\textbf
  {\bibinfo {volume} {12}},\ \bibinfo {pages} {011045} (\bibinfo {year}
  {2022})}\BibitemShut {NoStop}%
\bibitem [{\citenamefont {Lu}\ and\ \citenamefont
  {Grover}(2021)}]{lu_spacetime_2021}%
  \BibitemOpen
  \bibfield  {author} {\bibinfo {author} {\bibfnamefont {Tsung-Cheng}\
  \bibnamefont {Lu}}\ and\ \bibinfo {author} {\bibfnamefont {Tarun}\
  \bibnamefont {Grover}},\ }\bibfield  {title} {\enquote {\bibinfo {title}
  {Spacetime duality between localization transitions and measurement-induced
  transitions},}\ }\href {\doibase 10.1103/PRXQuantum.2.040319} {\bibfield
  {journal} {\bibinfo  {journal} {PRX Quantum}\ }\textbf {\bibinfo {volume}
  {2}},\ \bibinfo {pages} {040319} (\bibinfo {year} {2021})}\BibitemShut
  {NoStop}%
\bibitem [{\citenamefont {Dehghani}\ \emph {et~al.}(2023)\citenamefont
  {Dehghani}, \citenamefont {Lavasani}, \citenamefont {Hafezi},\ and\
  \citenamefont {Gullans}}]{dehghani_neural-network_2023}%
  \BibitemOpen
  \bibfield  {author} {\bibinfo {author} {\bibfnamefont {Hossein}\ \bibnamefont
  {Dehghani}}, \bibinfo {author} {\bibfnamefont {Ali}\ \bibnamefont
  {Lavasani}}, \bibinfo {author} {\bibfnamefont {Mohammad}\ \bibnamefont
  {Hafezi}}, \ and\ \bibinfo {author} {\bibfnamefont {Michael~J.}\ \bibnamefont
  {Gullans}},\ }\bibfield  {title} {\enquote {\bibinfo {title} {Neural-network
  decoders for measurement induced phase transitions},}\ }\href {\doibase
  10.1038/s41467-023-37902-1} {\bibfield  {journal} {\bibinfo  {journal}
  {Nature Communications}\ }\textbf {\bibinfo {volume} {14}},\ \bibinfo {pages}
  {2918} (\bibinfo {year} {2023})}\BibitemShut {NoStop}%
\bibitem [{\citenamefont {Garratt}\ \emph {et~al.}(2023)\citenamefont
  {Garratt}, \citenamefont {Weinstein},\ and\ \citenamefont
  {Altman}}]{garratt_measurements_2023}%
  \BibitemOpen
  \bibfield  {author} {\bibinfo {author} {\bibfnamefont {Samuel~J.}\
  \bibnamefont {Garratt}}, \bibinfo {author} {\bibfnamefont {Zack}\
  \bibnamefont {Weinstein}}, \ and\ \bibinfo {author} {\bibfnamefont {Ehud}\
  \bibnamefont {Altman}},\ }\bibfield  {title} {\enquote {\bibinfo {title}
  {Measurements {Conspire} {Nonlocally} to {Restructure} {Critical} {Quantum}
  {States}},}\ }\href {\doibase 10.1103/PhysRevX.13.021026} {\bibfield
  {journal} {\bibinfo  {journal} {Physical Review X}\ }\textbf {\bibinfo
  {volume} {13}},\ \bibinfo {pages} {021026} (\bibinfo {year}
  {2023})}\BibitemShut {NoStop}%
\bibitem [{\citenamefont {Garratt}\ and\ \citenamefont
  {Altman}(2023)}]{garratt_probing_2023}%
  \BibitemOpen
  \bibfield  {author} {\bibinfo {author} {\bibfnamefont {Samuel~J.}\
  \bibnamefont {Garratt}}\ and\ \bibinfo {author} {\bibfnamefont {Ehud}\
  \bibnamefont {Altman}},\ }\bibfield  {title} {\enquote {\bibinfo {title}
  {Probing post-measurement entanglement without post-selection},}\ }\href
  {\doibase 10.48550/arXiv.2305.20092} {\  (\bibinfo {year} {2023}),\
  10.48550/arXiv.2305.20092}\BibitemShut {NoStop}%
\bibitem [{\citenamefont {Huang}(2022)}]{huang_learning_2022}%
  \BibitemOpen
  \bibfield  {author} {\bibinfo {author} {\bibfnamefont {Hsin-Yuan}\
  \bibnamefont {Huang}},\ }\bibfield  {title} {\enquote {\bibinfo {title}
  {Learning quantum states from their classical shadows},}\ }\href {\doibase
  10.1038/s42254-021-00411-5} {\bibfield  {journal} {\bibinfo  {journal}
  {Nature Reviews Physics}\ }\textbf {\bibinfo {volume} {4}},\ \bibinfo {pages}
  {81--81} (\bibinfo {year} {2022})}\BibitemShut {NoStop}%
\bibitem [{\citenamefont {Struchalin}\ \emph {et~al.}(2021)\citenamefont
  {Struchalin}, \citenamefont {Zagorovskii}, \citenamefont {Kovlakov},
  \citenamefont {Straupe},\ and\ \citenamefont
  {Kulik}}]{struchalin_experimental_2021}%
  \BibitemOpen
  \bibfield  {author} {\bibinfo {author} {\bibfnamefont {G.I.}\ \bibnamefont
  {Struchalin}}, \bibinfo {author} {\bibfnamefont {Ya.~A.}\ \bibnamefont
  {Zagorovskii}}, \bibinfo {author} {\bibfnamefont {E.V.}\ \bibnamefont
  {Kovlakov}}, \bibinfo {author} {\bibfnamefont {S.S.}\ \bibnamefont
  {Straupe}}, \ and\ \bibinfo {author} {\bibfnamefont {S.P.}\ \bibnamefont
  {Kulik}},\ }\bibfield  {title} {\enquote {\bibinfo {title} {Experimental
  {Estimation} of {Quantum} {State} {Properties} from {Classical} {Shadows}},}\
  }\href {\doibase 10.1103/PRXQuantum.2.010307} {\bibfield  {journal} {\bibinfo
   {journal} {PRX Quantum}\ }\textbf {\bibinfo {volume} {2}},\ \bibinfo {pages}
  {010307} (\bibinfo {year} {2021})}\BibitemShut {NoStop}%
\bibitem [{\citenamefont {Chen}\ \emph {et~al.}(2021)\citenamefont {Chen},
  \citenamefont {Yu}, \citenamefont {Zeng},\ and\ \citenamefont
  {Flammia}}]{chen_robust_2021}%
  \BibitemOpen
  \bibfield  {author} {\bibinfo {author} {\bibfnamefont {Senrui}\ \bibnamefont
  {Chen}}, \bibinfo {author} {\bibfnamefont {Wenjun}\ \bibnamefont {Yu}},
  \bibinfo {author} {\bibfnamefont {Pei}\ \bibnamefont {Zeng}}, \ and\ \bibinfo
  {author} {\bibfnamefont {Steven~T.}\ \bibnamefont {Flammia}},\ }\bibfield
  {title} {\enquote {\bibinfo {title} {Robust {Shadow} {Estimation}},}\ }\href
  {\doibase 10.1103/PRXQuantum.2.030348} {\bibfield  {journal} {\bibinfo
  {journal} {PRX Quantum}\ }\textbf {\bibinfo {volume} {2}},\ \bibinfo {pages}
  {030348} (\bibinfo {year} {2021})}\BibitemShut {NoStop}%
\bibitem [{\citenamefont {Koh}\ and\ \citenamefont
  {Grewal}(2022)}]{koh_classical_2022}%
  \BibitemOpen
  \bibfield  {author} {\bibinfo {author} {\bibfnamefont {Dax~Enshan}\
  \bibnamefont {Koh}}\ and\ \bibinfo {author} {\bibfnamefont {Sabee}\
  \bibnamefont {Grewal}},\ }\bibfield  {title} {\enquote {\bibinfo {title}
  {Classical {Shadows} {With} {Noise}},}\ }\href {\doibase
  10.22331/q-2022-08-16-776} {\bibfield  {journal} {\bibinfo  {journal}
  {Quantum}\ }\textbf {\bibinfo {volume} {6}},\ \bibinfo {pages} {776}
  (\bibinfo {year} {2022})}\BibitemShut {NoStop}%
\bibitem [{\citenamefont {Huang}\ \emph {et~al.}(2021)\citenamefont {Huang},
  \citenamefont {Kueng},\ and\ \citenamefont
  {Preskill}}]{huang_efficient_2021}%
  \BibitemOpen
  \bibfield  {author} {\bibinfo {author} {\bibfnamefont {Hsin-Yuan}\
  \bibnamefont {Huang}}, \bibinfo {author} {\bibfnamefont {Richard}\
  \bibnamefont {Kueng}}, \ and\ \bibinfo {author} {\bibfnamefont {John}\
  \bibnamefont {Preskill}},\ }\bibfield  {title} {\enquote {\bibinfo {title}
  {Efficient {Estimation} of {Pauli} {Observables} by {Derandomization}},}\
  }\href {\doibase 10.1103/PhysRevLett.127.030503} {\bibfield  {journal}
  {\bibinfo  {journal} {Physical Review Letters}\ }\textbf {\bibinfo {volume}
  {127}},\ \bibinfo {pages} {030503} (\bibinfo {year} {2021})}\BibitemShut
  {NoStop}%
\bibitem [{\citenamefont {Acharya}\ \emph {et~al.}(2021)\citenamefont
  {Acharya}, \citenamefont {Saha},\ and\ \citenamefont
  {Sengupta}}]{acharya_shadow_2021}%
  \BibitemOpen
  \bibfield  {author} {\bibinfo {author} {\bibfnamefont {Atithi}\ \bibnamefont
  {Acharya}}, \bibinfo {author} {\bibfnamefont {Siddhartha}\ \bibnamefont
  {Saha}}, \ and\ \bibinfo {author} {\bibfnamefont {Anirvan~M.}\ \bibnamefont
  {Sengupta}},\ }\bibfield  {title} {\enquote {\bibinfo {title} {Shadow
  tomography based on informationally complete positive operator-valued
  measure},}\ }\href {\doibase 10.1103/PhysRevA.104.052418} {\bibfield
  {journal} {\bibinfo  {journal} {Physical Review A}\ }\textbf {\bibinfo
  {volume} {104}},\ \bibinfo {pages} {052418} (\bibinfo {year}
  {2021})}\BibitemShut {NoStop}%
\bibitem [{\citenamefont {Nguyen}\ \emph {et~al.}(2022)\citenamefont {Nguyen},
  \citenamefont {Bonsel}, \citenamefont {Steinberg},\ and\ \citenamefont
  {Guhne}}]{nguyen_optimizing_2022}%
  \BibitemOpen
  \bibfield  {author} {\bibinfo {author} {\bibfnamefont {H.~Chau}\ \bibnamefont
  {Nguyen}}, \bibinfo {author} {\bibfnamefont {Jan~Lennart}\ \bibnamefont
  {Bonsel}}, \bibinfo {author} {\bibfnamefont {Jonathan}\ \bibnamefont
  {Steinberg}}, \ and\ \bibinfo {author} {\bibfnamefont {Otfried}\ \bibnamefont
  {Guhne}},\ }\bibfield  {title} {\enquote {\bibinfo {title} {Optimizing
  {Shadow} {Tomography} with {Generalized} {Measurements}},}\ }\href {\doibase
  10.1103/PhysRevLett.129.220502} {\bibfield  {journal} {\bibinfo  {journal}
  {Physical Review Letters}\ }\textbf {\bibinfo {volume} {129}},\ \bibinfo
  {pages} {220502} (\bibinfo {year} {2022})}\BibitemShut {NoStop}%
\bibitem [{\citenamefont {Wan}\ \emph {et~al.}(2023)\citenamefont {Wan},
  \citenamefont {Huggins}, \citenamefont {Lee},\ and\ \citenamefont
  {Babbush}}]{wan_matchgate_2023}%
  \BibitemOpen
  \bibfield  {author} {\bibinfo {author} {\bibfnamefont {Kianna}\ \bibnamefont
  {Wan}}, \bibinfo {author} {\bibfnamefont {William~J.}\ \bibnamefont
  {Huggins}}, \bibinfo {author} {\bibfnamefont {Joonho}\ \bibnamefont {Lee}}, \
  and\ \bibinfo {author} {\bibfnamefont {Ryan}\ \bibnamefont {Babbush}},\
  }\bibfield  {title} {\enquote {\bibinfo {title} {Matchgate {Shadows} for
  {Fermionic} {Quantum} {Simulation}},}\ }\href {\doibase
  10.1007/s00220-023-04844-0} {\bibfield  {journal} {\bibinfo  {journal}
  {Communications in Mathematical Physics}\ }\textbf {\bibinfo {volume}
  {404}},\ \bibinfo {pages} {629--700} (\bibinfo {year} {2023})}\BibitemShut
  {NoStop}%
\bibitem [{\citenamefont {Hu}\ \emph {et~al.}(2023)\citenamefont {Hu},
  \citenamefont {Choi},\ and\ \citenamefont {You}}]{hu_classical_2023}%
  \BibitemOpen
  \bibfield  {author} {\bibinfo {author} {\bibfnamefont {Hong-Ye}\ \bibnamefont
  {Hu}}, \bibinfo {author} {\bibfnamefont {Soonwon}\ \bibnamefont {Choi}}, \
  and\ \bibinfo {author} {\bibfnamefont {Yi-Zhuang}\ \bibnamefont {You}},\
  }\bibfield  {title} {\enquote {\bibinfo {title} {Classical shadow tomography
  with locally scrambled quantum dynamics},}\ }\href {\doibase
  10.1103/PhysRevResearch.5.023027} {\bibfield  {journal} {\bibinfo  {journal}
  {Physical Review Research}\ }\textbf {\bibinfo {volume} {5}},\ \bibinfo
  {pages} {023027} (\bibinfo {year} {2023})}\BibitemShut {NoStop}%
\bibitem [{\citenamefont {Akhtar}\ \emph
  {et~al.}(2023{\natexlab{a}})\citenamefont {Akhtar}, \citenamefont {Hu},\ and\
  \citenamefont {You}}]{akhtar_scalable_2023}%
  \BibitemOpen
  \bibfield  {author} {\bibinfo {author} {\bibfnamefont {Ahmed~A.}\
  \bibnamefont {Akhtar}}, \bibinfo {author} {\bibfnamefont {Hong-Ye}\
  \bibnamefont {Hu}}, \ and\ \bibinfo {author} {\bibfnamefont {Yi-Zhuang}\
  \bibnamefont {You}},\ }\bibfield  {title} {\enquote {\bibinfo {title}
  {Scalable and {Flexible} {Classical} {Shadow} {Tomography} with {Tensor}
  {Networks}},}\ }\href {\doibase 10.22331/q-2023-06-01-1026} {\bibfield
  {journal} {\bibinfo  {journal} {Quantum}\ }\textbf {\bibinfo {volume} {7}},\
  \bibinfo {pages} {1026} (\bibinfo {year} {2023}{\natexlab{a}})}\BibitemShut
  {NoStop}%
\bibitem [{\citenamefont {Bertoni}\ \emph {et~al.}(2022)\citenamefont
  {Bertoni}, \citenamefont {Haferkamp}, \citenamefont {Hinsche}, \citenamefont
  {Ioannou}, \citenamefont {Eisert},\ and\ \citenamefont
  {Pashayan}}]{bertoni_shallow_2022}%
  \BibitemOpen
  \bibfield  {author} {\bibinfo {author} {\bibfnamefont {Christian}\
  \bibnamefont {Bertoni}}, \bibinfo {author} {\bibfnamefont {Jonas}\
  \bibnamefont {Haferkamp}}, \bibinfo {author} {\bibfnamefont {Marcel}\
  \bibnamefont {Hinsche}}, \bibinfo {author} {\bibfnamefont {Marios}\
  \bibnamefont {Ioannou}}, \bibinfo {author} {\bibfnamefont {Jens}\
  \bibnamefont {Eisert}}, \ and\ \bibinfo {author} {\bibfnamefont {Hakop}\
  \bibnamefont {Pashayan}},\ }\bibfield  {title} {\enquote {\bibinfo {title}
  {Shallow shadows: {Expectation} estimation using low-depth random {Clifford}
  circuits},}\ }\href {\doibase 10.48550/arXiv.2209.12924} {\  (\bibinfo {year}
  {2022}),\ 10.48550/arXiv.2209.12924}\BibitemShut {NoStop}%
\bibitem [{\citenamefont {Arienzo}\ \emph {et~al.}(2022)\citenamefont
  {Arienzo}, \citenamefont {Heinrich}, \citenamefont {Roth},\ and\
  \citenamefont {Kliesch}}]{arienzo_closed-form_2022}%
  \BibitemOpen
  \bibfield  {author} {\bibinfo {author} {\bibfnamefont {Mirko}\ \bibnamefont
  {Arienzo}}, \bibinfo {author} {\bibfnamefont {Markus}\ \bibnamefont
  {Heinrich}}, \bibinfo {author} {\bibfnamefont {Ingo}\ \bibnamefont {Roth}}, \
  and\ \bibinfo {author} {\bibfnamefont {Martin}\ \bibnamefont {Kliesch}},\
  }\bibfield  {title} {\enquote {\bibinfo {title} {Closed-form analytic
  expressions for shadow estimation with brickwork circuits},}\ }\href
  {\doibase 10.48550/arXiv.2211.09835} {\  (\bibinfo {year} {2022}),\
  10.48550/arXiv.2211.09835}\BibitemShut {NoStop}%
\bibitem [{\citenamefont {Ippoliti}\ \emph {et~al.}(2023)\citenamefont
  {Ippoliti}, \citenamefont {Li}, \citenamefont {Rakovszky},\ and\
  \citenamefont {Khemani}}]{ippoliti_operator_2023}%
  \BibitemOpen
  \bibfield  {author} {\bibinfo {author} {\bibfnamefont {Matteo}\ \bibnamefont
  {Ippoliti}}, \bibinfo {author} {\bibfnamefont {Yaodong}\ \bibnamefont {Li}},
  \bibinfo {author} {\bibfnamefont {Tibor}\ \bibnamefont {Rakovszky}}, \ and\
  \bibinfo {author} {\bibfnamefont {Vedika}\ \bibnamefont {Khemani}},\
  }\bibfield  {title} {\enquote {\bibinfo {title} {Operator {Relaxation} and
  the {Optimal} {Depth} of {Classical} {Shadows}},}\ }\href {\doibase
  10.1103/PhysRevLett.130.230403} {\bibfield  {journal} {\bibinfo  {journal}
  {Physical Review Letters}\ }\textbf {\bibinfo {volume} {130}},\ \bibinfo
  {pages} {230403} (\bibinfo {year} {2023})}\BibitemShut {NoStop}%
\bibitem [{\citenamefont {Ippoliti}(2023)}]{ippoliti_classical_2023}%
  \BibitemOpen
  \bibfield  {author} {\bibinfo {author} {\bibfnamefont {Matteo}\ \bibnamefont
  {Ippoliti}},\ }\bibfield  {title} {\enquote {\bibinfo {title} {Classical
  shadows based on locally-entangled measurements},}\ }\href {\doibase
  10.48550/arXiv.2305.10723} {\  (\bibinfo {year} {2023}),\
  10.48550/arXiv.2305.10723}\BibitemShut {NoStop}%
\bibitem [{\citenamefont {Tran}\ \emph {et~al.}(2023)\citenamefont {Tran},
  \citenamefont {Mark}, \citenamefont {Ho},\ and\ \citenamefont
  {Choi}}]{tran_measuring_2023}%
  \BibitemOpen
  \bibfield  {author} {\bibinfo {author} {\bibfnamefont {Minh~C.}\ \bibnamefont
  {Tran}}, \bibinfo {author} {\bibfnamefont {Daniel~K.}\ \bibnamefont {Mark}},
  \bibinfo {author} {\bibfnamefont {Wen~Wei}\ \bibnamefont {Ho}}, \ and\
  \bibinfo {author} {\bibfnamefont {Soonwon}\ \bibnamefont {Choi}},\ }\bibfield
   {title} {\enquote {\bibinfo {title} {Measuring {Arbitrary} {Physical}
  {Properties} in {Analog} {Quantum} {Simulation}},}\ }\href {\doibase
  10.1103/PhysRevX.13.011049} {\bibfield  {journal} {\bibinfo  {journal}
  {Physical Review X}\ }\textbf {\bibinfo {volume} {13}},\ \bibinfo {pages}
  {011049} (\bibinfo {year} {2023})}\BibitemShut {NoStop}%
\bibitem [{\citenamefont {Shivam}\ \emph {et~al.}(2023)\citenamefont {Shivam},
  \citenamefont {von Keyserlingk},\ and\ \citenamefont
  {Sondhi}}]{shivam_classical_2023}%
  \BibitemOpen
  \bibfield  {author} {\bibinfo {author} {\bibfnamefont {Saumya}\ \bibnamefont
  {Shivam}}, \bibinfo {author} {\bibfnamefont {Curt~W.}\ \bibnamefont {von
  Keyserlingk}}, \ and\ \bibinfo {author} {\bibfnamefont {Shivaji~L.}\
  \bibnamefont {Sondhi}},\ }\bibfield  {title} {\enquote {\bibinfo {title} {On
  classical and hybrid shadows of quantum states},}\ }\href {\doibase
  10.21468/SciPostPhys.14.5.094} {\bibfield  {journal} {\bibinfo  {journal}
  {SciPost Physics}\ }\textbf {\bibinfo {volume} {14}},\ \bibinfo {pages} {094}
  (\bibinfo {year} {2023})}\BibitemShut {NoStop}%
\bibitem [{\citenamefont {McGinley}\ and\ \citenamefont
  {Fava}(2023)}]{mcginley_shadow_2023}%
  \BibitemOpen
  \bibfield  {author} {\bibinfo {author} {\bibfnamefont {Max}\ \bibnamefont
  {McGinley}}\ and\ \bibinfo {author} {\bibfnamefont {Michele}\ \bibnamefont
  {Fava}},\ }\bibfield  {title} {\enquote {\bibinfo {title} {Shadow
  {Tomography} from {Emergent} {State} {Designs} in {Analog} {Quantum}
  {Simulators}},}\ }\href {\doibase 10.1103/PhysRevLett.131.160601} {\bibfield
  {journal} {\bibinfo  {journal} {Physical Review Letters}\ }\textbf {\bibinfo
  {volume} {131}},\ \bibinfo {pages} {160601} (\bibinfo {year}
  {2023})}\BibitemShut {NoStop}%
\bibitem [{\citenamefont {Dall'Arno}(2014)}]{dallarno_accessible_2014}%
  \BibitemOpen
  \bibfield  {author} {\bibinfo {author} {\bibfnamefont {Michele}\ \bibnamefont
  {Dall'Arno}},\ }\bibfield  {title} {\enquote {\bibinfo {title} {Accessible
  information and informational power of quantum 2-designs},}\ }\href {\doibase
  10.1103/PhysRevA.90.052311} {\bibfield  {journal} {\bibinfo  {journal}
  {Physical Review A}\ }\textbf {\bibinfo {volume} {90}},\ \bibinfo {pages}
  {052311} (\bibinfo {year} {2014})}\BibitemShut {NoStop}%
\bibitem [{\citenamefont {Dall'Arno}\ \emph {et~al.}(2014)\citenamefont
  {Dall'Arno}, \citenamefont {Buscemi},\ and\ \citenamefont
  {Ozawa}}]{dallarno_tight_2014}%
  \BibitemOpen
  \bibfield  {author} {\bibinfo {author} {\bibfnamefont {Michele}\ \bibnamefont
  {Dall'Arno}}, \bibinfo {author} {\bibfnamefont {Francesco}\ \bibnamefont
  {Buscemi}}, \ and\ \bibinfo {author} {\bibfnamefont {Masanao}\ \bibnamefont
  {Ozawa}},\ }\bibfield  {title} {\enquote {\bibinfo {title} {Tight bounds on
  accessible information and informational power},}\ }\href {\doibase
  10.1088/1751-8113/47/23/235302} {\bibfield  {journal} {\bibinfo  {journal}
  {Journal of Physics A: Mathematical and Theoretical}\ }\textbf {\bibinfo
  {volume} {47}},\ \bibinfo {pages} {235302} (\bibinfo {year}
  {2014})}\BibitemShut {NoStop}%
\bibitem [{\citenamefont {Hausladen}\ and\ \citenamefont
  {Wootters}(1994)}]{hausladen_pretty_1994}%
  \BibitemOpen
  \bibfield  {author} {\bibinfo {author} {\bibfnamefont {Paul}\ \bibnamefont
  {Hausladen}}\ and\ \bibinfo {author} {\bibfnamefont {William~K.}\
  \bibnamefont {Wootters}},\ }\bibfield  {title} {\enquote {\bibinfo {title} {A
  '{Pretty} {Good}' {Measurement} for {Distinguishing} {Quantum} {States}},}\
  }\href {\doibase 10.1080/09500349414552221} {\bibfield  {journal} {\bibinfo
  {journal} {Journal of Modern Optics}\ }\textbf {\bibinfo {volume} {41}},\
  \bibinfo {pages} {2385--2390} (\bibinfo {year} {1994})}\BibitemShut {NoStop}%
\bibitem [{\citenamefont {Jozsa}\ \emph {et~al.}(1994)\citenamefont {Jozsa},
  \citenamefont {Robb},\ and\ \citenamefont {Wootters}}]{jozsa_lower_1994}%
  \BibitemOpen
  \bibfield  {author} {\bibinfo {author} {\bibfnamefont {Richard}\ \bibnamefont
  {Jozsa}}, \bibinfo {author} {\bibfnamefont {Daniel}\ \bibnamefont {Robb}}, \
  and\ \bibinfo {author} {\bibfnamefont {William~K.}\ \bibnamefont
  {Wootters}},\ }\bibfield  {title} {\enquote {\bibinfo {title} {Lower bound
  for accessible information in quantum mechanics},}\ }\href {\doibase
  10.1103/PhysRevA.49.668} {\bibfield  {journal} {\bibinfo  {journal} {Physical
  Review A}\ }\textbf {\bibinfo {volume} {49}},\ \bibinfo {pages} {668--677}
  (\bibinfo {year} {1994})}\BibitemShut {NoStop}%
\bibitem [{\citenamefont {Petz}(1986)}]{petz_sufficient_1986}%
  \BibitemOpen
  \bibfield  {author} {\bibinfo {author} {\bibfnamefont {Denes}\ \bibnamefont
  {Petz}},\ }\bibfield  {title} {\enquote {\bibinfo {title} {Sufficient
  subalgebras and the relative entropy of states of a von {Neumann} algebra},}\
  }\href {\doibase 10.1007/BF01212345} {\bibfield  {journal} {\bibinfo
  {journal} {Communications in Mathematical Physics}\ }\textbf {\bibinfo
  {volume} {105}},\ \bibinfo {pages} {123--131} (\bibinfo {year}
  {1986})}\BibitemShut {NoStop}%
\bibitem [{\citenamefont {Barnum}\ and\ \citenamefont
  {Knill}(2002)}]{barnum_reversing_2002}%
  \BibitemOpen
  \bibfield  {author} {\bibinfo {author} {\bibfnamefont {H.}~\bibnamefont
  {Barnum}}\ and\ \bibinfo {author} {\bibfnamefont {E.}~\bibnamefont {Knill}},\
  }\bibfield  {title} {\enquote {\bibinfo {title} {Reversing quantum dynamics
  with near-optimal quantum and classical fidelity},}\ }\href {\doibase
  10.1063/1.1459754} {\bibfield  {journal} {\bibinfo  {journal} {Journal of
  Mathematical Physics}\ }\textbf {\bibinfo {volume} {43}},\ \bibinfo {pages}
  {2097--2106} (\bibinfo {year} {2002})}\BibitemShut {NoStop}%
\bibitem [{\citenamefont {Wilde}(2015)}]{wilde_recoverability_2015}%
  \BibitemOpen
  \bibfield  {author} {\bibinfo {author} {\bibfnamefont {Mark~M.}\ \bibnamefont
  {Wilde}},\ }\bibfield  {title} {\enquote {\bibinfo {title} {Recoverability in
  quantum information theory},}\ }\href {\doibase 10.1098/rspa.2015.0338}
  {\bibfield  {journal} {\bibinfo  {journal} {Proceedings of the Royal Society
  A: Mathematical, Physical and Engineering Sciences}\ }\textbf {\bibinfo
  {volume} {471}},\ \bibinfo {pages} {20150338} (\bibinfo {year}
  {2015})}\BibitemShut {NoStop}%
\bibitem [{\citenamefont {Penington}\ \emph {et~al.}(2022)\citenamefont
  {Penington}, \citenamefont {Shenker}, \citenamefont {Stanford},\ and\
  \citenamefont {Yang}}]{penington_replica_2022}%
  \BibitemOpen
  \bibfield  {author} {\bibinfo {author} {\bibfnamefont {Geoff}\ \bibnamefont
  {Penington}}, \bibinfo {author} {\bibfnamefont {Stephen~H.}\ \bibnamefont
  {Shenker}}, \bibinfo {author} {\bibfnamefont {Douglas}\ \bibnamefont
  {Stanford}}, \ and\ \bibinfo {author} {\bibfnamefont {Zhenbin}\ \bibnamefont
  {Yang}},\ }\bibfield  {title} {\enquote {\bibinfo {title} {Replica wormholes
  and the black hole interior},}\ }\href {\doibase 10.1007/JHEP03(2022)205}
  {\bibfield  {journal} {\bibinfo  {journal} {Journal of High Energy Physics}\
  }\textbf {\bibinfo {volume} {2022}},\ \bibinfo {pages} {205} (\bibinfo {year}
  {2022})}\BibitemShut {NoStop}%
\bibitem [{\citenamefont {Vasseur}\ \emph {et~al.}(2019)\citenamefont
  {Vasseur}, \citenamefont {Potter}, \citenamefont {You},\ and\ \citenamefont
  {Ludwig}}]{vasseur_entanglement_2019}%
  \BibitemOpen
  \bibfield  {author} {\bibinfo {author} {\bibfnamefont {Romain}\ \bibnamefont
  {Vasseur}}, \bibinfo {author} {\bibfnamefont {Andrew~C.}\ \bibnamefont
  {Potter}}, \bibinfo {author} {\bibfnamefont {Yi-Zhuang}\ \bibnamefont {You}},
  \ and\ \bibinfo {author} {\bibfnamefont {Andreas W.~W.}\ \bibnamefont
  {Ludwig}},\ }\bibfield  {title} {\enquote {\bibinfo {title} {Entanglement
  transitions from holographic random tensor networks},}\ }\href {\doibase
  10.1103/PhysRevB.100.134203} {\bibfield  {journal} {\bibinfo  {journal}
  {Physical Review B}\ }\textbf {\bibinfo {volume} {100}},\ \bibinfo {pages}
  {134203} (\bibinfo {year} {2019})}\BibitemShut {NoStop}%
\bibitem [{\citenamefont {Jian}\ \emph {et~al.}(2020)\citenamefont {Jian},
  \citenamefont {You}, \citenamefont {Vasseur},\ and\ \citenamefont
  {Ludwig}}]{jian_measurement-induced_2020}%
  \BibitemOpen
  \bibfield  {author} {\bibinfo {author} {\bibfnamefont {Chao-Ming}\
  \bibnamefont {Jian}}, \bibinfo {author} {\bibfnamefont {Yi-Zhuang}\
  \bibnamefont {You}}, \bibinfo {author} {\bibfnamefont {Romain}\ \bibnamefont
  {Vasseur}}, \ and\ \bibinfo {author} {\bibfnamefont {Andreas W.~W.}\
  \bibnamefont {Ludwig}},\ }\bibfield  {title} {\enquote {\bibinfo {title}
  {Measurement-induced criticality in random quantum circuits},}\ }\href
  {\doibase 10.1103/PhysRevB.101.104302} {\bibfield  {journal} {\bibinfo
  {journal} {Physical Review B}\ }\textbf {\bibinfo {volume} {101}},\ \bibinfo
  {pages} {104302} (\bibinfo {year} {2020})}\BibitemShut {NoStop}%
\bibitem [{\citenamefont {Bu}\ \emph {et~al.}(2022)\citenamefont {Bu},
  \citenamefont {Koh}, \citenamefont {Garcia},\ and\ \citenamefont
  {Jaffe}}]{bu_classical_2022}%
  \BibitemOpen
  \bibfield  {author} {\bibinfo {author} {\bibfnamefont {Kaifeng}\ \bibnamefont
  {Bu}}, \bibinfo {author} {\bibfnamefont {Dax~Enshan}\ \bibnamefont {Koh}},
  \bibinfo {author} {\bibfnamefont {Roy~J.}\ \bibnamefont {Garcia}}, \ and\
  \bibinfo {author} {\bibfnamefont {Arthur}\ \bibnamefont {Jaffe}},\ }\bibfield
   {title} {\enquote {\bibinfo {title} {Classical shadows with
  {Pauli}-invariant unitary ensembles},}\ }\href {\doibase
  10.48550/arXiv.2202.03272} {\  (\bibinfo {year} {2022}),\
  10.48550/arXiv.2202.03272}\BibitemShut {NoStop}%
\bibitem [{\citenamefont {Barratt}\ \emph
  {et~al.}(2022{\natexlab{b}})\citenamefont {Barratt}, \citenamefont {Agrawal},
  \citenamefont {Gopalakrishnan}, \citenamefont {Huse}, \citenamefont
  {Vasseur},\ and\ \citenamefont {Potter}}]{barratt_field_2022}%
  \BibitemOpen
  \bibfield  {author} {\bibinfo {author} {\bibfnamefont {Fergus}\ \bibnamefont
  {Barratt}}, \bibinfo {author} {\bibfnamefont {Utkarsh}\ \bibnamefont
  {Agrawal}}, \bibinfo {author} {\bibfnamefont {Sarang}\ \bibnamefont
  {Gopalakrishnan}}, \bibinfo {author} {\bibfnamefont {David~A.}\ \bibnamefont
  {Huse}}, \bibinfo {author} {\bibfnamefont {Romain}\ \bibnamefont {Vasseur}},
  \ and\ \bibinfo {author} {\bibfnamefont {Andrew~C.}\ \bibnamefont {Potter}},\
  }\bibfield  {title} {\enquote {\bibinfo {title} {Field {Theory} of {Charge}
  {Sharpening} in {Symmetric} {Monitored} {Quantum} {Circuits}},}\ }\href
  {\doibase 10.1103/PhysRevLett.129.120604} {\bibfield  {journal} {\bibinfo
  {journal} {Physical Review Letters}\ }\textbf {\bibinfo {volume} {129}},\
  \bibinfo {pages} {120604} (\bibinfo {year} {2022}{\natexlab{b}})}\BibitemShut
  {NoStop}%
\bibitem [{\citenamefont {Buchhold}\ \emph {et~al.}(2021)\citenamefont
  {Buchhold}, \citenamefont {Minoguchi}, \citenamefont {Altland},\ and\
  \citenamefont {Diehl}}]{buchhold_effective_2021}%
  \BibitemOpen
  \bibfield  {author} {\bibinfo {author} {\bibfnamefont {M.}~\bibnamefont
  {Buchhold}}, \bibinfo {author} {\bibfnamefont {Y.}~\bibnamefont {Minoguchi}},
  \bibinfo {author} {\bibfnamefont {A.}~\bibnamefont {Altland}}, \ and\
  \bibinfo {author} {\bibfnamefont {S.}~\bibnamefont {Diehl}},\ }\bibfield
  {title} {\enquote {\bibinfo {title} {Effective {Theory} for the
  {Measurement}-{Induced} {Phase} {Transition} of {Dirac} {Fermions}},}\ }\href
  {\doibase 10.1103/PhysRevX.11.041004} {\bibfield  {journal} {\bibinfo
  {journal} {Physical Review X}\ }\textbf {\bibinfo {volume} {11}},\ \bibinfo
  {pages} {041004} (\bibinfo {year} {2021})}\BibitemShut {NoStop}%
\bibitem [{\citenamefont {Akhtar}\ \emph
  {et~al.}(2023{\natexlab{b}})\citenamefont {Akhtar}, \citenamefont {Hu},\ and\
  \citenamefont {You}}]{akhtar_measurement-induced_2023}%
  \BibitemOpen
  \bibfield  {author} {\bibinfo {author} {\bibfnamefont {Ahmed~A.}\
  \bibnamefont {Akhtar}}, \bibinfo {author} {\bibfnamefont {Hong-Ye}\
  \bibnamefont {Hu}}, \ and\ \bibinfo {author} {\bibfnamefont {Yi-Zhuang}\
  \bibnamefont {You}},\ }\bibfield  {title} {\enquote {\bibinfo {title}
  {Measurement-{Induced} {Criticality} is {Tomographically} {Optimal}},}\
  }\href {\doibase 10.48550/arXiv.2308.01653} {\  (\bibinfo {year}
  {2023}{\natexlab{b}}),\ 10.48550/arXiv.2308.01653}\BibitemShut {NoStop}%
\bibitem [{\citenamefont {Cotler}\ \emph {et~al.}(2023)\citenamefont {Cotler},
  \citenamefont {Mark}, \citenamefont {Huang}, \citenamefont {Hernandez},
  \citenamefont {Choi}, \citenamefont {Shaw}, \citenamefont {Endres},\ and\
  \citenamefont {Choi}}]{cotler_emergent_2023}%
  \BibitemOpen
  \bibfield  {author} {\bibinfo {author} {\bibfnamefont {Jordan~S.}\
  \bibnamefont {Cotler}}, \bibinfo {author} {\bibfnamefont {Daniel~K.}\
  \bibnamefont {Mark}}, \bibinfo {author} {\bibfnamefont {Hsin-Yuan}\
  \bibnamefont {Huang}}, \bibinfo {author} {\bibfnamefont {Felipe}\
  \bibnamefont {Hernandez}}, \bibinfo {author} {\bibfnamefont {Joonhee}\
  \bibnamefont {Choi}}, \bibinfo {author} {\bibfnamefont {Adam~L.}\
  \bibnamefont {Shaw}}, \bibinfo {author} {\bibfnamefont {Manuel}\ \bibnamefont
  {Endres}}, \ and\ \bibinfo {author} {\bibfnamefont {Soonwon}\ \bibnamefont
  {Choi}},\ }\bibfield  {title} {\enquote {\bibinfo {title} {Emergent {Quantum}
  {State} {Designs} from {Individual} {Many}-{Body} {Wave} {Functions}},}\
  }\href {\doibase 10.1103/PRXQuantum.4.010311} {\bibfield  {journal} {\bibinfo
   {journal} {PRX Quantum}\ }\textbf {\bibinfo {volume} {4}},\ \bibinfo {pages}
  {010311} (\bibinfo {year} {2023})}\BibitemShut {NoStop}%
\bibitem [{\citenamefont {Choi}\ \emph {et~al.}(2023)\citenamefont {Choi},
  \citenamefont {Shaw}, \citenamefont {Madjarov}, \citenamefont {Xie},
  \citenamefont {Finkelstein}, \citenamefont {Covey}, \citenamefont {Cotler},
  \citenamefont {Mark} \emph {et~al.}}]{choi_preparing_2023}%
  \BibitemOpen
  \bibfield  {author} {\bibinfo {author} {\bibfnamefont {Joonhee}\ \bibnamefont
  {Choi}}, \bibinfo {author} {\bibfnamefont {Adam~L.}\ \bibnamefont {Shaw}},
  \bibinfo {author} {\bibfnamefont {Ivaylo~S.}\ \bibnamefont {Madjarov}},
  \bibinfo {author} {\bibfnamefont {Xin}\ \bibnamefont {Xie}}, \bibinfo
  {author} {\bibfnamefont {Ran}\ \bibnamefont {Finkelstein}}, \bibinfo {author}
  {\bibfnamefont {Jacob~P.}\ \bibnamefont {Covey}}, \bibinfo {author}
  {\bibfnamefont {Jordan~S.}\ \bibnamefont {Cotler}}, \bibinfo {author}
  {\bibfnamefont {Daniel~K.}\ \bibnamefont {Mark}},  \emph {et~al.},\
  }\bibfield  {title} {\enquote {\bibinfo {title} {Preparing random states and
  benchmarking with many-body quantum chaos},}\ }\href {\doibase
  10.1038/s41586-022-05442-1} {\bibfield  {journal} {\bibinfo  {journal}
  {Nature}\ }\textbf {\bibinfo {volume} {613}},\ \bibinfo {pages} {468--473}
  (\bibinfo {year} {2023})}\BibitemShut {NoStop}%
\bibitem [{\citenamefont {Ippoliti}\ and\ \citenamefont
  {Ho}(2022)}]{ippoliti_solvable_2022}%
  \BibitemOpen
  \bibfield  {author} {\bibinfo {author} {\bibfnamefont {Matteo}\ \bibnamefont
  {Ippoliti}}\ and\ \bibinfo {author} {\bibfnamefont {Wen~Wei}\ \bibnamefont
  {Ho}},\ }\bibfield  {title} {\enquote {\bibinfo {title} {Solvable model of
  deep thermalization with distinct design times},}\ }\href {\doibase
  10.22331/q-2022-12-29-886} {\bibfield  {journal} {\bibinfo  {journal}
  {Quantum}\ }\textbf {\bibinfo {volume} {6}},\ \bibinfo {pages} {886}
  (\bibinfo {year} {2022})}\BibitemShut {NoStop}%
\bibitem [{\citenamefont {Ng}\ and\ \citenamefont
  {Mandayam}(2010)}]{ng_simple_2010}%
  \BibitemOpen
  \bibfield  {author} {\bibinfo {author} {\bibfnamefont {Hui~Khoon}\
  \bibnamefont {Ng}}\ and\ \bibinfo {author} {\bibfnamefont {Prabha}\
  \bibnamefont {Mandayam}},\ }\bibfield  {title} {\enquote {\bibinfo {title}
  {Simple approach to approximate quantum error correction based on the
  transpose channel},}\ }\href {\doibase 10.1103/PhysRevA.81.062342} {\bibfield
   {journal} {\bibinfo  {journal} {Physical Review A}\ }\textbf {\bibinfo
  {volume} {81}},\ \bibinfo {pages} {062342} (\bibinfo {year}
  {2010})}\BibitemShut {NoStop}%
\bibitem [{\citenamefont {Cotler}\ \emph {et~al.}(2019)\citenamefont {Cotler},
  \citenamefont {Hayden}, \citenamefont {Penington}, \citenamefont {Salton},
  \citenamefont {Swingle},\ and\ \citenamefont
  {Walter}}]{cotler_entanglement_2019}%
  \BibitemOpen
  \bibfield  {author} {\bibinfo {author} {\bibfnamefont {Jordan}\ \bibnamefont
  {Cotler}}, \bibinfo {author} {\bibfnamefont {Patrick}\ \bibnamefont
  {Hayden}}, \bibinfo {author} {\bibfnamefont {Geoffrey}\ \bibnamefont
  {Penington}}, \bibinfo {author} {\bibfnamefont {Grant}\ \bibnamefont
  {Salton}}, \bibinfo {author} {\bibfnamefont {Brian}\ \bibnamefont {Swingle}},
  \ and\ \bibinfo {author} {\bibfnamefont {Michael}\ \bibnamefont {Walter}},\
  }\bibfield  {title} {\enquote {\bibinfo {title} {Entanglement {Wedge}
  {Reconstruction} via {Universal} {Recovery} {Channels}},}\ }\href {\doibase
  10.1103/PhysRevX.9.031011} {\bibfield  {journal} {\bibinfo  {journal}
  {Physical Review X}\ }\textbf {\bibinfo {volume} {9}},\ \bibinfo {pages}
  {031011} (\bibinfo {year} {2019})}\BibitemShut {NoStop}%
\bibitem [{\citenamefont {Fuchs}\ and\ \citenamefont
  {Sasaki}(2003)}]{fuchs_squeezing_2003}%
  \BibitemOpen
  \bibfield  {author} {\bibinfo {author} {\bibfnamefont {Christopher~A.}\
  \bibnamefont {Fuchs}}\ and\ \bibinfo {author} {\bibfnamefont {Masahide}\
  \bibnamefont {Sasaki}},\ }\bibfield  {title} {\enquote {\bibinfo {title}
  {Squeezing quantum information through a classical channel: measuring the
  "quantumness" of a set of quantum states},}\ }\href@noop {} {\bibfield
  {journal} {\bibinfo  {journal} {Quantum Information \& Computation}\ }\textbf
  {\bibinfo {volume} {3}},\ \bibinfo {pages} {377--404} (\bibinfo {year}
  {2003})}\BibitemShut {NoStop}%
\bibitem [{\citenamefont {Fuchs}(2004)}]{fuchs_quantumness_2004}%
  \BibitemOpen
  \bibfield  {author} {\bibinfo {author} {\bibfnamefont {Christopher~A.}\
  \bibnamefont {Fuchs}},\ }\bibfield  {title} {\enquote {\bibinfo {title} {On
  the quantumness of a hilbert space},}\ }\href@noop {} {\bibfield  {journal}
  {\bibinfo  {journal} {Quantum Information \& Computation}\ }\textbf {\bibinfo
  {volume} {4}},\ \bibinfo {pages} {467--478} (\bibinfo {year}
  {2004})}\BibitemShut {NoStop}%
\bibitem [{\citenamefont {Kostenberger}(2021)}]{kostenberger_weingarten_2021}%
  \BibitemOpen
  \bibfield  {author} {\bibinfo {author} {\bibfnamefont {Georg}\ \bibnamefont
  {Kostenberger}},\ }\bibfield  {title} {\enquote {\bibinfo {title} {Weingarten
  {Calculus}},}\ }\href {\doibase 10.48550/arXiv.2101.00921} {\  (\bibinfo
  {year} {2021}),\ 10.48550/arXiv.2101.00921}\BibitemShut {NoStop}%
\end{thebibliography}%

\end{document}